\documentclass[%
prb,twocolumn,amsmath,amssymb,aps,superscriptaddress.nofootinbib]{revtex4-1}

\usepackage{graphicx}
\usepackage{dcolumn}
\usepackage{bm}
\usepackage{graphicx}
\usepackage{braket}
\usepackage{xcolor}
\usepackage{comment}
\usepackage{hyperref}
\usepackage{multirow}
\usepackage{tensor}

\newcommand{\vol}{V}
\begin{document}

\title{Viscoelastic response of quantum Hall fluids in a tilted field}

\author{Bendeguz Offertaler}
\affiliation{Department of Physics, Princeton University, Princeton, New Jersey 08544, USA}
\author{Barry Bradlyn}%
\email{bbradlyn@illinois.edu}
\affiliation{Department of Physics and Institute for Condensed Matter Theory, University of Illinois at Urbana-Champaign, Urbana, IL, 61801-3080, USA}%
\date{\today}
\begin{abstract}
In this paper, we examine the viscoelastic properties of integer quantum Hall (IQH) states in a tilted magnetic field. In particular, we explore to what extent the tilted-field system behaves like a two-dimensional electron gas with anisotropic mass in the presence of strain deformations. We first review the Kubo formalism for viscosity in an external magnetic field, paying particular attention to the role of rotational symmetry and contact terms. Next, we compute the conductivity, stress, and viscosity tensors for IQH states in the presence of a tilted field and vertical confining potential. By comparing our results with the recently developed bimetric formalism, we show that, at the level of the contracted Hall viscosity tensor, the mapping between tilted field and effective mass anisotropy holds only if we simultaneously modify the background perpendicular magnetic field; in other words, a simultaneous measurement of the density, contracted Hall viscosity, and Hall conductivity at fixed particle number can distinguish between tilted field and effective mass anisotropy. Additionally, we show that in the presence of a tilted magnetic field, the stress tensor acquires an unusual anisotropic ground state average, leading to anomalous elastic response functions. We develop a formalism for projecting a three-dimensional Hamiltonian with confining potential and magnetic field to a two-dimensional Hamiltonian in order to further address the phenomenology of the tilted-field IQH fluid. We find that the projected fluid couples non-minimally to geometric deformations, indicating the presence of internal geometric degrees of freedom.

\end{abstract}

\maketitle

\section{\label{sec:intro}Introduction}

One of the most striking features of topological phases of matter is the presence of quantized, nondissipative linear response coefficients. The paradigmatic example, the integer or fractional quantized Hall conductance in two dimensions, serves as the defining feature of the (integer or fractional) quantum Hall fluid. The Hall conductance, however, does not uniquely determine the topological properties of a quantum Hall state. Further information about the nature of a quantum Hall state can be determined from its response to \emph{geometric deformations}\cite{1995-AvronSeilerZograf,Levay1995}: the Hall viscosity, defined as the non-dissipative response of the stress tensor of the quantum Hall fluid to a time-varying shear\cite{read2009non,read2011hall,bradlyn2012kubo,Gromov20141,2007-TokatlyVignale,2009-TokatlyVignale-JPhys,wiegmann2014anomalous}, probes the value of the shift $\mathcal{S}$ on the sphere\cite{Haldane1983}. In rotationally invariant systems, the value of the Hall viscosity is quantized in units of the electron density $\rho$,
\begin{equation}
\eta^H=\frac{1}{4}\rho\hbar\mathcal{S},
\end{equation}
 and can be used to distinguish quantum Hall fluids with the same Hall conductance, such as the competing proposals for the $\nu=5/2$ quantum Hall state\cite{read2009non,Zaletel-PhysRevLett.110.236801}.

While the Hall viscosity has served as an excellent analytical and numerical tool, experimental measurement is lacking. Most proposals for experiments involve either direct manipulation of the fluid\cite{1998-Avron} which are hard to carry out for quantum Hall systems, or else rely on the connection between viscosity and conductivity present in Galilean invariant systems\cite{scaffidi2017hydrodynamic,Hoyos2012,bradlyn2012kubo,delacretaz2017transport}. In anisotropic systems, however, the connection between the viscosity and the conductivity can break down, and the Hall viscosity ceases to be quantized. The shift, however, still functions as a robust topological invariant (in the absence of translation symmetry breaking)\cite{haldane2015geometry}, and it, along with a new geometric invariant--the ``anisospin'' $\varsigma$--determine the Hall viscosity\cite{gromov2017investigating}. The anisospin has recently been shown to influence the evolution of quantum Hall states after a geometric quench\cite{liu2018geometric,lapa2018geometric}.

Previous work on anisotropic Hall viscosity has focused on systems with anisotropic kinetic energy or dielectric functions\cite{haldane2009hall,haldane2015geometry,gromov2017investigating}, or quantum Hall states near a nematic transition\cite{fradkin1999liquid,fradkin2010nematic,you2014theory,maciejko2013field}. However, one experimentally tunable source of anisotropy comes from a tilted magnetic field\cite{eisensteintiltedfqh,eisensteintiltedinteger,shayegantilted,pantilted,csathytiltsecondll,girvintilt1999}. Because a quantum Hall fluid lives embedded in our three-dimensional space, it is more properly modeled as a three-dimensional system in a strong potential, which we can take to be a function of the $x_3=z$ coordinate only. The quantum Hall fluid thus couples perturbatively to small in-plane (x and y) components of an applied magnetic field, which introduces an anisotropy to the fluid. This is commonly achieved by tilting the two-dimensional electron gas relative to a fixed external magnetic field. It is well understood that the wavefunctions for a quantum Hall system in an in-plane magnetic field map to those of a system with an anisotropic effective mass tensor\cite{maan1984combined,halonen1990subband,wang2003spontaneous,papic2013fractional}. Because tilted-field and effective mass anisotropy have very different physical origins, however, it is not immediately clear which measurable quantities--and in particular which response functions--in the tilted-field system will map to those in a system with anisotropic effective mass. Knowing these mappings could be particularly relevant for understanding the various anisotropic phases near the $\nu=5/2$ quantum Hall plateau.\cite{eisensteinhigher,fradkin1999liquid,xia52tilt,fradkin2010nematic} Furthermore, recent work\cite{yang2017anisotropic} has explored the shortcomings of the mapping of the tilted field system to one with anisotropic mass, as they pertain to pseudopotential interactions. To complement this, we explore here the nature of the effective two-dimensional fluid of electrons in a tilted field, to see in what ways it differs from an ordinary (an)isotropic electron gas.

To address these questions, we focus in this work on momentum and current transport in a non-interacting integer quantum Hall system with a tilted magnetic field and harmonic confining potential. The exact solvability of this model will allow us to directly compute the stress tensor, conductivity tensor, and viscosity tensor, and compare the results with those for a system with only mass tensor anisotropy. Owing to the fact that an in-plane field and a mass tensor couple to geometric deformations in different ways, we will find several complications in the mapping between the tilted field and an anisotropic mass tensor. While the Hall conductivity is insensitive to anisotropy, we will see that the ground state average stress tensor can distinguish between different sources of rotational symmetry breaking. Going further, this leads to the appearance of exotic elastic moduli in the response of the tilted-field system to rotational strains. We will show also that the Hall viscosity of the tilted field system maps onto the Hall viscosity of a system with mass anisotropy, provided one allows the value of the perpendicular magnetic field to change in the mapping. Demanding that the perpendicular field have its experimentally tuned value, on the other hand, yields a description of a fluid with both mass anisotropy and a non-quantized coupling to bimetric geometry. Note that in contrast to Refs.~\onlinecite{gromov2017bimetric,liu2018geometric,lapa2018geometric}, here we use a classical background source of anisotropy, rather than the dynamical metric found in fractional quantum Hall systems. Finally, to tie these observations together we formalize a projection procedure for mapping the intrinsically three-dimensional confined quantum Hall fluid to an effective two-dimensional system. In doing so, we uncover the origin of the exotic couplings to background geometry that distinguish the tilted-field system from an ordinary anisotropic fluid of point particles.

The structure of this paper is as follows. In Section~\ref{sec:linear response functions} we review some salient features of the Kubo formalism for conductivity and viscosity, paying particular attention to the role of rotational symmetry in fixing the viscosity and elastic moduli. In doing so, we shall rederive the contact terms in the Kubo formula of Ref.~\onlinecite{bradlyn2012kubo} in a new way, which sheds some light on their physical interpretation. Next, in Section~\ref{sec:2D band mass}, we review the Hall conductivity, stress tensor, and viscosity tensor for an IQH system with an anisotropic mass tensor. This serves as a warm up and point of comparison for our corresponding analysis in Sec.~\ref{sec:3D tilted field} of the \emph{three-dimensional} IQH system with perpendicular confining potential and an in-plane component of the magnetic field. We compute the Hall conductivity, average stress tensor, and Hall viscosity, expanding to leading order in the in-plane component of the field. These can be found in Eqs.~(\ref{eq:TF Hall conductivity}), (\ref{eq:TF stress tensor ground state xx}--\ref{eq:TF stress tensor ground state yy}), and (\ref{eq:TF Hall viscosity 1112}--\ref{eq:TF Hall elastic modulus}) respectively, and are a main result of this work. Finally, in Section~\ref{sec:effective anisotropy}, we carry out a formal mapping from the three-dimensional tilted-field system to a two-dimensional system with an effective anisotropy tensor. We show in what cases it is valid to approximate the tilted-field system by a two-dimensional system with mass anisotropy. Furthermore, we find that the projected system corresponds to a fluid that couples non-minimally to the background geometry. The Hamiltonian, current operator, and stress tensor for the projected system are given in Eqs.~(\ref{eq: tilde projected Hamiltonian}--\ref{eq: tilde T_yy}), and are our second main result. We conclude with a discussion of the theoretical outlook and experimental implications of our results. Along the way, we shall relegate the more technical details of our derivations to the appendix.

Before moving on, let us briefly comment on our notational conventions. We work in units where $\hbar=c=e=1$. The electron has charge $-|e|=-1$. We use Roman indices $a,b=1,2$ to label directions in two-dimensional space, and we reserve Greek indices $\mu,\nu=1,2,3,\dots$ to denote directions in three-dimensional space, or when the number of dimensions is unspecified. Repeated indices are always summed over unless otherwise specified. We use $x_\mu$ and $p_\mu$ to denote the position and canonical momentum operators, and reserve $r_\mu$ for the coordinate and $\pi_\mu\equiv p_\mu+A_\mu$ for the physical momentum.

When we discuss three-dimensional systems with a confining potential, the confinement will always be in the $x_3=z$-direction. Thus, the ``perpendicular magnetic field'' will always refer to $B_z=\vec{B}\cdot\hat{z}$, while in-plane or parallel field will always refer to the $x$ and $y$ components of the field. Finally, in both systems we consider, we look at response functions in the filling factor $\nu=1$ ground state.

\section{\label{sec:linear response functions}Linear response functions}

Let us begin by reviewing the most pertinent results from linear response theory in the context of a time-reversal asymmetric fluid. We write down the general Kubo formula, with an eye towards features relevant to topological phases of matter. In particular, we will review the Kubo formula for the (Hall) conductivity, and re-derive the Kubo formulas for the (Hall) viscosity. We will focus especially on the role of rotational symmetry and the interpretation of ``contact'' (diamagnetic) terms in the response functions. Further background can be found in Refs.~\onlinecite{forster1975hydrodynamic,bradlyn2012kubo,bradlyn2015linear,abanov2014}.

\subsection{Review of linear response theory}\label{sec:linear response theory}\label{sec:properties of response functions}

The typical linear response set-up\cite{fetter2012quantum} starts with a system described by an unperturbed (time-independent) Hamiltonian $H_0$, an unperturbed density matrix $\rho_0$ that commutes with $H_0$, and a perturbing Hamiltonian $\Delta H=f_n(t)B_n e^{\epsilon t}$. In this work, we will take $\rho_0$ to be the density matrix for the ground state of $H_0$.
In the perturbing Hamiltonian, $f_n(t)$ are a set of tunable external fields (e.g., they could be electric or magnetic fields), $B_n$ are the operators to which they couple (e.g., they could be current or spin operators) and $e^{\epsilon t}$ ensures that the perturbations turn off adiabatically as $t\to -\infty$, which is necessary for regularization. We are interested in the evolution of the expectation values of a set of operators, $A_m$, in the presence of the perturbation.

The perturbed density matrix in the interaction picture takes the form $\rho=\lim_{t_0\to -\infty}\mathcal{U}_I(t,t_0) \rho_0 \mathcal{U}^\dagger_I(t,t_0)$ where $\mathcal{U}_I(t,t_0)=\text{T exp}\left(-i\int_{t_0}^t dt' H_I(t')\right)$ is the interaction picture evolution operator. Working to linear order in the external fields, one arrives at the principal result of linear response theory:
\begin{align}\label{eq:perturbed observable 1}
\delta \braket{A_m}(t)&=\text{Tr}(A_m(\rho-\rho_0))\\&=\int_{-\infty}^\infty dt' \chi_{mn}(t-t')f_n(t')\label{eq:perturbed observable 2},
\end{align}
where the linear response function $\chi_{mn}(t-t')$ is given by the generalized Kubo formula
\begin{align}\label{eq:response function definition}
\chi_{mn}(t)=-i\Theta(t)\lim_{\epsilon\to 0^+}\braket{[A_m(t),B_n(0)]}_0e^{-\epsilon t}.
\end{align}
We emphasize that the time dependence of $A_m(t)$ in the above expression is evaluated in the interaction picture: $A_m(t)=e^{iH_0t}A_m(0)e^{-iH_0t}$. Note also the short hand $\braket{\mathcal{O}}_0\equiv \text{Tr}(\rho_0\mathcal{O})$.

In Eq.~(\ref{eq:perturbed observable 1}) we assumed that the perturbing fields $f_n(t)$ directly affect only the density matrix $\rho$; however, in many cases the observables themselves depend on the fields. If $A_m(f_n)=A_m^0+A^1_{mn}f_n(t)+O(f^2)$, then Eq.~(\ref{eq:perturbed observable 2}) still holds as long as the response function is modified to
\begin{align}\label{eq:perturbed observable w/ field dependence}
\chi_{mn}(t)\to \chi_{mn}(t)+\braket{A^1_{mn}}_0\delta(t).
\end{align}
Thus, the explicit dependence of the observables on the fields gives rise to ``contact terms" in the response function (so named because of the delta function).

In the event that the perturbing Hamiltonian is described by local interactions and the observables $A_m$ are local operators, we may write
\begin{align}\label{eq:perturbed observable position dependent}
\delta \braket{A_m}(\vec{r},t)&=\int_{-\infty}^\infty dt' \int d^dr' \chi_{mn}(\vec{r},\vec{r}',t-t')f_n(\vec{r}',t'),
\end{align}
where the intensive response function is given by
\begin{align}\label{eq:response function position dependent}
\chi_{mn}(\vec{r},\vec{r}',t-t')&=-i\Theta(t)\braket{[A^0_m(\vec{r},t),B_n(\vec{r}',0)]}_0e^{-\epsilon t}\nonumber\\&+
\braket{A^1_{mn}(\vec{r})}_0\delta(\vec{r}-\vec{r}')\delta(t-t').
\end{align}
 
Because Eqs.~(\ref{eq:perturbed observable 2}) and (\ref{eq:perturbed observable position dependent}) involve convolutions in time (and in space for (\ref{eq:perturbed observable position dependent}) if the system is translation invariant), it is often useful to work in the frequency and wavevector domain. From Eqs.~(\ref{eq:perturbed observable 2}), (\ref{eq:response function definition}) and (\ref{eq:perturbed observable w/ field dependence}), we have
\begin{align}
\delta \braket{A_m}(\omega)&=\chi_{mn}(\omega)f_n(\omega)\label{eq:perturbed expectation Fourier}\\ \chi_{mn}(\omega)&=-i\lim_{\epsilon\to 0^+}\int_0^\infty dt e^{i\omega^+t}\braket{[A^0_m(t),B_n(0)]}_0\nonumber\\&+\braket{A_{mn}^1}_0.\label{eq:response function Fourier}
\end{align}
From Eqs.~(\ref{eq:perturbed observable position dependent}) and (\ref{eq:response function position dependent}), in the case of a homogeneous unperturbed state and uniform perturbation, we have
\begin{align}
\delta\braket{A_m}(\vec{q},\omega)&=\frac{1}{\vol }\chi_{mn}(\vec{q},-\vec{q},\omega)f_n(\vec{q},\omega),\label{eq:perturbed expectation spatial Fourier}\\\chi_{mn}(\vec{q},\vec{q}',\omega)&=-i\int_0^\infty dt e^{i\omega^+t}\braket{[A^0_m(\vec{q},t),B_n(\vec{q}',0)]}_0\nonumber\\&+\braket{A_{mn}^1{(\vec{q}+\vec{q}')}}_0.\label{eq:response function spatial Fourier}
\end{align}
Here $\vol $ is the volume of the system. In order to carefully track the volume dependence of various quantities, we will be explicit that  $\chi_{mn}(q,q',\omega)$ is a function of two wavevectors. To make contact with the standard linear response exposition, we also introduce:
\begin{align}
\chi_{mn}(\vec{q},\omega)\equiv\frac{1}{\vol }\chi_{mn}(\vec{q},-\vec{q},\omega).
\end{align}

If the system is homogeneous and the perturbation is uniform in space, then $\braket{A_m}(\vec{q}=0,\omega)=\vol \braket{A_m}(\vec{r},\omega)$ and $f_n(\vec{q}=0,\omega)=\vol f_n(\vec{r},\omega)$, where $\vec{r}$ is any point in the space. It follows from (\ref{eq:perturbed expectation spatial Fourier}) that
\begin{align}\label{eq:perturbed expectation value of local observable homogeneous}
\delta \braket{A_m}(\vec{r},\omega)&=\left(\frac{1}{\vol}\chi_{mn}(\vec{q}=0,\vec{q}'=0,\omega)\right)f_n(\vec{r},\omega).
\end{align}
Thus, we identify $\frac{1}{\vol }\chi_{mn}(\vec{q}=0,\vec{q}'=0,\omega)$ as the intensive response function at zero wavevector. Alternatively, the response function of the integrated observable, 
\begin{equation}
\int{d^dr\; A_m(\vec{r})}=A^0_m(\vec{q}=0)+A_{mn}^1(\vec{q}=0)f_n(t),
\end{equation}
 given the perturbation 
\begin{equation}
 \Delta H=f_n(t)\int d^drB_n(\vec{r})=B_n(\vec{q}'=0)f_n(t),
\end{equation}
 is then simply $\chi_{mn}(\vec{q}=0,\vec{q}'=0)$. Following accepted conventions, we will refer to local response functions with the suffix ``-ity,'' while we refer to extensive response functions with the suffix ``-ance.''\cite{gromovtalk,sriramhydro} For a homogeneous system, Eq.~(\ref{eq:perturbed expectation value of local observable homogeneous}) shows that the intensive response function (conductivity, viscosity) is equal to the extensive response function (conductance, viscosance) divided by the volume.

In this paper, we will be computing response functions for both two and three-dimensional systems. As is well understood, the intensive response functions for two and three-dimensional systems cannot be directly compared, since they have different units. In order to make a comparison, we can integrate a three-dimensional intensive response function, $\chi_{mn}^{3D}$, along one spatial direction to obtain an effective two-dimensional response function, $\chi_{mn}^{2D}$. In particular, for a three-dimensional system with length $L_x, L_y$ and $L_z$ in the $x,y$ and $z$ directions respectively, we can integrate over $z$ to obtain
\begin{equation}
\chi^{2D}_{mn}\equiv L_z\chi_{mn}^{3D}=\frac{1}{L_xL_y}\chi_{mn}.\label{eq:effective 2d response}
\end{equation}
We will make use of this in Secs.~\ref{sec:3D tilted field} and \ref{sec:effective anisotropy} to compare the conductivity and viscosity for the tilted field system to that of a 2D system with mass anisotropy.

Before moving on to a discussion of specific linear response functions, we remark that we are primarily interested in the response of gapped (i.e.~topological) phases. Because an energy gap precludes any dissipative response, the nonvanishing linear response functions we focus on at zero frequency will be antisymmetric, which we will refer to as ``Hall coefficients.'' For the Hall conductivity and Hall viscosity in particular to be nonzero, recall that time-reversal symmetry must be broken\cite{forster1975hydrodynamic,1995-AvronSeilerZograf}.

\subsection{Hall conductivity}\label{subsec:Hall conductivity}

Let us recall the Kubo formula for the Hall conductivity tensor. We are interested in the response of the current density $\langle j_\mu^A\rangle$ given the perturbing Hamiltonian $\Delta H=-\int d^dr j_\mu^A (\vec{r},t)A_\mu (\vec{r},t)$. Here, the superscript $A$ emphasizes that the current density may depend on the perturbing vector field. The appropriate response function is the conductivity tensor, which for a homogeneous system and uniform electric perturbations 
can be written
\begin{align}\label{eq:expression for conductivity}
\sigma_{\mu\nu}(\omega)&\equiv \sigma_{\mu\nu}(\omega,\vec{q}=0)\\&=\frac{iC_{\mu\nu}}{\omega_+}+\frac{1}{\vol \omega^+}\int_0^\infty dt e^{i\omega^+t}\braket{[J_\mu(t),J_\nu(0)]}_0.\nonumber
\end{align}
Here, $J_\mu\equiv \int d^dr j_\mu^{A=0}(\vec{r})$ is the unperturbed current operator and the tensor $C_{\mu\nu}$ is the contact term. The form of $C_{\mu\nu}$ depends on the specific system being considered. For a non-relativistic system with isotropic kinetic term, $C_{\mu\nu}=\frac{n}{m}\delta_{\mu\nu}$ where $n$ is the particle density and $m$ is the mass; for a non-relativistic system with mass anisotropy, $C_{\mu\nu}=n\tilde{m}_{\mu\nu}$, where $\tilde{m}_{\mu\nu}$ is the inverse mass tensor. In all cases, the conductivity contact term is symmetric (since it comes from directly varying a quadratic function of the vector potential) and does not contribute to the Hall conductivity, which we define as
\begin{align}\label{eq:Hall conductivity definition}
\sigma_{\mu\nu}^H\equiv \frac{1}{2}\lim_{\omega\to 0}\left[\sigma_{\mu\nu}(\omega)-\sigma_{\nu\mu}(\omega)\right].
\end{align}
For $N$ non-interacting electrons in a degenerate level with energy $E_0$, we can extract a simple explicit form for $\sigma_{\mu\nu}^H$\cite{niu1985quantized} when the current matrix elements depend only on the energy of the states (as for a Landau level). Combining Eqs.~(\ref{eq:expression for conductivity}) and (\ref{eq:Hall conductivity definition}) and inserting a complete set of energy eigenstates $\ket{\alpha,\beta}$ (where $\alpha$ labels the energy eigenspace and $\beta$ labels different states within each eigenspace), we find
\begin{align}\label{eq:Hall conductivity with CSS inserted}
\sigma_{\mu\nu}^H&=\frac{2N}{\vol }\sum_{\alpha\neq 0} \frac{\text{Im}(\braket{\alpha=0|J_\mu^{(1)}|\alpha}\braket{\alpha|J_\nu^{(1)}|\alpha=0})}{(E_0-E_\alpha)^2}.
\end{align}
Here $\text{Im}(v)\equiv\frac{1}{2i}(v-v^*)$, and the superscript on $J_\mu^{(1)}$ emphasizes that it is the single particle current operator. We have suppressed the $\beta$ label in the states $\ket{\alpha=0,\beta}$ and $\ket{\alpha,\beta}$ because by assumption the result is the same for any consistent choice of $\beta$.

\subsection{\label{subsec:Hall viscosity}Hall viscosity}
We next define and derive a useful form for the Hall viscosity of a quantum mechanical system. The Hall viscosity is the antisymmetric component of the viscosity tensor, which characterizes how the stress tensor responds to a weak time-dependent strain\cite{1998-Avron}. It has alternatively been referred to as ``odd viscosity,'' ``Lorentz shear modulus\cite{2007-TokatlyVignale,2009-TokatlyVignale-JPhys},'' or ``anomalous viscosity\cite{wiegmann2014anomalous}.'' While the majority of the following discussion can be found in Ref.~\onlinecite{bradlyn2012kubo}, we repeat it here both to establish our notation, and to emphasize certain features of the Kubo formula relevant for anisotropic systems. Additionally, we present a physically motivated discussion of the contact terms in the Kubo formula.

We first define the stress tensor and strain generators. Let us consider the Hamiltonian describing $N$ interacting electrons moving in an electromagnetic background, given by
\begin{align}\label{eq:unstrained Hamiltonian}
    H_0=\frac{1}{2m}\sum_{i=1}^N\pi_\mu^i\pi_\mu^i+\frac{1}{2}\sum_{i\neq j}V(\vec{x}^i-\vec{x}^j).
\end{align}
Here, $\pi_\mu^i\equiv p_\mu^i+A_\mu(\vec{x}^i)$ is the physical momentum of the $i$th electron, and has the commutation relations $[x_\mu^i,\pi_\nu^j]=i\delta_{\mu\nu}\delta_{ij}$ and $[\pi_\mu^i,\pi_\nu^j]=-i\delta_{ij}\epsilon_{\mu\nu\rho}B_\rho(\vec{x}^i)$.

In the presence of a time-varying uniform strain, the perturbed Hamiltonian takes the form
\begin{equation}\label{eq:strained Hamiltonian}
    H_\Lambda(t)=\frac{1}{2m}g^{\mu\nu}(t)\sum_{i=1}^N\pi^i_\mu\pi^i_\nu +\frac{1}{2}\sum_{i\neq j}V(\Lambda^T(t)(\vec{x}^i-\vec{x}^j)),
\end{equation}
where $\Lambda\in GL(d,\mathbb{R})$ is an invertible matrix, $g^{\mu\nu}(t)=\Lambda_{\alpha\mu}^{-1}(t)\Lambda_{\alpha\nu}^{-1}(t)$; it is natural to also define its inverse $g_{\mu\nu}(t)=\Lambda_{\mu \alpha}(t)\Lambda_{\nu \alpha}(t)$.

Next, we introduce Hermitian strain generators $J_{\mu\nu}$. Given $\Lambda(t)=e^{\lambda(t)}$, we require that the strain operators $S(t)\equiv e^{-i\lambda_{\mu\nu}(t)J_{\mu\nu}}$ satisfy
\begin{align}
S(t)x_\mu^i S^{-1}(t)&=\Lambda_{\mu\nu}^T(t)x_\nu^i,\label{eq:strain transformation rule 1}\\S(t)\pi_\mu^i S^{-1}(t)&=\Lambda^{-1}_{\mu\nu}(t)\pi_\nu^i,\label{eq:strain transformation rule 2}
\end{align}
from which it follows that 
\begin{align}
S(t)H_0S^{-1}(t)=H_\Lambda(t)\label{eq:strained Hamiltonian with strain ops}.
\end{align}
Expanding the transformation rules (\ref{eq:strain transformation rule 1}) and (\ref{eq:strain transformation rule 2}) to linear order in $\lambda_{\mu\nu}(t)$, we arrive at the commutation relations
\begin{align}
[x_\mu^i,J_{\nu\rho}]&=-i\delta_{\mu\rho}x_\nu^i,\label{eq:J and x commutator}\\
[\pi_\mu^i,J_{\nu\rho}]&=i\delta_{\mu\nu}\pi_\rho^i.\label{eq:J and pi commutator}
\end{align}
Making use of the Jacobi identity, we also deduce the commutation relations between the strain generators,
\begin{align}
[J_{\mu\nu},J_{\rho\sigma}]=i\delta_{\nu\rho}J_{\mu\sigma}-i\delta_{\mu\sigma}J_{\rho\nu},\label{eq:J and J commutator}
\end{align}
which define the Lie algebra $\mathfrak{gl}(d,\mathbb{R})$ of the general linear group. Specific forms of the strain generators will not be needed here, but can be found in Ref.~\onlinecite{bradlyn2012kubo}.

We define the intensive stress tensor, $\tau_{\mu\nu}(\vec{r},t)$, via the continuity equation for the momentum density $g_\mu(\vec{r},t)$:
\begin{align}
0&=\frac{\partial g_\mu(\vec{r},t)}{\partial t}+\frac{\partial \tau_{\nu\mu}(\vec{r},t)}{\partial r_\nu},\label{eq:momentum continuity eq}
\end{align}
where the momentum density is defined by
\begin{align}
g_\mu(\vec{r},t)&\equiv\frac{1}{2}\sum_{i=1}^N\{\pi_\mu^i,\delta(\vec{r}-\vec{x}^i)\}.\label{eq:momentum density}
\end{align}
Taking the spatial Fourier transform of the continuity equation, and expanding about the long-wavelength limit, 
we arrive at the stress-strain Ward identity
\begin{align}
T_{\mu\nu}=-\partial_t J_{\mu\nu}=-i[H_0,J_{\mu\nu}],\label{eq:integrated stress tensor}
\end{align}
where we have introduced a new symbol for the the spatially integrated (extensive) stress tensor,
\begin{equation}
T_{\mu\nu}\equiv\tau_{\mu\nu}(\vec{q}=0)=\int d^dr \tau(\vec{r}).
\end{equation} 
From Eqs.~(\ref{eq:strained Hamiltonian with strain ops}) and (\ref{eq:integrated stress tensor}), we can deduce an alternative expression for the integrated stress tensor in the unstrained system:
\begin{align}
T_{\mu\nu}&=-\frac{\partial H_\Lambda}{\partial \lambda_{\mu\nu}}\biggr\rvert_{\lambda_{\mu\nu}=0}.\label{eq:integrated stress tensor alternative}
\end{align}
This form of the definition for the stress tensor is familiar from field theory, where the stress tensor is defined in terms of the variation of the action with respect to the vielbein.
Note that for a system without rotational symmetry, the dependence of the Hamiltonian on $\lambda$ need not be through the metric $g$, and so the stress tensor generally need not be symmetric. 

Defining the stress tensor in the strained system is a matter of appropriately generalizing Eqs.~(\ref{eq:integrated stress tensor}) and (\ref{eq:integrated stress tensor alternative}). It may seem natural to define the strained integrated stress tensor to be $T_{\mu\nu}^\Lambda=-i[H_\Lambda,J_{\mu\nu}]=-\frac{\partial H_\Lambda}{\partial \lambda_{\mu\nu}}$. However, it is better to define it to be
\begin{align}
T^\Lambda_{\mu\nu}&\equiv-\Lambda_{\rho\mu}\Lambda^{-1}_{\nu\sigma} \frac{\partial H_\Lambda}{\partial \lambda_{\rho\sigma}}.\label{eq:strained integrated stress tensor}
\end{align}
To motivate this definition of $T^\Lambda_{\mu\nu}$, recall from our definition
\begin{equation}\label{eq:metric in terms of Lambdas}
g_{\mu\nu}=(\Lambda\Lambda^T)_{\mu\nu}=\Lambda_{\mu\alpha}\Lambda_{\nu\alpha}
\end{equation}
that the first and last index of $\Lambda$ have very different meanings: the first ($\mu,\nu$) index refers to a direction in ambient space, while the second ($\alpha$) index refers to a direction in an auxiliary ``internal'' space; in other words, $\Lambda_{\mu\alpha}$ is a spatial vielbein $e^\alpha_\mu$ in the language of Refs.~\onlinecite{bradlyn2014low,Gromov20141,son2013newton}. The Kubo formalism for viscosity developed in Refs.~\onlinecite{luttinger1964,bradlyn2012kubo} is given in a gauge-fixed form, where the internal and ambient basis directions are tied together. Disentangling the indices, we see that the strain generators and the stress tensor derived from Eq.~(\ref{eq:integrated stress tensor}) have two spacetime indices, where the first index is coordinate like (an upper index), and the second index is momentum-like (a lower index)\cite{haldane2015geometry}$^{,}$\footnote{we follow the standard terminology here, where indices are raised and lowered with the metric $g_{\mu\nu}$}.

We can, alternatively, examine the stress tensor with different types of indices. To do so, we can repeat our derivation of the Ward identity using a strained configuration as our starting point. We seek the canonical transformation which implements the generalization of Eqs.~(\ref{eq:strain transformation rule 1}) and (\ref{eq:strain transformation rule 2}) to an already strained background, i.e.~
\begin{align}
S(t)\Lambda_{\nu\mu}x_\nu S^{-1}(t)&=\Lambda^{(2)}_{\alpha\mu}\Lambda_{\nu \alpha}x_\nu,\\
S(t)\Lambda^{-1}_{\mu\nu}\pi_\nu S^{-1}(t)&=(\Lambda^{(2)})^{-1}_{\mu\alpha}\Lambda^{-1}_{\alpha \nu}\pi_\nu.
\end{align}

To avoid overburdening notation, we will continue in our fixed gauge for the internal indices. The generator of this transformation is, of course, the strain generator expressed in the ``big $\mathbf{X}$'' variables\cite{bradlyn2012kubo}
\begin{align}
\mathbf{X}&=\Lambda^T\mathbf{x}, \label{eq:big X}\\
\mathbf{\Pi}&=\Lambda^{-1}\boldsymbol{\pi}, \label{eq:big P}
\end{align} 
and the Ward identity in these variables yields the stress tensor $T^\Lambda_{\mu\nu}$ of Eq.~(\ref{eq:strained integrated stress tensor}). This gives us a new interpretation of the canonical transformation employed above and in Ref.~\onlinecite{bradlyn2012kubo}: keeping track of the meaning of the various indices, we see then this is the stress tensor with \emph{two internal indices}. We can expect that using this stress tensor will simplify the form of contact terms in the Kubo formula, since the internal directions (in contrast to the ambient directions) do not deform under strain perturbations. Furthermore, the Ward identity associated with rotational symmetry naturally places constraints on this form of the stress tensor\cite{bertlmann2000anomalies}. Since our focus will be on the effects of rotational symmetry breaking, the ``all-internal-index'' $T^\Lambda_{\mu\nu}$ serves as a natural starting point. Note that once we specialize to flat space ($\Lambda=0$) at the end of our calculations, the distinction between the indices is unobservable.

We have thus identified the linear response observables, $T^\Lambda_{\mu\nu}$, and can now examine their dependence on strain. Expanding to linear order in the strain fields, we find
\begin{align}
T^\Lambda_{\mu\nu}&=T_{\mu\nu}+\lambda_{\rho\mu}T_{\rho\nu}-\lambda_{\nu\sigma}T_{\mu\sigma}\nonumber\\&-i[J_{\mu\nu},T_{\rho\sigma}]\lambda_{\rho\sigma}+O(\lambda^2).\label{eq:leading order strained stress tensor}
\end{align}
Meanwhile, the perturbed Hamiltonian to leading order is
\begin{align}
H_\Lambda=H_0-\lambda_{\mu\nu}T_{\mu\nu}+O(\lambda^2).\label{eq:leading order strained Hamiltonian}
\end{align}
Since our goal is to define Hall viscosity as a  measure of a system's response to a {time varying} strain, we treat $\partial_t\lambda_{\mu\nu}$ as the perturbing fields. To do so, we must clarify what we mean when $\partial_t$ acts on operators. Just as in section \ref{sec:linear response theory}, it is simplest to work in the interaction picture, so that the perturbing Hamiltonian is $\Delta H_I=-\lambda_{\mu\nu}(t)T_{\mu\nu}(t)$ and derivation of an operator means $\partial_t\mathcal{O}=i[H_0,\mathcal{O}]$ for $\mathcal{O}=T_{\mu\nu}$ and $\mathcal{O}=J_{\mu\nu}$. Using $T_{\mu\nu}=-\partial_tJ_{\mu\nu}$, we rewrite the perturbing Hamiltonian as
\begin{align}\label{eq:leading order strained Hamiltonian 2}
\Delta H_I&=\partial_t(\lambda_{\mu\nu}J_{\mu\nu})-J_{\mu\nu}\partial_t\lambda_{\mu\nu}.
\end{align}
We can almost directly apply the formalism from section \ref{sec:linear response theory}, but must accommodate the total-derivative term appearing in the Hamiltonian. We begin by writing the evolution operator in the interaction picture:
\begin{align}\label{eq:evolution operator with strain perturbation}
\mathcal{U}_I(t)&=1+i\int_{t_0}^t J_{\mu\nu}\partial_{t'}\lambda_{\mu\nu}dt'\\&-i\int_{t_0}^t \partial_{t'}(\lambda_{\mu\nu}J_{\mu\nu})dt'+O(\lambda^2),\nonumber
\end{align}
where we leave the limit $t_0\to -\infty$ at the end implicit to avoid notational clutter. The first integral term is what generally appears in the derivation of the Kubo formula while the second term can be evaluated:
\begin{align}\label{eq:total derivative term of strained evolution operator}
-i\int_{t_0}^t\partial_{t'}(\lambda_{\mu\nu}J_{\mu\nu})dt'&=-i\lambda_{\mu\nu}J_{\mu\nu}.
\end{align}
Carrying out the linear response calculation, we see that the total derivative term just contributes a fourth contact term to Eq.~(\ref{eq:leading order strained stress tensor}), while the remainder of the evolution operator yields the convolution term. Explicitly, we find

\begin{align}
\braket{T^\Lambda_{\mu\nu}}&=\braket{T_{\mu\nu}}_0+\braket{T_{\rho\nu}}_0\lambda_{\rho\mu}-\braket{T_{\mu\sigma}}_0\lambda_{\nu\sigma}\nonumber\\&-i\braket{[J_{\mu\nu},T_{\rho\sigma}]}_0\lambda_{\rho\sigma}+i\braket{[J_{\rho\sigma},T_{\mu\nu}]}_0\lambda_{\rho\sigma}\nonumber\\&-\int_{-\infty}^\infty dt'X_{\mu\nu\rho\sigma}(t-t')\frac{\partial \lambda_{\rho\sigma}}{\partial t'}+O(\lambda^2),\label{eq:stress tensor linear response}
\end{align}
where the response function $X_{\mu\nu\rho\sigma}$ is given by:
\begin{align}
X_{\mu\nu\rho\sigma}(\omega)&=-i\int_0^\infty dt e^{i\omega^+t}\braket{[T_{\mu\nu}(t),J_{\rho\sigma}(0)]}_0.\label{eq:stress-strain linear response}
\end{align}
Though seemingly onerous, Eq.~(\ref{eq:stress tensor linear response}) simplifies considerably thanks to the Jacobi identity. Using $T_{\mu\nu}=-i[H_0,J_{\mu\nu}]$ and Eq.~(\ref{eq:J and J commutator}), we deduce
\begin{align}
  [T_{\rho\sigma},J_{\mu\nu}]-[T_{\mu\nu},J_{\rho\sigma}]&=i(\delta_{\mu\sigma}T_{\rho\nu}-\delta_{\nu\rho}T_{\mu\sigma}).\label{eq:stress-strain commutator identity}
\end{align}
Therefore, Eq.~(\ref{eq:stress tensor linear response}) reduces to:
\begin{align}
\braket{T_{\mu\nu}^\Lambda}-\braket{T_{\mu\nu}}_0&=-\int_{-\infty}^\infty dt'X_{\mu\nu\rho\sigma}(t-t')\frac{\partial \lambda_{\rho\sigma}}{\partial t'}.\label{eq:simplified stress-strain linear response}
\end{align}
The expression for $X_{\mu\nu\rho\sigma}$ given in Eq.~(\ref{eq:stress-strain linear response}) explicitly contains both the stress tensor and the strain generator. It is possible, using $T_{\mu\nu}=-\partial_tJ_{\mu\nu}$ and integration by parts, to put Eq.~(\ref{eq:stress-strain linear response}) in a stress-stress form:
\begin{align}\label{eq:stress-stress linear response function}
    X_{\mu\nu\rho\sigma}(\omega)&=\frac{1}{\omega^+}\braket{[T_{\mu\nu}(0),J_{\rho\sigma}(0)]}_0\nonumber\\&+\frac{1}{\omega^+}\int_0^\infty dte^{i\omega^+t}\braket{[T_{\mu\nu}(t),T_{\rho\sigma}(0)]}_0.
\end{align}
It is similarly possible to write the response function in a strain-strain form\cite{bradlyn2012kubo,bradlyn2015linear}, but it is not useful to do so in this paper.

At last, having derived an explicit form for the linear response function between integrated stress and strain, we can define the viscosity tensor, $\eta_{\mu\nu\rho\sigma}$. First, we recognize, through its dependence on $T_{\mu\nu}$, that $X_{\mu\nu\rho\sigma}$ is an extensive quantity. Viscosity, meanwhile, is conventionally defined in terms of local quantities. Thus, it proves useful to introduce a linear response function $\chi_{\mu\nu\rho\sigma}$ that is analogous to $X_{\mu\nu\rho\sigma}$ and captures the behavior of $\tau_{\mu\nu}$ subject to a strain parametrized by $\lambda_{\mu\nu}$. Assuming the system is homogeneous, so that $\braket{\tau_{\mu\nu}(\vec{r})}=\braket{\tau_{\mu\nu}(\vec{0})}\equiv \braket{\tau_{\mu\nu}}$, we take
\begin{align}\label{eq:intrinsic stress-strain response function}
\braket{\tau_{\mu\nu}^\Lambda}-\braket{\tau_{\mu\nu}}_0=-\int dt'\chi_{\mu\nu\rho\sigma}(t-t')\frac{\partial \lambda_{\rho\sigma}}{\partial t'},
\end{align}
as the definition of $\chi_{\mu\nu\rho\sigma}$.

Although $X_{\mu\nu\rho\sigma}$ and $\chi_{\mu\nu\rho\sigma}$ are the response functions for quantities related by a multiple of the volume $\vol $, the relation between $X_{\mu\nu\rho\sigma}$ and $\chi_{\mu\nu\rho\sigma}$ is not simply $X_{\mu\nu\rho\sigma}=\vol \chi_{\mu\nu\rho\sigma}$ because the volume of the system is itself dependent on the perturbing strain. In particular 
\begin{align}\label{eq:relation between integrated and intrinsic stress tensor exp val}
\braket{T_{\mu\nu}^\Lambda}=\vol \rvert_{\Lambda=I}\text{det}\left(\Lambda\right)\braket{\tau_{\mu\nu}^\Lambda},
\end{align}
and because $\text{det}(\Lambda)=1+\lambda_{\rho\rho}+O(\lambda^2)$, we deduce

\begin{align}\label{eq:X and chi relation 2}
\frac{X_{\mu\nu\rho\sigma}}{\vol }&=\chi_{\mu\nu\rho\sigma}+\frac{\braket{\tau_{\mu\nu}}_0\delta_{\rho\sigma}}{i\omega}.
\end{align}
This guarantees that, for a homogeneous system, we can ultimately calculate the intensive viscosity $\eta_{\mu\nu\rho\sigma}$ using the extensive response function $X_{\mu\nu\rho\sigma}$, for which we have the convenient expression, Eq.~(\ref{eq:stress-stress linear response function}). First, however, we need a relation between $\chi_{\mu\nu\rho\sigma}$ and $\eta_{\mu\nu\rho\sigma}$, which we deduce as follows.

We begin by expressing Eq.~(\ref{eq:intrinsic stress-strain response function}) in frequency space:
\begin{align}\label{eq:intrinsic linear response function fourier}
\delta\braket{\tau_{\mu\nu}}(\omega)&=-i\omega \chi_{\mu\nu\rho\sigma}(\omega)\lambda_{\rho\sigma}(\omega).
\end{align}
We see that the stress tensor does not respond to static perturbations (corresponding to $\omega=0$) unless $\chi_{\mu\nu\rho\sigma}(\omega)$ has a pole at $\omega=0$, a consequence of our decision to use $\partial_t\lambda_{\rho\sigma}$ rather than $\lambda_{\rho\sigma}$ as the perturbing fields in Eq.~(\ref{eq:intrinsic stress-strain response function}). However, we know that the stress tensor  \emph{does} respond to static strains. For example, a static compression of a substance generally increases its pressure and a rigid rotation of an anisotropic substance can rotate one component of the stress tensor into a different unequal component\cite{parodi}.

Therefore, we must account for static strains and understand the poles of $\chi_{\mu\nu\rho\sigma}(\omega)$. Let us restrict ourselves to incompressible fluids, which are gapped and have at most a simple pole in the response function. 
We write the linear response function as
\begin{align}\label{eq:viscosity and elastic moduli definition}
\chi_{\mu\nu\rho\sigma}(\omega)&=-\eta_{\mu\nu\rho\sigma}(\omega)-\frac{\kappa^{-1}_{\mu\nu\rho\sigma}}{i\omega}.
\end{align}
where the viscosity tensor $\eta_{\mu\nu\rho\sigma}$ is defined to be the analytic part of the Laurent-like expansion for $\chi_{\mu\nu\rho\sigma}$ about $\omega=0$, and the elastic modulus tensor $\kappa^{-1}_{\mu\nu\rho\sigma}$ is defined to be the coefficient of the pole. By splitting the response function in this way, we 
match our intuition that viscosity measures the dynamic rather than static behavior of the system.

Finally, we define the Hall viscosity tensor analogously to the Hall conductivity; it is the antisymmetric component of the viscosity in the $\omega\to 0$ limit:
\begin{align}\label{eq:Hall viscosity definition}
    \eta^H_{\mu\nu\rho\sigma}&\equiv \frac{1}{2}\lim_{\omega_+\to 0}(\eta_{\mu\nu\rho\sigma}(\omega)-\eta_{\rho\sigma\mu\nu}(\omega)).
\end{align}
For completeness, let us also define the Hall elastic modulus:
\begin{align}\label{eq:Hall elastic modulus definition}
\kappa^{-H}_{\mu\nu\rho\sigma}&\equiv \frac{1}{2}\left(\kappa_{\mu\nu\rho\sigma}^{-1}-\kappa_{\rho\sigma\mu\nu}^{-1}\right).
\end{align}
As we will see shortly, for isotropic fluids $\kappa^{-H}_{\mu\nu\rho\sigma}=0$, which explains why it is not as commonly discussed as the Hall viscosity.

We can derive explicit expressions for the Hall viscosity and the Hall elastic moduli. Focusing first on the integral term in Eq.~(\ref{eq:stress-stress linear response function}), which we denote by $X_{\mu\nu\rho\sigma}^{\int}(\omega)$, we can insert a complete set of states to find (under a similar set of assumptions as in Sec.~\ref{subsec:Hall conductivity}) 
\begin{align}\label{eq:Hall viscosity with CSS inserted}
    &\frac{1}{2}\lim_{\omega^+\to 0}\left(X_{\mu\nu\rho\sigma}^{\int}(\omega)-X_{\rho\sigma\mu\nu}^{\int}(\omega)\right)\nonumber\\&=\frac{2N}{\vol }\sum_{\alpha\neq 0}\frac{\text{Im}\left(\braket{0|T_{\mu\nu}^{(1)}|\alpha}\braket{\alpha|T_{\rho\sigma}^{(1)}|0}\right)}{(E_0-E_\alpha)^2}\nonumber \\
    &\equiv-\eta^H_{\mu\nu\rho\sigma}.
\end{align}

We see that the zero-frequency limit of the antisymmetric part of the integral term is perfectly well-behaved in a gapped system, and so can be identified with the (negative of the) Hall viscosity. The Hall elastic modulus is then given by the sum of the antisymmetric components of the contact term in Eq.~(\ref{eq:stress-stress linear response function}) and the second term in Eq.~(\ref{eq:X and chi relation 2}):
\begin{align}\label{eq:Hall elastic modulus}
\kappa^{-H}_{\mu\nu\rho\sigma}&=-\braket{\tau_{\mu\sigma}}_0\delta_{\nu\rho}+\braket{\tau_{\rho\nu}}_0\delta_{\mu\sigma}\nonumber\\&-\braket{\tau_{\rho\sigma}}_0\delta_{\mu\nu}+\braket{\tau_{\mu\nu}}_0\delta_{\rho\sigma}.
\end{align}
We note that $\kappa^{-1}_{\mu\nu\rho\sigma}$ is symmetric under $\mu\leftrightarrow \rho$, but antisymmetric under $\nu\leftrightarrow \sigma$. More importantly, when the fluid is isotropic, $\braket{\tau_{\mu\nu}}_0\propto \delta_{\mu\nu}$ and the Hall elastic moduli vanish (unless the fluid is active\cite{vitelli2017odd}).
Eqs.~(\ref{eq:Hall viscosity with CSS inserted}) and (\ref{eq:Hall elastic modulus}) is the main result of the present section.

It is evident from Eq.~(\ref{eq:Hall viscosity with CSS inserted}) that the Hall viscosity tensor satisfies $\eta^H_{\mu\nu\rho\sigma}=-\eta^H_{\rho\sigma\mu\nu}$. Furthermore, if the stress tensor is symmetric, then the Hall viscosity tensor also satisfies $\eta^H_{\mu\nu\rho\sigma}=\eta^H_{\nu\mu\rho\sigma}=\eta^H_{\mu\nu\rho\sigma}$. Thus, in a two-dimensional space with symmetric stress tensor, the Hall viscosity has three independent components and it can be condensed into a two-component symmetric tensor. This is done in Ref.~\onlinecite{haldane2015geometry}. We follow their example and define the contracted Hall viscosity to be
\begin{align}\label{eq:contracted Hall viscosity}
    \eta^H_{ab}\equiv \frac{1}{2}\epsilon_{ac}\epsilon_{bd}\epsilon_{ef}\eta^H_{cedf}.
\end{align}
Even when the stress tensor is not symmetric, it is useful to examine the anisotropy via the contracted rather than the full Hall viscosity. For instance, Ref.~\onlinecite{gromov2017investigating} outlines a bimetric approach to characterizing anisotropy in the 2D quantum Hall fluid, one which we will use to frame and better understand our results; its point of contact with the Hall viscosity is through $\eta_{ab}^H$. Additionally, we shall see that for the system with an anisotropic mass tensor--which does not have a symmetric stress tensor--both $\eta_{abcd}^H$ and $\eta^H_{ab}$ are dependent only on the mass tensor. Thus, for both systems we examine $\eta_{ab}^H$ fully determines $\eta_{abcd}^H$.

We have now derived the general expressions related to the Hall viscosity that we will use to analyze the IQH system with either mass anisotropy or tilted field anisotropy. We will now move on to analyze these systems in detail
Since we will derive expressions for the stress tensors directly from the momentum density continuity equations, we will not need the explicit expressions for $J_{\mu\nu}$.

\section{\label{sec:2D band mass}2D System with Mass Anisotropy}
Let us set the stage by analyzing the linear response functions for a quantum Hall system with mass anisotropy (sometimes called band mass anisotropy\cite{yang2012band}). The first-quantized $N$ particle Hamiltonian is given by
\begin{equation}\label{eq:band mass Hamiltonian}
    H_{AM}=\frac{1}{2}\sum_{i=1}^N\tilde{m}_{ab}\pi^i_a\pi^i_b,\hspace{0.5cm}\nabla\times \vec{A}=B\hat{z},
\end{equation}
where the inverse mass tensor $\tilde{m}_{ab}$ is symmetric and invertible. The results derived here provide context for the results derived for the tilted field system. For simplicity, we consider the single particle (i.e., $N=1$) case, but it is easy to subsequently determine the multiparticle results: because we neglect interactions and because the relevant expectation values are all constant within the highly degenerate lowest energy eigenspace, the ground state observables we examine all scale linearly with the number of electrons.

We begin by diagonalizing the single particle Hamiltonian, following Ref.~\onlinecite{qiu2012model}. The relevant commutators between the position and the physical momenta are
\begin{align}
[x_a,\pi_b]&=i\delta_{ab},\label{eq:BM x and pi commutator} \\\relax [\pi_a,\pi_b]&=-iB\epsilon_{ab}.\label{eq:BM pi and pi commutator}
\end{align}
We can decompose the mass tensor $m_{ab}$ and its inverse $\tilde{m}_{ab}$, satisfying $m_{ab}\tilde{m}_{bc}=\tilde{m}_{ab}m_{bc}=\delta_{ac}$, in terms of complex vectors $\vec{\mu}=(\mu_x,\mu_y)$ and $\vec{\nu}=(\nu_x,\nu_y)$ by writing $m_{ab}=m(\mu_a^*\mu_b+\mu_b^*\mu_a)$ and $\tilde{m}_{ab}=\frac{1}{m}(\nu_a^*\nu_b+\nu_b^*\nu_a)$. In terms of the components of $m_{ab}$, $\vec{\mu}$ and $\vec{\nu}$ are given by
\begin{align}
(\mu_x,\mu_y)&=\frac{e^{i\phi}}{\sqrt{2m}}\left(\sqrt{m_{11}},\sqrt{m_{22}}e^{i\varphi}\right),\nonumber\\(\nu_x,\nu_y)&=\frac{e^{i\theta}}{2i\text{Im}(\mu_x\mu_y^*)}(-\mu_y,\mu_x).\label{eq:expression for nu vector}
\end{align}
where $\varphi=\arccos{\frac{m_{12}}{\sqrt{m_{11}m_{22}}}}$. The parameters $m$, $\phi$ and $\theta$ are redundancies in the description and we use the freedom to set $\theta=\phi=0$ and $m=\sqrt{\text{det}(m_{ab})}$, the latter of which ensures that the matrices $\frac{1}{m}m_{ab}$ and $m\tilde{m}_{ab}$ are the mass and inverse mass tensors rescaled to have unit determinant. We can now write down some useful, interrelated identities:
\begin{align}
    \text{Im}(\mu_a\mu_b^*)&=\text{Im}(\nu_a\nu_b^*)=-\frac{1}{2}\epsilon_{ab},\label{eq:omega nu identity 1}\\\text{Re}(\mu_a\nu_b^*)&=\frac{1}{2}\delta_{ab},\label{eq:omega nu identity 2}\\\nu_a&=-i\epsilon_{ab}\mu_b,\label{eq:omega nu identity 3}\\\mu_a&=-i\epsilon_{ab}\nu_b,\label{eq:omega nu identity 4}\\\mu_a\nu_a&=-i\epsilon_{ab}\mu_a\mu_b=-i\epsilon_{ab}\nu_a\nu_b=0,\label{eq:omega nu identity 5}\\\mu_a\nu_a^*&=i\epsilon_{ab}\mu_a\mu_b^*=i\epsilon_{ab}\nu_a\nu_b^*=1.\label{eq:omega nu identity 6}
\end{align}
Next, we define $b\equiv \frac{1}{\sqrt{B}}\nu_a^*\pi_a$ so that the Hamiltonian becomes 
\begin{align}\label{eq:BM Hamiltonian diagonalized}
    H_{AM}=\frac{\omega_\circ}{2}(bb^\dagger +b^\dagger b),
\end{align}
where $\omega_\circ\equiv \frac{B}{m}$ is the cyclotron frequency. From Eqs.~(\ref{eq:BM pi and pi commutator}) and (\ref{eq:omega nu identity 1}), we deduce $[b,b^\dagger]=1$. The Hamiltonian is therefore diagonalized in terms of the raising/lowering operators $(b^\dagger,b)$.

A second lowering operator is given by $a=\sqrt{B}\mu_aR_a$, where $R_a=x_a-\frac{\epsilon_{ab}}{B}\pi_b$ is the guiding center coordinate. It satisfies $[R_a,x_b]=\frac{i}{B}\epsilon_{ab}$, $[R_a,\pi_b]=0$ and $[R_a,R_b]=\frac{i}{B}\epsilon_{ab}$, from which it follows that $[a,a^\dagger]=1$ and $[a,b]=[a,b^\dagger]=0$. The energy eigenstates are then given by
\begin{align}\label{eq:BM eigenstates}
    \ket{m,n}=\frac{(a^\dagger)^m(b^\dagger)^n}{\sqrt{m!n!}}\ket{0},
\end{align}
where the ground state $\ket{0}$ is annihilated by $a$ and $b$. The corresponding energies are $E_{mn}=\frac{\omega_\circ}{2}\left(n+\frac{1}{2}\right)$.

The operators $(b^\dagger,b)$ and $(a^\dagger,a)$ have particularly simple time evolution, which is handy when working with equation  (\ref{eq:expression for conductivity}). Specifically, solving $\partial_tb(t)=i[H,b]=-i\omega_\circ b$ and $\partial_ta(t)=i[H,b]=0$, we find
\begin{align}
a(t)&=a(0),&a^\dagger(t)&=a^\dagger(0),\label{eq:time dependence of a, BM}\\b(t)&=b(0)e^{-i\omega_\circ t},&b^\dagger(t)&=b^\dagger(0)e^{i\omega_\circ t}.\label{eq:time dependence of b, BM}
\end{align}
Finally, from the definition of $b$ and Eq.~(\ref{eq:omega nu identity 2}), we see that the momenta can be expressed simply in terms of the $(b^\dagger,b)$ operators:
\begin{align}\label{eq:BM momentum in terms of ladder ops}
    \pi_a=\sqrt{B}(\mu_ab+\mu_a^*b^\dagger).
\end{align}

\subsection{Hall conductivity}\label{sec:BM Hall conductivity}
We next apply the linear response formalism to calculate the Hall conductivity for the mass anisotropic system. Given the integrated current $J_a=-\dot{x}_a=-i[H_{AM},x_a]=-\tilde{m}_{ab}\pi_b$, Eq.~(\ref{eq:expression for conductivity}) yields
\begin{align}\label{eq:BM conductivity}
    \sigma_{ab}(\omega)&=\frac{\tilde{m}_{ac}\tilde{m}_{bd}}{\omega^+L^2}\int_0^\infty dt e^{i\omega^+t}\braket{[\pi_c(t),\pi_d(0)]}_0\nonumber\\&=\frac{B\tilde{m}_{ac}\tilde{m}_{bd}}{\omega^+L^2}\int_0^\infty dt e^{i\omega^+t}\left(\mu_c\mu_d^*e^{-i\omega_\circ t}-\text{h.c.}\right)\nonumber\\&=\frac{B\tilde{m}_{ac}\tilde{m}_{bd}}{i\omega^+L^2}\left[\frac{\mu_c^*\mu_d}{\omega^++\omega_\circ}-\frac{\mu_c\mu_d^*}{\omega^+-\omega_\circ}\right].
\end{align}
And therefore the Hall conductivity takes the form
\begin{align}\label{eq:BM Hall conductivity}
    \sigma_{ab}^H(\omega)&=\frac{B}{iL^2}\tilde{m}_{ac}\tilde{m}_{bd}(\mu_c^*\mu_d-\mu_c\mu_d^*)\frac{1}{\omega^+}\frac{\omega^+}{(\omega^+)^2-\omega_\circ^2}\nonumber\\&=-\frac{1}{BL^2}m^2\tilde{m}_{ac}\tilde{m}_{bd}\epsilon_{cd}\left(1-\frac{(\omega^+)^2}{\omega_\circ^2}\right)^{-1}\nonumber\\&=-\frac{1}{BL^2}\epsilon_{ab}\left(1-\frac{(\omega^+)^2}{\omega_\circ^2}\right)^{-1}.
\end{align}

In getting from the second to third line, we used the identity  $\epsilon_{a_1\ldots a_n}M_{a_1b_1}\ldots M_{a_nb_n}=\text{det}(M)\epsilon_{b_1\ldots b_n}$ for any matrix $M$ and the fact that $m\tilde{m}_{ab}$ by definition has unit determinant.

If we take the limit $\omega^+\to 0$, reinsert factors of $e$ and include the contributions from all $N$ electrons, we find that the zero-frequency Hall conductivity is given by
\begin{align}\label{eq:BM Hall conductivity full expression}
    \sigma_{ab}^H=-\frac{e\rho}{B}\epsilon_{ab},
\end{align}
where $\rho\equiv \frac{N}{L^2}$ is the electron density. Using the fact that there is one electron state per quantum of magnetic flux per energy level, the ground state electron density, $\rho=\frac{Be}{2\pi \hbar}$ at $\nu=1$ filling, is independent of the anisotropy. Therefore, the conductivity in the ground state is independent of the anisotropy.

One can also directly compute the symmetric part of the Hall conductivity:
\begin{align}
\sigma_{(ab)}(\omega)&=\frac{B\tilde{m}_{ac}\tilde{m}_{bd}(\mu_c^*\mu_d+\mu_c\mu_d^*)}{L^2}\frac{1}{i\omega^+}\frac{\omega_\circ}{-(\omega^+)^2+\omega_\circ^2}\nonumber\\&+\frac{i\tilde{m}_{ab}}{\omega^+ L^2}\nonumber\\&=\frac{\tilde{m}_{ab}}{L^2}\frac{1}{i\omega^+}\left(\frac{1}{1-\frac{\omega_+^2}{\omega_\circ^2}}-1\right),
\end{align}
where we were careful to include the contact term from Eq.~\ref{eq:expression for conductivity}. Taking the $\omega^+\to 0$ limit, we find $\sigma_{(ab)}=0$, as expected for a dissipative response function in a gapped system at zero frequency.

\subsection{\label{subsec:BM stress tensor and Hall viscosity}Stress tensor and Hall viscosity}
To calculate the Hall viscosity via Eq.~(\ref{eq:Hall viscosity with CSS inserted}), we seek an operator $\tau_{ab}$ that in the Heisenberg picture obeys the momentum continuity equation:
\begin{align}\label{eq:BM continuity equation}
    \frac{\partial g_a}{\partial t}+\frac{\partial\tau_{ba}}{\partial r_b}=f^{L}_a,
\end{align}
where the Lorentz force density is given by $f^L_a=B\epsilon_{ab}j_b=-B\epsilon_{ab}\tilde{m}_{bc}g_c$. After a few lines of simplifying commutation relations, we find
\begin{align}\label{eq:BM continuity equation 2}
    \frac{\partial g_a(\vec{r},t)}{\partial t}&=i[H_{AM},g_a(\vec{r},t)]=-B\tilde{m}_{bc}\epsilon_{ab}g_c(\vec{r},t)\nonumber\\&-\frac{\partial}{\partial r_c}\left(\frac{1}{4}\tilde{m}_{bc}\{\{\pi_b,\delta(\vec{r}-\vec{x})\},\pi_a\}\right).
\end{align}
It follows therefore that $\tau_{ab}$ is given by
\begin{align}\label{eq:BM intrinsic stress tensor}
    \tau_{ab}=\frac{\tilde{m}_{ac}}{4}\{\pi_b,\{\pi_c,\delta(\vec{r}-\vec{x})\}\},
\end{align}
up to a term whose divergence is zero. Integrating over space, we arrive at
\begin{align}\label{eq:BM integrated stress tensor}
    T_{ab}=\frac{\tilde{m}_{ac}}{2}\{\pi_b,\pi_c\}.
\end{align}
Using the equations of motion, we could alternatively express this in terms of the velocity $v_a$ as $T_{ab}=\{v_a,\pi_b\}/2$. The expectation value of the integrated stress tensor in the ground state is then
\begin{align}\label{eq:BM integrated stress tensor is ground state}
    \braket{T_{ab}}_0=\frac{\tilde{m}_{ac}}{2}\braket{0|\{\pi_c,\pi_b\}|0}=\frac{\omega_\circ}{2}\delta_{ab}.
\end{align}
Reinserting factors of $\hbar$, including the contribution of all $N$ electrons, and dividing by the area to get the intensive stress tensor, we find
\begin{align}\label{eq:BM intrinsic stress tensor in ground state full expression}
\braket{\tau_{ab}}_0=\frac{\hbar\omega_\circ\rho}{2}\delta_{ab}.
\end{align}
This result is also independent of the anisotropy.

Because the ground state expectation value of the stress tensor is rotationally invariant, the Hall elastic modulus is zero. Meanwhile, we compute the Hall viscosity from Eq.~(\ref{eq:Hall viscosity with CSS inserted}). After some straightforward computation\cite{gromov2017investigating,haldane2015geometry,bradlyn2012kubo}, we find
\begin{align}\label{eq:BM Hall viscosity}
    \eta_{abcd}^H&=-\frac{\tilde{m}_{ae}\tilde{m}_{cf}}{2L^2}\sum_{n=1 }^\infty\frac{\text{Im}\left(\braket{0|\{\pi_b,\pi_e\}|n}\braket{n|\{\pi_d,\pi_f\}|0}\right)}{(\mu_cn)^2}\nonumber\\&=-\frac{1}{L^2}\text{Im}(\nu_a\mu_b\nu_c^*\mu_d^*).
\end{align}

By using the diagonalizability of $\tilde{m}_{ab}$, we can put Eq.~(\ref{eq:BM Hall viscosity}) in a more familiar form. Assuming, without loss of generality, that $\tilde{m}_{ab}=m\text{ diag}(\alpha,1/\alpha)$, in which case $\vec{\mu}=\frac{1}{\sqrt{2}}\left(\sqrt{\alpha},\frac{i}{\sqrt{\alpha}}\right)$ and $\vec{\nu}=\frac{1}{\sqrt{2}}\left(\frac{1}{\sqrt{\alpha}},\sqrt{\alpha}i\right)$, we find
\begin{align}\label{eq:omega nu extra identity}
    \mu_a\nu_b^*=\left(\begin{array}{cc}1 & -i\alpha\\ i/\alpha & 1\end{array}\right)_{ab}=\frac{1}{2}\left(\delta_{ab}-i\epsilon_{ac}(m\tilde{m}_{bc})\right).
\end{align}
Therefore, including factors of $\hbar$ and the contributions from all $N$ electrons, we find
\begin{align}\label{eq:BM Hall viscosity full expression}
    \eta_{abcd}^H=\frac{\hbar\rho}{4}\left[\delta_{ad}\epsilon_{be}(m\tilde{m}_{ce})+\delta_{bc}\epsilon_{ed}(m\tilde{m}_{ae})\right].
\end{align}
It is also straightforward to compute the contracted Hall viscosity:
\begin{align}\label{eq:BM contracted Hall viscosity full expression}
    \eta_{ab}^H=\frac{\hbar\rho}{4}\left(\frac{1}{m}m_{ab}\right).
\end{align} 
Notice that the determinant of $\eta_{ab}^H$ determines the electron density.

\section{\label{sec:3D tilted field}3D system with tilted field anisotropy}
\begin{figure}[t]
\includegraphics[width=0.4\textwidth]{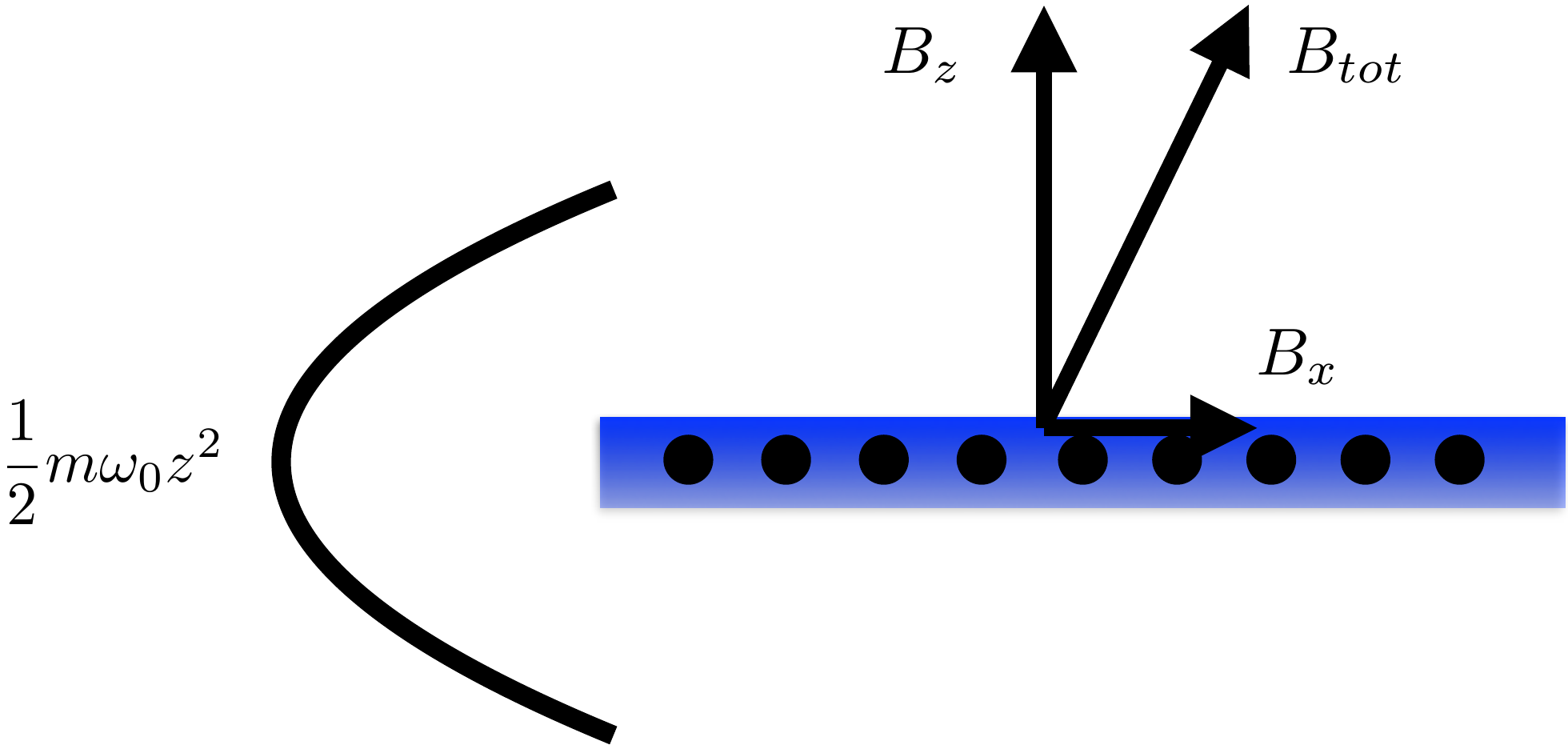}
\caption{Setup for the tilted-field system. We consider electrons (black circles) in a quantum well (blue region) moving in the background of a perpendicular magnetic field $B_z$, along with an in-plane field $B_x$. The electrons are confined to the quantum well by a harmonic potential $V=1/2m\omega_0^2z^2$, depicted on the left.}\label{fig:3D setup}
\end{figure}

Now that we have seen how the conductivity, stress tensor, and viscosity emerge with mass anisotropy, we are ready to tackle our main objective: characterizing the quantum Hall system in a tilted magnetic field. The physical system we consider is that of an electron moving in a confining potential $V(x,y,z)=\frac{1}{2}m\omega_0^2z^2$ and a background magnetic field $\vec{B}=B_x\hat{x}+B_z\hat{z}$. The $N$ particle Hamiltonian is given by
\begin{align}\label{eq:tilted field Hamiltonian}
    H_{TF}&=\frac{1}{2m}\sum_{i=1}^N\pi^i_\mu\pi_\mu^i+\frac{1}{2}m\omega_0^2(z^i)^2,\nonumber
    \\&\hspace{2cm}\nabla\times\vec{A}=B_x\hat{x}+B_z\hat{z}.
\end{align}
Note, because the system is fully three-dimensional, $\mu$ takes the values $1,2,3$. Additionally, as in the mass anisotropic system, we let $N=1$ to simplify calculations and include the multiple electron contributions to the response functions at the end. We show the setup schematically in Fig.~\ref{fig:3D setup}.

It is convenient to introduce the cyclotron frequencies $\omega_x\equiv \frac{B_x}{m}$ and $\omega_z\equiv\frac{B_z}{m}$. Furthermore, since we are interested in the regime of weak tilt and strong confinement, we take $\omega_x\ll \omega_z\ll\omega_0$. We will thus expand many of our results in powers of 
\begin{equation}
\ell \equiv \frac{\omega_z}{\omega_0}\;\; \mathrm{and}\;\; k\equiv \frac{\omega_x}{\omega_z}.
\end{equation}

We will see that it is convenient to keep only terms of order $k^2\ell^2$ and lower. Assuming $k$ and $\ell$ to be small in an experimental setting is reasonable. First, the the angle of the magnetic field can relative to the sample can be tuned to make $k$ small. Second, the typical depth of a GaAs quantum well is on the order of $\sim 100 meV$\cite{adachi1985gaas,gaaswell}, and the magnetic field necessary to generate a cyclotron frequency of the same magnitude is on the order of $\sim 100T$, well above the fields needed to observe the quantum Hall effect even in the lowest Landau level
\cite{klitzing1980new, tsui1982two}. Typical experiments will thus have $\ell \sim 10^{-2}$.

Our first order of business is to diagonalize the tilted field Hamiltonian. The relevant commutators between the position operators and physical momenta are
\begin{align}
[x_\mu,\pi_\nu]&=i\delta_{\mu\nu},\label{eq:TF commutation position and momenta relations 1}\\\relax [\pi_\mu,\pi_\nu]&=-i\epsilon_{\mu\nu\rho}B_\rho.\label{eq:TF commutation position and momenta relations 2}
\end{align}
Once the Hamiltonian is diagonalized, we proceed to calculating the Hall conductivity, stress tensor and Hall viscosity, just as in the anisotropic mass case.

\subsection{\label{subsec:Hamiltonian diagonalization with ladder ops}Diagonalizing the Hamiltonian with gauge-invariant ladder operators}

We will diagonalize the tilted field Hamiltonian using manifestly gauge invariant operators. Our discussion will closely follow that of Ref.~\onlinecite{yang2017anisotropic}, which, since it is central to the narrative of our work, we recapitulate here. We will use this review to establish notation, and to make some additional remarks which will prove important moving forward.

We begin by writing the Hamiltonian as
\begin{align}
H_{TF}&=\frac{1}{2m}(\pi_x^2+\pi_y^2+\pi_z^2+(m\omega_0z)^2)\label{eq:TF gauge indep Hamiltonian 1}\\&=\frac{\omega_z}{2}(a^\dagger a+aa^\dagger)+\frac{\omega_0}{2}(b^\dagger b+bb^\dagger).\label{eq:TF gauge indep Hamiltonian 2}
\end{align}
In the second line, we have introduced the lowering operators
\begin{align}
    a&=\frac{1}{\sqrt{2B_z}}(\pi_x-i\pi_y), \label{eq:TF gauge indep lowering operator a}\\b&=\frac{1}{\sqrt{2m\omega_0}}(\pi_z-im\omega_0z).\label{eq:TF gauge indep lowering operator b}
\end{align}
They obey $[a,a^\dagger]=[b,b^\dagger]=1$ and $[a,b]=[a,b^\dagger]=-\frac{1}{2}\frac{B_x}{\sqrt{B_zm\omega_0}}=-\frac{1}{2}\frac{\omega_x}{\sqrt{\omega_z\omega_0}}$. To work with two decoupled oscillators, we define
\begin{align}
    \alpha&=a+\frac{\omega_x}{2\sqrt{\omega_z\omega_0}}(b-b^\dagger),\label{eq:TF gauge indep lowering operator alpha}
\end{align}
which satisfies $[\alpha,\alpha^\dagger]=1$ and $[\alpha,b]=[\alpha,b^\dagger]=0$. The cost of working with $\alpha$ instead of $a$ is that the Hamiltonian is not diagonal in terms of $\alpha$ and $b$:
\begin{align}
H_{TF}&=\frac{\omega_z}{2}\{\alpha,\alpha^\dagger\}+\frac{1}{2}\left(\omega_0+\frac{\omega_x^2}{2\omega_0}\right)\{b,b^\dagger\}\nonumber\\&-\frac{\omega_x^2}{4\omega_0}(b^2+\text{h.c.})+\frac{\omega_x}{2}\sqrt{\frac{\omega_z}{\omega_0}}(\alpha^\dagger b^\dagger-\alpha^\dagger b+\text{h.c.}).\label{eq:TF gauge indep Hamiltonian 3}
\end{align}

To diagonalize the Hamiltonian, we use the Bogoliubov transformation
\begin{equation}\label{eq:TF defining X and Y ops}
[b^\dagger,\alpha^\dagger,b,\alpha]^T=U[X^\dagger,Y^\dagger,X,Y]^T,
\end{equation}
which is effected by the matrix
\begin{align}
U&=\frac{1}{2\sqrt{\omega_2^2-\omega_1^2}}\left(\begin{array}{cc}U_1^\dagger & -U_2^T \\ -U_2^\dagger & U_1^T\end{array}\right),\label{eq:TF U}\\
U^{-1}&=\frac{1}{2\sqrt{\omega_2^2-\omega_1^2}}\left(\begin{array}{cc}U_1 & U_2\\ U_2^*&  U_1^*\end{array}\right),\label{eq:TF U inverse}
\end{align}
where the sub-matrices $U_1$ and $U_2$ are
\begin{align}
    U_1&=\left(\begin{array}{cc}i(\omega_z+\omega_2)\sqrt{\frac{\omega_z^2-\omega_1^2}{\omega_z\omega_2}} & i(\omega_z+\omega_1)\sqrt{\frac{\omega_2^2-\omega_z^2}{\omega_z\omega_1}}\\ (\omega_z+\omega_1)\sqrt{\frac{\omega_2^2-\omega_z^2}{\omega_z\omega_1}} & -(\omega_z+\omega_2)\sqrt{\frac{\omega_z^2-\omega_1^2}{\omega_z\omega_2}}\end{array}\right),\label{eq:TF U1}\\U_2&=\left(\begin{array}{cc}i(\omega_2-\omega_z)\sqrt{\frac{\omega_z^2-\omega_1^2}{\omega_z\omega_2}} & i(\omega_z-\omega_1)\sqrt{\frac{\omega_2^2-\omega_z^2}{\omega_z\omega_1}}\\ -(\omega_z-\omega_1)\sqrt{\frac{\omega_2^2-\omega_z^2}{\omega_z\omega_1}} & (\omega_2-\omega_z)\sqrt{\frac{\omega_z^2-\omega_1^2}{\omega_z\omega_2}}\end{array}\right).\label{eq:TF U2}
\end{align}

In terms of the newly introduced gauge invariant ladder operators $X,Y$, which satisfy the usual ladder commutation relations $[X,X^\dagger]=[Y,Y^\dagger]=1$, $[X,Y]=[X,Y^\dagger]=0$, the Hamiltonian takes the desired form:
\begin{align}\label{eq:TF gauge indep Hamiltonian diagonalized}
H_{TF}&=\frac{\omega_1}{2}(X^\dagger X+XX^\dagger)+\frac{\omega_2}{2}(Y^\dagger Y+YY^\dagger).
\end{align}
We have introduced the tilted field eigenfrequencies, 
\begin{align}
    \omega_1^2&=\frac{1}{2}\left(\epsilon_1-\sqrt{\epsilon_1^2-\epsilon_2^2}\right),\label{eq:TF gauge indep eigenfrequency 1}\\\omega_2^2&=\frac{1}{2}\left(\epsilon_1+\sqrt{\epsilon_1^2-\epsilon_2^2}\right).\label{eq:TF gauge indep eigenfrequency 2}
\end{align}
where $\epsilon_1=\omega_z^2+\omega_0^2+\omega_x^2$, $\epsilon_2=2\omega_0\omega_z$. We note that when $k,\ell\ll 1$, $\omega_1\approx \omega_z\left(1-\frac{k^2\ell^2}{2}\right)$ and $\omega_2\approx \omega_0\left(1+\frac{k^2\ell^2}{2}\right)$. In the regime we are primarily interested in, therefore, $X^\dagger$ and $X$ have the interpretation of being the raising and lowering operators in the plane, while $Y^\dagger$ and $Y$ have the interpretation of being the raising and lowering operators along the $z$-axis. We confirm this interpretation by writing $X$ and $Y$ in terms of $\pi_x$, $\pi_y$, $\pi_z$ and $z$; the explicit expressions can be found in Appendix~\ref{app:important results}. Although $X$ has contributions from $\pi_z$ and $z$, they vanish as $k\to 0$. Likewise, $Y$ has contributions from $\pi_x$ and $\pi_y$, which also vanish as $k\to 0$. In later sections, we will focus on the planar limit of the tilted field results; viewing $X$ as the in-plane ladder operator and $Y$ as the out-of-plane ladder operator will prove useful.

Typically, the energy eigenstates of a scalar particle in three dimensions are labeled by three quantum numbers. Thus, we expect there to be a third lowering operator alongside $X$ and $Y$. Indeed, one is given by
\begin{align}\label{eq:TF third gauge indep lowering op}
    c&=\alpha^\dagger-i\sqrt{\frac{B_z}{2}}\left(x+iy\right).
\end{align}
It obeys $[c,c^\dagger]=1$, $[c,\alpha]=[c,\alpha^\dagger]=[c,b]=[c,b^\dagger]=0$ and therefore also $[c,X]=[c,X^\dagger]=[c,Y]=[c,Y^\dagger]=0$.

Letting $\ket{0}$ be the state annihilated by $X$, $Y$ and $c$, the energy eigenstates of the tilted field Hamiltonian are given by
\begin{align}\label{eq:TF gauge indep energy eigenstates}
    \ket{m,n,p}=\frac{(X^\dagger)^m(Y^\dagger)^n(c^\dagger)^p}{\sqrt{m!n!p!}}\ket{0},
\end{align}
with corresponding energy eigenvalues $E_{mnp}=\omega_1\left(m+\frac{1}{2}\right)+\omega_2\left(n+\frac{1}{2}\right)$.

Just as in the anisotropic mass case, the ladder operators have particularly simple time dependence, which simplifies subsequent calculations:
\begin{align}
c(t)&=c(0),&c^\dagger(t)&=c^\dagger(0),\label{eq:time dependence of c, TF}\\X(t)&=X(0)e^{-i\omega_1 t},&X^\dagger(t)&=X^\dagger(0)e^{i\omega_1 t},\label{eq:time dependence of X, TF}\\Y(t)&=Y(0)e^{-i\omega_2 t},&Y^\dagger(t)&=Y^\dagger(0)e^{i\omega_2 t}.\label{eq:time dependence of Y, TF}
\end{align}

Furthermore, we will make frequent use of the expressions for the physical momenta in terms of the $(\alpha^\dagger, \alpha)$ and $(b^\dagger,b)$ operators or, more often, the $(X^\dagger,X)$ and $(Y^\dagger,Y)$ operators. From Eqs.~(\ref{eq:TF gauge indep lowering operator a})-(\ref{eq:TF gauge indep lowering operator alpha}), we find

\begin{align}\label{eq:TF relating pis and alpha,b}
\left(\frac{\pi_x}{\sqrt{B_z}},\frac{\pi_y}{\sqrt{B_z}},\frac{\pi_z}{\sqrt{B_z}},\sqrt{B_z}z\right)^T&=V(b^\dagger,\alpha^\dagger,b,\alpha)^T,
\end{align}
where
\begin{align}\label{eq:TF V}
V&=\frac{1}{\sqrt{2}}\left(\begin{array}{cccc}0&1&0&1\\ik\ell^{\frac{1}{2}}&-i&-ik\ell^{\frac{1}{2}}&i\\\ell^{-\frac{1}{2}}&0&\ell^{-\frac{1}{2}}&0\\-i\ell^{\frac{1}{2}}&0&i\ell^{\frac{1}{2}}&0\end{array}\right),\\V^{-1}&=\frac{1}{\sqrt{2}}\left(\begin{array}{cccc}0&0&\ell^{\frac{1}{2}}&i\ell^{-\frac{1}{2}}\\1&i&0&ik\\0&0&\ell^{\frac{1}{2}}&-i\ell^{-\frac{1}{2}}\\1&-i&0&-ik\end{array}\right).\label{eq:TF V inverse}
\end{align}
We may therefore combine Eqs.~(\ref{eq:TF defining X and Y ops}) and (\ref{eq:TF relating pis and alpha,b}) to relate the physical momenta and $z$ to the $X$ and $Y$ operators. Namely,
\begin{align}\label{eq:TF relating pis and X,Y}
\left(\frac{\pi_x}{\sqrt{B_z}},\frac{\pi_y}{\sqrt{B_z}},\frac{\pi_z}{\sqrt{B_z}},\sqrt{B_z}z\right)^T&=W(X^\dagger,Y^\dagger,X,Y)^T,
\end{align}
where we have introduced $W\equiv UV$.

Before moving on, let us derive the Landau level degeneracy in the presence of a tilted field. Consider a system with periodic boundary conditions in the $x-$ and $y-$ directions, with length $L_x$ and $L_y$ respectively. Our strategy will be to identify a pair of unitary, non-commuting, spatially periodic magnetic translation operators\cite{zak1964magnetic} $W_x$ and $W_y$, satisfying
\begin{align}
W_xW_y&=e^{2\pi i/M}W_yW_x,\\
[H,W_x]&=[H,W_y]=0.
\end{align}
Working in a basis of $W_x$ eigenstates, we see that $W_y$ multiplies the $W_x$ eigenvalue by $e^{2\pi i/M}$; we can do this $M$ times before we return to the original state due to the periodicity of the exponential, and so we conclude that the degeneracy of each level is $M$\footnote{strictly speaking, all we can deduce is that the degeneracy of each level is an integer multiple of $M$, since $W_x$ may have degenerate eigenstates. However, in the current case of interest, the eigenstates of $W_x$ will be nondegenerate.}.

In our case, we can identify $W_x$ and $W_y$ with the exponentials of the magnetic translation generators constructed out of the $c,c^\dag$. These generators, denoted $w_x$ and $w_y$ can be written:
\begin{align}
w_x&=\sqrt{\frac{B_z}{2}}\left(c+c^\dag\right),\\
w_y&=i\sqrt{\frac{B_z}{2}}\left(c^\dag-c\right).
\end{align}
The key observation is that these differ from the ordinary two-dimensional magnetic translation generators only by terms involving $z$. This reflects the fact that the tilted field system still retains magnetic translation symmetry in the $x-y$ plane. Note that $w_x$ and $w_y$ are, up to a factor of $B_z$, the guiding center coordinates for the titled field system. Imposing periodic boundary conditions, we see that $w_x$ and $w_y$ themselves are not well-defined operators on our system. However, we can define the manifestly periodic exponentiated translations\cite{fradkin2013field}
\begin{align}
W_x&=e^{2\pi iw_x/(B_zL_y)}, \\
W_y&=e^{2\pi iw_y/(B_zL_x)},
\end{align}
which ensure that translations across the entire system commute with all smaller translations. It follows from the commutation relations then that
\begin{equation}
W_xW_y=W_yW_xe^{-4\pi^2i/\Phi},
\end{equation}
where we have defined the flux $\Phi=B_zL_xL_y$. We thus see from our previous argument that the tilted-field Landau level degeneracy is
\begin{equation}\label{eq:ground state degeneracy}
M=\frac{\Phi}{\Phi_0}=\frac{B_zL_xL_y}{2\pi},
\end{equation}
independent of $B_x$, just as in the untilted case.

\subsection{\label{subsec:TF Hall conductivity}Hall conductivity}
Having diagonalized the Hamiltonian, we may apply the linear response formalism to determine the Hall conductivity. In particular, we will be interested in computing the effective 2D conductivity obtained by integrating along the anisotropic $z$ direction. As in the isotropic case, the current operator is given by $J_\mu=-\dot{x}_\mu=-\frac{1}{m}\pi_\mu$. Using Eqs.~(\ref{eq:effective 2d response}), (\ref{eq:Hall conductivity with CSS inserted}) and (\ref{eq:TF relating pis and X,Y}) , we have
\begin{align}\label{eq:TF Hall conductivity calculation}
\sigma_{\mu\nu}^H&=\frac{2}{m^2L_xL_y}\nonumber\\&\hspace{0.5cm}\times\sum_{(n_X,n_Y)\neq (0,0)}^\infty\frac{\text{Im}\left(\braket{0|\pi_{\mu}|n_X,n_Y}\braket{n_X,n_Y|\pi_\nu|0}\right)}{(n_X\omega_1+n_Y\omega_2)^2}\nonumber\\&=\frac{2B_z}{m^2L^3}\text{Im}\left(\frac{W_{\mu3}W_{\nu1}}{\omega_1^2}+\frac{W_{\mu4}W_{\nu2}}{\omega_2^2}\right),
\end{align}
By manipulating the exact expressions for the components of $W$ and $\omega_1$, $\omega_2$, we find that the Hall conductivity simplifies greatly. Inserting factors of $e$ and including the contributions of all $N$ electrons, we find
\begin{align}\label{eq:TF Hall conductivity}
 \sigma_{\mu\nu}^H&=\frac{e\rho}{B_z}\left(\begin{array}{ccc}0 & -1& 0\\1 & 0 & 0\\ 0 & 0 & 0\end{array}\right),
\end{align}
where $\rho\equiv \frac{N}{L_xL_y}$. This is an exact result, valid for any tilt angle. Evidently, the Hall conductivity does not depend on the tilted component of the magnetic field, $B_x$, nor on the confining potential, $\omega_0$. And, as we determined in Eq.~(\ref{eq:ground state degeneracy}), $\rho=B_z/\phi_0$ at $\nu=1$ filling, so the $\nu=1$ ground state Hall conductivity does not depend on the perpendicular magnetic field strength either, which is the usual result. Note also that $\sigma_{(\mu\nu)}=0$ at zero frequency because the tilted field system is gapped; hence $\sigma_{\mu\nu}=\sigma_{\mu\nu}^H$.

If there were no confining potential, the Hall conductivity would be that of an isotropic system, suitably rotated about the $y$-axis. Namely:
\begin{align}\label{eq:TF Hall conductivity no confinement}
    \sigma_{\mu\nu}^H(\omega_0=0)&=\frac{e\rho}{B_x^2+B_z^2}\left(\begin{array}{ccc}0 & -B_z& 0\\B_z & 0 & B_x\\ 0 & -B_x & 0\end{array}\right).
\end{align}
Comparing (\ref{eq:TF Hall conductivity}) and (\ref{eq:TF Hall conductivity no confinement}), we see that the Hall conductivity tensor is discontinuous at $\omega_0=0$ (except when $B_x=0$), which makes Eq.~(\ref{eq:TF Hall conductivity}) at first glance a somewhat surprising result.

Nonetheless, we can present a qualitative argument why $\sigma_{xz}^H$ and $\sigma_{yz}^H$ should be zero in the presence of the confining potential, no matter how weak it is. These components of the conductivity tensor capture the current in the $z$-direction in response to an applied electric field in the $x$ or $y$ direction. In order for a state to carry current in the $z$ direction, it must be extended in that direction (or more precisely, there must exist a gauge in which it is extended). However, it is not possible for a state with finite energy to be extended in a quadratic well, regardless of the well's shallowness. Consider the simple case of the one-dimensional quantum harmonic oscillator. The average energy of a state $\ket{\psi}$ is then $\braket{E}\sim \braket{x^2}/2+\braket{p^2}/2$. Since $\Delta^2=\braket{x^2}-\braket{x}^2$ is the variance of the position of the particle, we see that if the variance is infinite then the average energy is as well.

The discontinuity of the Hall conductivity at $\omega_0=0$ also reflects the fact that the energy gap above the ground state closes as $\omega_0\to 0$. The number of states available to the ground state electrons is fixed while the gap is open and increases dramatically when the gap closes; only when the number of states increases does the conductivity change. It is simple to explicitly check that $\omega_1\to 0$ as $\omega_0\to 0$.

\subsection{\label{subsec:TF stress tensor}Stress tensor}
In order to discuss the Hall viscosity, we next derive an expression for the integrated stress tensor, again via the continuity equation for the momentum density. This time it is given by
\begin{align}\label{eq:TF continuity equation}
    \frac{\partial g_\mu}{\partial t}+\frac{\partial \tau_{\nu\mu}}{\partial r_\nu}=f^L_\mu+f^C_\mu,
\end{align}
where the Lorentz force density is $f^L_\mu=\epsilon_{\mu\nu\rho}j_\nu B_\rho=-\frac{1}{m}\epsilon_{\mu\nu\rho}g_\nu B_\rho$ and the confining force density is $f_\mu^C=-m\omega_0^2z\delta_{\mu3}\delta(\vec{r}-\vec{x})$. Using the Heisenberg equation of motion, we find
\begin{align}\label{eq:TF continuity equation 2}
    \frac{\partial g_\mu(\vec{r},t)}{\partial t}&=i[H_{TF},g_\mu(\vec{r},t)]\nonumber\\&=-\frac{\partial}{\partial r_\nu}\left(\frac{1}{4m}\{\{\pi_\nu,\delta(\vec{r}-\vec{x})\},\pi_\mu\}\right)\nonumber\\&-\frac{1}{m}\epsilon_{\mu\nu\rho}g_\nu(\vec{r},t) B_\rho-m\omega_0^2z\delta_{\mu 3}\delta(\vec{r}-\vec{x}).
\end{align}
We recognize the latter two terms as the Lorentz force and the confining force. Therefore, up to a term whose divergence is zero, the intensive stress tensor is given by
\begin{align}\label{eq:TF intrinsic stress tensor}
    \tau_{\mu\nu}&=\frac{1}{4m}\{\{\pi_\mu,\delta(\vec{r}-\vec{x})\},\pi_\nu\}.
\end{align}
Integrating over space gives the integrated stress tensor
\begin{align}\label{eq:TF integrated stress tensor}
    T_{\mu\nu}=\frac{1}{2m}\{\pi_\mu,\pi_\nu\}.
\end{align}
This is the same result as for an isotropic three-dimensional system.

Using Eq.~(\ref{eq:TF relating pis and X,Y}), we can determine the ground state expectation value for the integrated stress tensor. If we reinsert factors of $\hbar$ and include contributions from all $N$ electrons, we find that the in-plane components of the ground state stress tensor are given by
\begin{align}
    \braket{T_{xx}}_0&=\frac{N\hbar \omega_z}{2}\left(1-\frac{k^2\ell^2}{2}\right),\label{eq:TF stress tensor ground state xx}\\\braket{T_{xy}}_0&=0,\label{eq:TF stress tensor ground state xy}\\\braket{T_{yy}}_0&=\frac{N\hbar\omega_z}{2}\left(1+k^2\ell-\frac{3}{2}k^2\ell^2\right).\label{eq:TF stress tensor ground state yy}
\end{align}
Although we do not say much about them in the present paper, we also give the leading order expressions for $\braket{T_{xz}}$ $\braket{T_{yz}}$ and $\braket{T_{zz}}$ in Appendix~\ref{app:important results}. These out-of-plane components of the stress tensor may be interesting subjects for future work.

Let us make a few observations about the in-plane components of the stress tensor. Firstly, restricting to the quasi-two-dimensional (as revealed from the Hall conductivity) $x-y$ plane, we see that the ground state expectation value of the stress tensor is evidently not that of an isotropic fluid: $\braket{T_{xx}}_0\neq \braket{T_{yy}}_0$. This differentiates the in-plane behavior of the tilted field stress tensor from the stress tensor in the system with mass anisotropy. Secondly, the sub-leading correction to $\braket{T_{yy}}_0$ is of order $k^2\ell=\frac{\omega_x^2}{\omega_z\omega_0}$. Unlike $k^2\ell^2=\frac{\omega_x^2}{\omega_0^2}$, it depends on the strength of the perpendicular magnetic field. Because the term is unusual, it is worth exploring its origin in detail. We will revisit this in Sec.~\ref{sec:projected the tilted field system}.

\subsection{\label{subsec:TF Hall viscosity}Hall viscosity}
Having determined the stress tensor of the titled field system in Eq.~(\ref{eq:TF integrated stress tensor}), we can apply Eqs.~(\ref{eq:effective 2d response}) and (\ref{eq:Hall viscosity with CSS inserted}) to determine the effective 2D Hall viscosity (integrating over the z direction). We find
\begin{align}\label{eq:TF Hall viscosity calculation}
&-L_xL_y\eta^H_{\mu\nu\rho\sigma}\nonumber\\=&\sum_{(n_X,n_Y)\neq (0,0)}^\infty\!\!\!\text{Im}\frac{\left(\braket{0|\{\pi_\mu,\pi_\nu\}|n_X,n_Y}\braket{n_X,n_Y|\{\pi_\rho,\pi_\sigma\}|0}\right)}{2m^2(n_X\omega_1+n_Y\omega_2)^2}\nonumber\\
&=\frac{2B_z^2}{m^2}\text{Im}\left[\frac{\left(W_{\mu 3}W_{\nu 4}+W_{\mu 4}W_{\nu 3}\right)\left(W_{\rho 1}W_{\sigma 2}+W_{\rho 2}W_{\sigma 1}\right)}{(\omega_1+\omega_2)^2}\right.\nonumber\\
&\left.\hspace{1cm}+\frac{W_{\mu 3}W_{\nu 3}W_{\rho 1}W_{\sigma 1}}{2\omega_1^2}+\frac{W_{\mu 4}W_{\nu 4}W_{\rho 2}W_{\sigma 2}}{2\omega_2^2}\right].
\end{align}
Unlike the Hall conductivity, the Hall viscosity does not reduce to a particularly nice exact form. Instead, we look at the leading order expansion in $k$ and $\ell$. We are particularly interested in the in-plane components of the Hall viscosity tensor. With factors of $\hbar$ and the contributions from all $N$ electrons included, the in plane components are given by
\begin{align}
    \eta_{1122}^H&=0,\label{eq:TF Hall viscosity 1122}\\
    \eta_{1112}^H&=\frac{\hbar\rho}{4}\left(1-\frac{1}{2}k^2\ell^2\right),\label{eq:TF Hall viscosity 1112}\\
    \eta_{1222}^H&=\frac{\hbar\rho}{4}\left(1-\frac{3}{2}k^2\ell^2\right).\label{eq:TF Hall viscosity 1222}
\end{align}

At $\nu=1$ filling, the perpendicular component of the magnetic field determines the magnitude of the Hall viscosity through $\rho$, while the strength of the parallel component of the magnetic field relative to the confining potential determines the deviation of the Hall viscosity from the isotropic form (because, recall, $k^2\ell^2=\omega_x^2/\omega_0^2$). Meanwhile, the contracted form of the in-plane Hall viscosity tensor is
 \begin{align}
     \eta^H_{ab}=\frac{\hbar\rho}{4}\left(\begin{array}{cc}1-\frac{3}{2}k^2\ell^2&0\\0&1-\frac{1}{2}k^2\ell^2\end{array}\right).\label{eq:TF contracted Hall viscosity}
 \end{align}
 
Finally, because the stress tensor in the ground state is not isotropic, the Hall elastic modulus given in Eq.~(\ref{eq:Hall elastic modulus}) is non-zero. The in-plane component is
\begin{align}\label{eq:TF Hall elastic modulus}
    \kappa^{-H}_{1221}(\omega)&=\braket{\tau_{yy}}_0-\braket{\tau_{xx}}_0=\frac{\hbar\omega_z\rho}{2}k^2\ell(1-\ell).
\end{align}
Note that this Hall elastic modulus can alternatively be derived by considering the change in $\langle\tau_{\mu\nu}\rangle$ in response to a static rotation, and so is the susceptibility that directly probes the anisotropy of the stress tensor. The non-zero Hall elastic modulus is a noteworthy result with experimental implications; fortunately, the Hall elastic modulus is proportional to $k^2\ell$ rather than $k^2\ell^2$, which, given $\ell \sim 10^{-2}$ in a lab, makes it significantly larger than other anisotropic corrections to the Hall viscosity. Experimental verification of the Hall elastic modulus would provide physical confirmation of the anisotropy of the stress tensor in the ground state, and hence of the unusual constant term $\frac{\hbar \omega_z\rho}{2}\frac{\omega_x^2}{\omega_0}$ appearing in $\braket{T_{yy}}_0$ in Eq.~(\ref{eq:TF stress tensor ground state yy}). Perhaps most importantly, the Hall elastic modulus represents a directly measurable property that differentiates the tilted field system from the system with mass anisotropy.

Finally, for completeness we also include the leading order behavior of the out-of-plane components of the Hall viscosity in Appendix~\ref{app:important results}. Three-dimensional contributions to the Hall viscosity analogous to these have recently appeared in the study of topological semimetals.\cite{landsteiner-weyl-visc,vozmediano-rotational-strain}

\section{\label{sec:effective anisotropy}Effective anisotropy of the tilted field system}
Now that we have computed the current and stress response functions for the tilted field system, we will attempt to view the system, in the limit of small tilt and strong confinement, as a two-dimensional fluid. The natural question arises, is there an intrinsically two-dimensional system that recreates the behavior of the strongly confined tilted field system?

We answer this question in two steps. First, we try to find a mass tensor that captures the effect of the tilted field anisotropy on the stress and viscosity tensors. We will do this by comparing the tilted field viscosity tensor to the predictions of the recently developed bimetric theory of quadrupolar anisotropy of Ref.~\onlinecite{gromov2017investigating}. Second, we develop a method to directly project a quantum system from three dimensions to two dimensions using the ladder operators and apply it to the tilted field system.

With regards to the first step, we find that the tilted field system can be mapped to a two-dimensional system with mass anisotropy in such a way that the Hall conductivity and contracted Hall viscosity are reproduced. Other observables, like the electron density, the stress tensor, the antisymmetric components of the Hall viscosity, and the Hall elastic modulus, however, are not. By contrast, the Hamiltonian we derive from our projection method reproduces all the in-plane properties of the tilted field system that we looked at, but does so at the cost of requiring a coupling to the metric that is exotic for point particles.

\subsection{\label{subsec:anisotropy via Hall viscosity}Effective mass anisotropy from Hall viscosity}

In order to better understand the 2D behavior of the linear response functions of the tilted field system, we turn to a bimetric framework for understanding anisotropy in the quantum Hall effect, which was developed in Ref.~\onlinecite{gromov2017investigating} for a general 2D system. We begin by briefly summarizing the key ideas of the bimetric approach. Given that anisotropy often takes the form of a symmetric tensor, like an effective mass tensor, a dielectric tensor, or a quadrupolar interaction tensor, it seems natural to distill the cumulative effect of multiple weak anisotropies into a single symmetric two-component tensor field, denoted $v_{ab}$. Since rescaling coordinates rescales the tensor but does not change the degree of anisotropy, we may without loss of generality further assume the two-component tensor field has unit determinant. Using an effective field theory approach, the contracted Hall viscosity takes the form
\begin{align}\label{eq:bimetric Hall viscosity}
\eta_{ab}^H&=\frac{\hbar\rho}{2}\left[sg_{ab}+\varsigma v_{ab}+\xi \left(v_{ac}v_{bd}g^{cd}-g_{ab}\right)\right].
\end{align}
While the viscosity can vary continuously (even at fixed density) in the anisotropic system, the quantity $\mathcal{S}=2(s+\varsigma)$ is the shift of the quantum Hall state, and remains quantized.

Armed with Eq.~(\ref{eq:bimetric Hall viscosity}) and the fact that the shift $\mathcal{S}=1$ for $\nu=1$, we try to extract an effective anisotropy tensor $v_{ab}$ from Eq.~(\ref{eq:TF contracted Hall viscosity}). We work in flat background, so $g_{ab}=g^{ab}=\delta_{ab}$.

There are three equations-- corresponding to the three independent components of $\eta^H_{ab}$-- but four unknowns-- corresponding to $\xi$, $s-\varsigma$, and the two independent components of $v_{ab}$. We get rid of the additional degree of freedom by setting $s=0$. There is a simple physical interpretation for this choice, as is explained in Ref.~\onlinecite{gromov2017investigating}: Given a Hamiltonian with anisotropy in both the effective mass tensor $m_{ab}$ and the interaction tensor $\varepsilon_{ab}$, e.g.,
\begin{align}\label{eq:Hamiltonian with band mass and quadrupolar anisotropy}
H=\frac{1}{2}\tilde{m}_{ab}\pi_a\pi_b+\sum_{i\neq j} V(|\vec{x}_i-\vec{x}_j|;\varepsilon_{ab}),
\end{align}
it is possible to move the anisotropy entirely into the kinetic term or entirely into the potential term via suitable canonical transformations. Such canonical transformations also shift the values of $s$ and $\varsigma$; in particular, $s=0$ when all the anisotropy is moved to the kinetic term. Thus, by fixing $s=0$ (which leads to $\varsigma=\frac{1}{2}$ for the $\nu=1$ ground state), we pose the question: is there an effective mass tensor which replicates the contracted Hall viscosity determined for the tilted field system?

We find to leading order in $k$ and $\ell$ that the anisotropy tensor takes the form 
\begin{align}\label{eq:bimetric effective anisotropy tensor}
v_{ab}=\left(\begin{array}{cc}1\pm \sqrt{2}k\ell& 0 \\ 0 & 1\mp \sqrt{2}k\ell\end{array}\right),
\end{align}
and that $\xi$ is given by
\begin{align}\label{eq:bimetric xi}
\xi=-\frac{1}{4}\mp \frac{k\ell}{8\sqrt{2}}.
\end{align}
The $\pm/\mp$ in Eqs.~(\ref{eq:bimetric effective anisotropy tensor}) and (\ref{eq:bimetric xi}) indicates the existence of two anisotropy tensors corresponding to the tilted field system.

Note that this anisotropy tensor depends on $k\ell=\frac{\omega_x}{\omega_0}$ rather than on $k^2\ell^2=\frac{\omega_x^2}{\omega_0^2}$. Indeed, we explicitly dropped terms quadratic in $k\ell$ for $v$ and $\xi$ because they depend on cubic and higher order terms in $\eta_{ab}^H$. 
The fact that $v_{ab}$ and $\xi$ depend on $k\ell$ rather than $k^2\ell^2$ at the leading order is suggestive, because it means the effective anisotropy depends not just on the magnitude of the in-plane component of the magnetic field, but also on its sign. This indicates that perhaps an anisotropy \textit{vector} may be more successful at capturing the behavior of the tilted field system than an anisotropy \textit{tensor}. We leave the development of a theory of vector anisotropy to future work (though we note that some relevant results in this direction were given in Ref.~\onlinecite{gromov2016boundary}).

Note additionally that $\xi$ is perturbatively large for small values of the anisotropy (this presents no contradiction since $v_{ac}v_{bc}-\delta_{ab}\rightarrow 0$ as the anisotropy vanishes). Gromov et al.\cite{gromov2017investigating,gromov2017bimetric} emphasize that $\xi$ cannot be discarded on the basis of effective field theory. Although $\xi=0$ for the case of an integer quantum Hall fluid with mass anisotropy, there is numerical evidence of fractional quantum Hall states with effective mass anisotropy that yield a Hall viscosity corresponding to slightly non-zero $\xi$. For instance, the $\nu=1/3$ state yields $\xi=-0.06$, whose deviation from $0$, though small, is statistically significant. We have found that the tilted field system presents a second example of an anisotropic system for which $\xi$ can be taken to be non-zero.

In comparing Eq.~(\ref{eq:bimetric Hall viscosity}) to (\ref{eq:TF contracted Hall viscosity}) to extract an effective anisotropy tensor for the tilted field system, we implicitly assume that the electron surface density in the tilted field system is the same as the electron density in the effective two-dimensional system for which we identified the anisotropy tensor. It can prove enlightening, however, to allow the electron density in the effective 2D system to be different from the electron surface density in the tilted field system. In particular, if we allow the effective system's electron density to be $\rho^*=\rho(1-k^2\ell^2)$ and repeat the calculation to extract the anisotropy tensor, we find
\begin{align}\label{eq:bimetric effective anisotropy tensor 2}
v_{ab}=\left(\begin{array}{cc}1+\frac{1}{2}k^2\ell^2 & 0\\ 0 & 1-\frac{1}{2}k^2\ell^2\end{array}\right),
\end{align}
and $\xi=0$. In order to keep the Hall conductivity constant, we must then have that the magnetic field of the 2D system that the 3D tilted field system approximately maps to is $B_z^*=(1-k^2\ell^2)B_z$. Thus, the tilted field contracted Hall viscosity and Hall conductivity correspond to those of a system with an effective mass tensor,

\begin{align}\label{eq:effective mass tensor from Hall visc}
m^{\text{eff}}_{ab}&=m\left(\begin{array}{cc}1-\frac{1}{2}k^2\ell^2&0\\0&1+\frac{1}{2}k^2\ell^2\end{array}\right),
\end{align}
and a shifted magnetic field.

One pleasant feature of the effective mass tensor identified in Eq.~(\ref{eq:effective mass tensor from Hall visc}) is that it agrees with the mass tensor identified through alternative methods. For instance, the mass tensor deduced in Refs.~\onlinecite{papic2013fractional} and \onlinecite{yang2012band}, which is given for all values of $\omega_x$, $\omega_z$ and $\omega_0$, reduces to the one in Eq.~(\ref{eq:effective mass tensor from Hall visc}) when $\omega_x\ll \omega_z\ll \omega_0$. Furthermore, in section \ref{sec:projected the tilted field system}, we will see that planar behavior of the tilted field stress tensor is partly reproduced by the effective mass tensor in Eq.~(\ref{eq:effective mass tensor from Hall visc}). In particular, $T_{xx}$ and $T_{yx}$ expressed in terms of the $X$ and $c$ operators (with the $Y$ operators put into normal order and then set to zero) in the tilted field system match the forms of $T_{xx}$ and $T_{yx}$ expressed in terms of the two ladder operators in the two-dimensional system with effective mass given by Eq.~(\ref{eq:effective mass tensor from Hall visc}). The same does not hold for $T_{xy}$ and $T_{yy}$. Indeed, the effective mass tensor cannot fully reproduce the stress tensor of the tilted field system for two clear reasons: first, the tilted field stress tensor, unlike the anisotropic mass stress tensor, is symmetric; second, the tilted field stress tensor, unlike the anisotropic mass stress tensor, has anisotropic expectation value in the ground state. It follows that no anisotropic mass system can yield the same $T_{xx}$, $T_{yx}$ and $T_{yy}$ nor $T_{xx}$, $T_{yx}$ and $T_{xy}$ as the tilted field system. In other words, the effective mass tensor deduced from the contracted Hall viscosity in the regime of small tilt and strong confinement captures the planar behavior of the tilted field stress tensor to the maximal extent it can.

Note furthermore that since the two systems disagree on the form of $T_{xy}$, additional discrepancies appear if we look at the full Hall viscosity tensor. In particular, we have for the tilted field system that
\begin{equation}
\eta_{1112}=\eta_{1121},
\end{equation}
while the 2D system governed by the mass tensor Eq.~(\ref{eq:effective mass tensor from Hall visc}) has
\begin{equation}
\eta_{1112}-\eta_{1121}=\frac{\hbar\rho}{4m}(m_{22}-m_{11})\neq 0.
\end{equation}

To summarize, we find that a 2D quantum Hall system with magnetic field $B_z^*=(1-k^2\ell^2)B_z$ and effective mass tensor given in Eq.~(\ref{eq:effective mass tensor from Hall visc}) has a $\nu=1$ filled ground state whose Hall conductivity and contracted Hall viscosity match those of the tilted field system in the $\nu=1$ ground state in the regime of strong confinement and weak tilt. Thus, two readily observable properties of the tilted field system can be reproduced by the simpler two-dimensional system. The trade-off, however, is that two other readily observable properties, the magnetic field and the electron density, differ for the two systems. More importantly, perhaps, the system with an effective mass anisotropy fails to reproduce the subtlety of the tilted field stress tensor. This includes its constant term, its anisotropy, and its symmetry, which are responsible for the non-zero Hall elastic modulus and an absence of rotational contributions to the Hall viscosity, respectively.

\subsection{Projection of the tilted field system to an effective two-dimensional system}\label{sec:projection method}
To address the shortcomings of the previous analysis, we now utilize the ladder operators $X$, $Y$ and $c$ from section \ref{subsec:Hamiltonian diagonalization with ladder ops} to project the states, operators, and response functions of the tilted field system into two dimensions. We first develop the projection procedure in generality, and then push as far as possible the resulting correspondence between the tilted field system and an effective 2D system. 

\begin{figure}[t]
\includegraphics[width=0.4\textwidth]{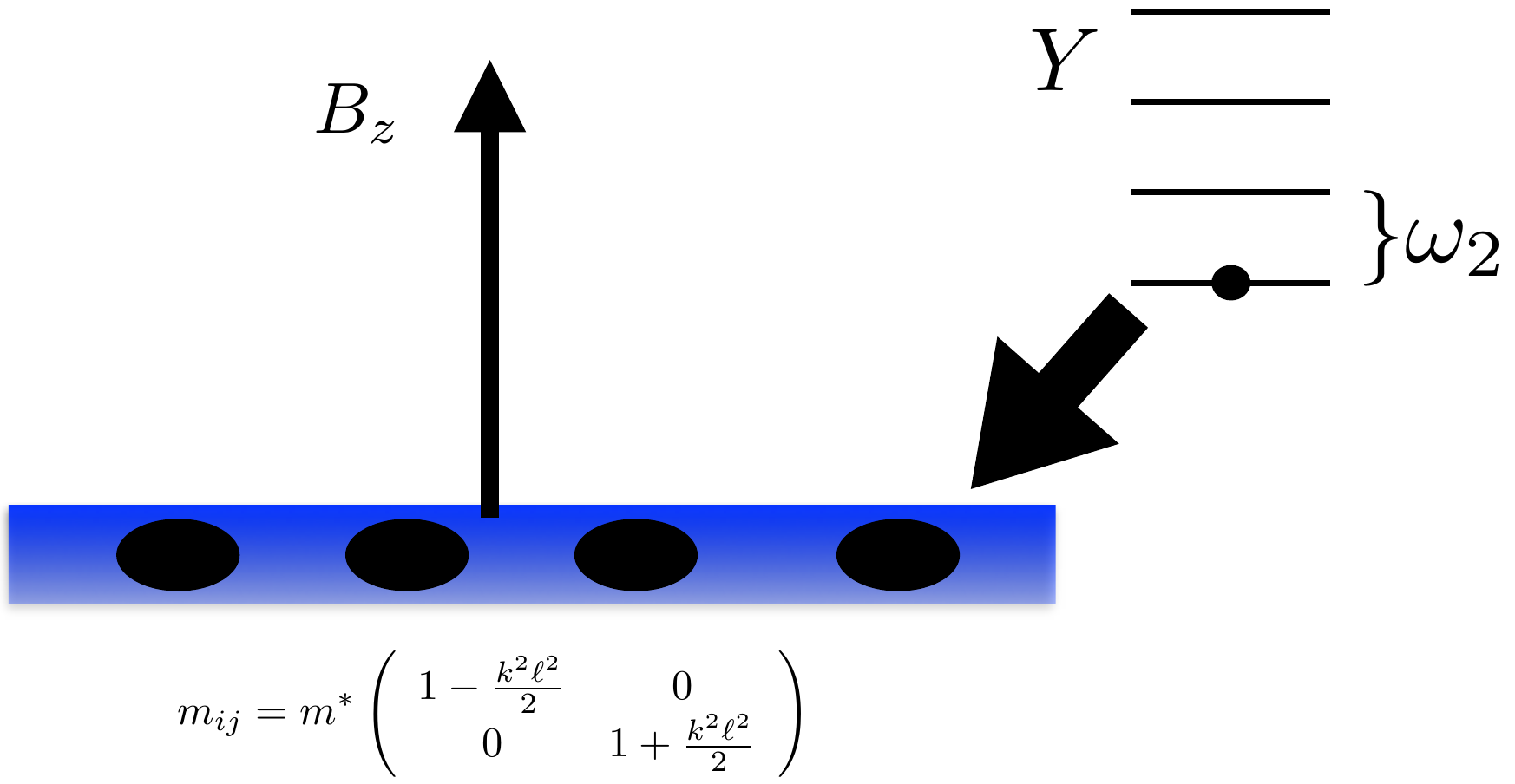}
\caption{Schematic of the results of the projection procedure for the tilted-field system. The effective two-dimensional system is composed of electrons (black ovals) in a true two-dimensional system (blue region), moving in background of a perpendicular magnetic field $B_z$. Each electron has a tower of ``internal'' states corresponding to the energy levels of the $Y$ harmonic oscillator of Sec.~\ref{sec:3D tilted field}, which we project into the ground state.}\label{fig:projection schematic}
\end{figure}

Our main result of this section is that the tilted field conductivity, stress tensor, and viscoelastic response functions are captured by the effective single-particle Hamiltonian
\begin{equation}
H^{\perp,\Lambda,A}=\frac{1}{2}\tilde{m}_{ab}\tilde{\pi}_a\tilde{\pi}_b+T^\perp_{ab}\lambda_{ab}+J_a^\perp \delta A_a+\frac{\omega_2}{2},
\end{equation}
where $\tilde{\pi}_a$ and $\tilde{x}_a$ are the kinetic momentum and position, respectively, and $\tilde{m}_{ab}$ is the inverse of the mass tensor we found in Eq.~(\ref{eq:effective mass tensor from Hall visc}), with the mass $m$ replaced by a shifted mass $m^*=(1+k^2\ell^2/2)m$. We also have $[\tilde{\pi}_a,\tilde{\pi}_b]=-iB_z\epsilon_{ab}$, with no rescaling of the magnetic field. While the current $J^\perp_a$ takes the usual minimal coupling form in terms of the $\tilde{\pi}_a$, the price we pay is that the stress tensor $T_{ab}^\perp$ couples nonminimally to metric perturbations $\lambda$. In particular, we find that as a single particle operator,
\begin{equation}
T^\perp_{xy}=T^\perp_{yx}=\frac{1}{2m^*}\left(1-\frac{k^2\ell^2}{2}\right)\{\tilde{\pi}_x,\tilde{\pi}_y\}, 
\end{equation}
is symmetric, while $T^\perp_{xx}$ and $T^\perp_{yy}$ have different scaling with $k\ell$ and take the form
\begin{align}
T_{xx}^\perp&=\left(1+\frac{k^2\ell^2}{2}\right)\frac{\tilde{\pi}_x^2}{m^*},\\
T_{yy}^\perp&=\left(1-\frac{3k^2\ell^2}{2}\right)\frac{\tilde{\pi}_y^2}{m^*}+\frac{\omega_x^2}{2\omega_0}.
\end{align}
In addition, $T_{yy}^\perp$ contains a c-number contribution, which we can interpret as a strain-dependence of the zero-point energy $\omega_2/2$. These features of the stress tensor conspire to reproduce all of the 2D response functions of the tilted field system discussed in Sec.~\ref{sec:3D tilted field}; the non-minimal coupling and exotic form of the stress tensor combine with the anisotropic mass $\tilde{m}_{ab}$ to give the Hall elastic modulus and full Hall viscosity tensor of the tilted field system without a rescaling of the magnetic field $B_z$. Taken together, our results suggest that the projection of the tilted field system is best thought of as a fluid of composite particles. We summarize the results of this projection in Fig.~\ref{fig:projection schematic}

\subsubsection{Theory of projection from 3D to 2D}\label{sec:projection from 3D to 2D}
Let us begin with a Hilbert space $\mathcal{H}$ and a diagonalized Hamiltonian
\begin{align}\label{eq:3Dto2D full diagonalized Hamiltonian}
H&=\frac{\omega_X}{2}(X^\dagger X+XX^\dagger)+\frac{\omega_Y}{2}(Y^\dagger Y+YY^\dagger)\nonumber\\&+\frac{\omega_Z}{2}(Z^\dagger Z+ZZ^\dagger),
\end{align}
where $X$, $Y$, and $Z$ form a set of three independent ladder operators: $[X,X^\dagger]=[Y,Y^\dagger]=[Z,Z^\dagger]=1$, with all other commutators being zero. Of course, the energy eigenstates are
\begin{align}\label{eq:3Dto2D energy eigenstates}
\ket{m,n,p}&=\frac{1}{\sqrt{m!n!p!}}(X^\dagger)^m(Y^\dagger)^n(Z^\dagger)^p\ket{0},
\end{align}
and the energies are $E_{mnp}=\omega_X\left(m+\frac{1}{2}\right)+\omega_Y\left(n+\frac{1}{2}\right)+\omega_Z\left(p+\frac{1}{2}\right)$.

Let us assume that $\omega_Y\gg \omega_X,\omega_Z$. Intuitively, this means that at low energies, the physics is dominated by the subspace
\begin{align}\label{eq:infrared Hilbert subspace}
\mathcal{H}^{\text{IR}}\equiv \text{Span}\{\ket{m,0,p}:m,p\in \{0,1,2,\ldots\}\},
\end{align}
where superscript IR stands for ``infrared.'' Clearly, the ground state is an element of $\mathcal{H}^{\text{IR}}$.

To clarify what we mean by ``the physics is dominated by $\mathcal{H}^{\text{IR}}$,'' it is convenient to define a versatile ``projection function" $\perp[\cdot]$ that can act on states in $\mathcal{H}$, operators acting on $\mathcal{H}$, expectation values, and linear response functions. Let us first introduce an operator $P$ which projects onto the ``low energy'' eigenstates:
\begin{align}\label{eq:projection operator}
P\equiv \sum_{m,p}\ket{m,0,p}\bra{m,0,p}.
\end{align}
The low-energy subspace of the full Hilbert space may then be denoted $\mathcal{H}^{\text{IR}}=P\mathcal{H}$. We can use $P$ to project any state $\ket{\psi}\in \mathcal{H}$ into $\mathcal{H}^{\text{IR}}$ and any dual state $\bra{\psi}\in \mathcal{H}^*$ into $\mathcal{H}^{{\text{IR}}*}$. This lets us define the action of $\perp$ on states and dual states:
\begin{align}\label{eq:projection of bras and kets}
\perp[\ket{\psi}]\equiv P\ket{\psi},\\
\perp[\bra{\psi}]\equiv \bra{\psi}P.
\end{align}
Because $P=P^\dagger$, we have $\perp[\ket{\psi}]^\dagger=(P\ket{\psi})^\dagger=\bra{\psi}P=\perp[\ket{\psi}^\dagger]$ and likewise $\perp[\bra{\psi}]^\dagger=\perp[\bra{\psi}^\dagger]$. Thus, acting with the projection function on a state commutes with taking the dual. Furthermore, projection of a state in $\mathcal{H}^{\text{IR}}$ leaves the state unchanged: if $\ket{\psi}\in \mathcal{H}^{\text{IR}}$, then $\perp[\ket{\psi}]=\ket{\psi}$. Likewise, if $\bra{\psi}\in \mathcal{H}^{{\text{IR}}*}$, then $\perp[\bra{\psi}]=\bra{\psi}$. 

We next use $P$ to define projection of an operator:
\begin{align}\label{eq:projection of operators}
	\perp[\mathcal{O}]&\equiv P\mathcal{O}P.
\end{align}
Just as with the projection of states, applying the projection function to an operator commutes with taking the adjoint of the operator.

Since
\begin{align}\label{eq:X commutes with P}
&X\sum_{m,p}\ket{m,0,p}\bra{m,0,p}\nonumber\\&=\sum_{m,p}\sqrt{m}\ket{m-1,0,p}\bra{m,0,p}\nonumber\\&=\sum_{m,p}\ket{m,0,p}\bra{m+1,0,p}\sqrt{m+1}\nonumber\\&=\sum_{m,p}\ket{m,0,p}\bra{m,0,p}X,
\end{align}
and analogously for $Z$, we see that $P$ commutes with the $X$ and $Z$ ladder operators. Furthermore, any operator $\mathcal{O}$ may be written as a series in  normal ordered powers of $Y$ and $Y^\dagger$ with the coefficients being arbitrary functions of the $X$ and $Z$ ladder operators. Namely, we may write
\begin{align}\label{eq:normal ordered series expansion of operator}
\mathcal{O}&=\sum_{m,n} A_{mn}(X,X^\dagger,Z,Z^\dagger) (Y^\dagger)^mY^n.
\end{align}
Since $YP=0$ and $PY^\dagger=0$, it follows that
\begin{align}\label{eq:projection of operator explicit form}
\perp[\mathcal{O}]&=A_{00}(X,X^\dagger,Z,Z^\dagger)P.
\end{align}
In other words, projecting an operator amounts to, first, putting the $Y$ and $Y^\dagger$ operators in normal order, second, dropping any term that still depends on $Y$ or $Y^\dagger$, and, third, multiplying by $P$ on the right (or on the left, or somewhere in the middle; the position of $P$ does not matter because it commutes with the other operators). Since $P$ acts as the identity when we focus on matrix elements between states in $\mathcal{H}^{\text{IR}}$, we will generally not write $P$ when deriving expressions for projected forms of operators.

Next, having defined projection of a state and projection of an operator, there are three natural ways to define the projection of a  matrix element:
\begin{align}
\perp[\braket{\psi|\mathcal{O}|\phi}]&\equiv \bra{\psi}\perp[\mathcal{O}]\ket{\phi},&\text{or}\label{eq:projection function exp value 1}\\\perp[\braket{\psi|\mathcal{O}|\phi}]&\equiv \perp[\bra{\psi}]\mathcal{O}\perp[\ket{\phi}],&\text{or}\label{eq:projection function exp value 2}\\ \perp[\braket{\psi|\mathcal{O}|\phi}]&\equiv \perp[\bra{\psi}]\perp[\mathcal{O}]\perp[\ket{\phi}].\label{eq:projection function exp value 3}
\end{align}
Because $P=P^2$, all three definitions are equivalent.

We can now state the main strength of the projection function, which is that it satisfies the following obvious but useful property:
\begin{align}\label{eq:expectation projection property}
\perp[\braket{\psi|\mathcal{O}|\phi}]&=\braket{\psi|\mathcal{O}|\phi}, \hspace{0.5cm} \text{if }\ket{\psi},\ket{\phi}\in \mathcal{H}^{\text{IR}},
\end{align}
Eq.~(\ref{eq:expectation projection property}) means, for instance, that projection does not affect expectation values in the ground state. Thus, when determining the ground state expectation value of an operator $\mathcal{O}$, we can work with $\perp[\mathcal{O}]$ instead of with the full operator.

Let us finally define the projection of a tensor, $T$. In the original system, let $T$ have $n$ indices $\alpha_1,\ldots,\alpha_n$ which take the values $1$, $2$, or $3$. We require that the projection of $T$ yield a tensor with $n$ indices $a_1,\ldots,a_n$ which take the values $1$ or $2$. For simplicity, let us assume the dependence of the components of the projected tensor on the components of the original tensor is linear, so we may write
\begin{align}\label{eq:definition of projection of tensor}
\perp[T]_{a_1\ldots a_n}=\tensor{U}{_{a_1\ldots a_n}^{\mu_1\ldots \mu_n}}T_{\mu_1\ldots \mu_n},
\end{align}
for some higher-dimensional non-square matrix $U$. We now face a decision. Different choices of $U$ correspond to projection functions that act differently on a given tensor. In the event that the projection procedure has an obvious physical interpretation, then there may be a natural choice for $U$. For the tilted field system, we saw in Sec.~\ref{subsec:Hamiltonian diagonalization with ladder ops} that both the magnetic translation generators and the ground state degeneracy take a form which singles out the $x-y$ plane. Additionally, the Hall conductivity tensor in Eq.~(\ref{eq:TF Hall conductivity}) reveals that any effective 2D quantum Hall system must reside in the $x-y$ plane. Thus, we will take
\begin{align}\label{eq:tensor projection matrix}
\tensor{U}{_{a_1\ldots a_n}^{\mu_1\ldots \mu_n}}&=\left\{\begin{array}{cc}1 & \forall i,a_i=\mu_i \\ 0 & \text{otherwise}\end{array}\right. .
\end{align}
This form of $U$ corresponds physically to focusing only on the in-plane components of tensors.

We have now fully defined the projection function $\perp$. Because the projection function acts linearly, it commutes with the action of taking the derivative with respect to an external parameter. Therefore, given an operator $\mathcal{O}_n$ that is equal to the derivative of the Hamiltonian $H(f_n)$ with respect to the external field $f_n$, we can compute its projection directly or by taking the derivative of the projected Hamiltonian. Namely,
\begin{align}\label{eq:projection commutes with derivation}
\perp[\mathcal{O}_n]\equiv \perp\left[\frac{\partial H(f_n)}{\partial f_n}\right]=\frac{\partial}{\partial f_n}\perp[H(f_n)].
\end{align}
This means that we can compute the ground state expectation value of ``variational operators,'' like the current or stress tensor, from the projected form of the Hamiltonian. Provided we compute the projection of the couplings to $f_m$, this will allow us to perform calculations with a system whose Hamiltonian is simpler and whose Hilbert space is smaller.

The projection function, however, also has its shortcomings, which is unsurprising given that it throws away some information. In particular, the projection of a product of operators is not equal to the product of the respective projections, (i.e., $\perp[\mathcal{O}_1\mathcal{O}_2]\neq \perp[\mathcal{O}_1]\perp[\mathcal{O}_2]$; this is obvious for the case $\mathcal{O}_1=Y$ and $\mathcal{O}_2=Y^\dagger$),
and therefore the projection of the commutator of operators is not equal to the commutator of the respective projections (i.e., $\perp[[\mathcal{O}_1,\mathcal{O}_2]]\neq [\perp[\mathcal{O}_1],\perp[\mathcal{O}_2]]$). This property of the projection onto a subspace is familiar in the study of topological phases: as was first formalized in Ref.~\onlinecite{simon1983homotopy}, this mechanism is responsible for the emergence of geometric phases in quantum systems. For recall that a projection operator $P(\mathbf{\lambda})$ onto a subspace of the full (and assumed topologically trivial) Hilbert space as a function of some parameter $\mathbf{\lambda}\in\mathcal{M}$ defines a vector bundle over $\mathcal{M}$, which may be nontrivial. On the one hand, one can show\cite{avron1995adiabatic} using
\begin{equation}
P(\mathbf{\lambda})^2=P
\end{equation}
that
\begin{equation}
P(\partial_{\lambda_i}P)P=0.
\end{equation}
Identifying $\partial_{\lambda_i}P$ with $\mathcal{O}_i$ above, we see then that
\begin{equation}
\left[P\partial_{\lambda_i}PP,P\partial_{\lambda_j}PP\right]=0.
\end{equation}
On the other hand, applying the projection after the commutator results in the Berry curvature
\begin{equation}
\Omega_{ij}=P[\partial_{\lambda_i} P,\partial_{\lambda_j} P]P.
\end{equation}

The failure of operator projection and operator commutation/product to commute has an important implication for the the projection of linear response functions. Let us consider the projection of a tensor linear response function $\chi_{\mu\nu}$. If $\chi_{\mu\nu}$ captures the response of $A_\mu$ to perturbations of the form $f_\nu A_\nu$ and $U$ is given by Eq.~(\ref{eq:tensor projection matrix}) as applied to $\chi_{\mu\nu}$ and $A_\mu$, then the projection of the response function, $\perp[\chi]_{ab}$, just drops the components of the full response function $\chi_{\mu\nu}$ for which $\mu=3$ or $\nu=3$. Explicitly, we may write
\begin{align}\label{eq:projected linear response function}
&\perp[\chi(\omega)]_{ab}\\&=-\frac{i}{\omega^+}\lim_{\epsilon\to 0^+}\int_0^\infty dt e^{i\omega^+t}\perp\left[\braket{0|[e^{iH t}A_ae^{-iH t},A_b]|0}\right]\nonumber\\&=-\frac{i}{\omega^+}\lim_{\epsilon\to 0^+}\int_0^\infty dt e^{i\omega^+t}\braket{0|\perp\left[[e^{iH t}A_ae^{-iH t},A_b]\right]|0}.\nonumber
\end{align}
We see then that the projection $\perp\chi$ of a response function $\chi$ plays the analogous role of the Berry curvature in adiabatic transport (c.~f.~Ref.~\onlinecite{read2011hall}).

There is, however, a second natural ``projected" linear response function that we should examine. To differentiate it from the linear response function one arrives at via the $\perp$ function, let us denote this second linear response function $\chi_{ab}^\perp$. We define it as follows: if we hand someone, call her Alice, the projected perturbed Hamiltonian $H^\perp(\{f_n\})\equiv \perp[H(\{f_n\})]$ and the projected Hilbert space $\mathcal{H}^{\text{IR}}$, she can determine its unperturbed energy eigenstates, $\ket{m,0,p}$, and compute the projected forms of the variational operators $A_a$. Alice then has all the necessary ingredients to do a linear response analysis. She can use Eq.~(\ref{eq:response function Fourier}) to compute what we define to be the second projected linear response function:
\begin{align}\label{eq:linear response function in projected system}
&\chi_{ab}^\perp(\omega)\\&\equiv \frac{1}{i\omega^+}\lim_{\epsilon\to 0^+}\int_0^\infty dt e^{i\omega^+t}\braket{0|[e^{iH^\perp t}\perp[A_a]e^{-iH^\perp t},\perp[A_b]]|0}.\nonumber
\end{align}
Note that Alice uses
\begin{align}\label{eq:Alice's projected Hamiltonian}
H^\perp\equiv \frac{\omega_X}{2}(X^\dagger X+XX^\dagger)+\frac{\omega_Z}{2}(Z^\dagger Z+ZZ^\dagger)+\frac{\omega_Y}{2}
\end{align}
rather than the Hamiltonian given in Eq.~(\ref{eq:3Dto2D full diagonalized Hamiltonian}) to compute the time dependence of the operator $\perp[A_a]$. Unsurprisingly, the projection of the Hamiltonian also yields a contribution from the zero point energy of the $Y$ ladder operators. 

It is quite clear from Eqs.~(\ref{eq:projected linear response function}) and (\ref{eq:linear response function in projected system}) that $\perp[\chi]_{ab}\neq \chi^\perp_{ab}$ because
\begin{align}\label{eq:failed projection and commutator identity}
\perp\left[[e^{iHt}A_ae^{-iHt},A_b]\right]\neq\left[e^{iH^\perp t}\perp[A_a]e^{-iH^\perp t},\perp[A_b]\right].
\end{align}
This appears to be a limitation of the projection method: since $\perp[\chi]_{ab}\neq \chi^\perp_{ab}$, we cannot compute the in-plane components of the linear response functions from the projected Hamiltonian. We must instead evaluate the response functions in the full 3D system and focus on the in-plane components only at the end of the computation.

There is a simple argument, however, that the antisymmetric part of the response function may obey $\perp[\chi^H]_{ab}\approx \chi_{ab}^{H\perp}$ with small corrections. Using reasoning analogous to that leading from Eq.~(\ref{eq:Hall conductivity definition}) to (\ref{eq:Hall conductivity with CSS inserted}), we can deduce explicit results for $\perp[\chi^H]_{ab}$ and $\chi^{H\perp}_{ab}$. We find:
\begin{align}\label{eq:projected Hall response function of first type with css inserted}
\perp[\chi^H]_{ab}&\equiv \frac{1}{2}\lim_{\omega^+\to 0}\left(\chi_{ab}(\omega)-\chi_{ba}(\omega)\right)\\&=2\sum_{(m,n,p)\neq (0,0,0)}\frac{\text{Im}\left(\braket{0|A_a|m,n,p}\braket{m,n,p|A_b|0}\right)}{(m\omega_X+n\omega_Y+p\omega_Z)^2},\nonumber
\end{align}
and
\begin{align}\label{eq:projected Hall response function of second type with css inserted}
\chi^{H\perp}_{ab}&\equiv \frac{1}{2}\lim_{\omega^+\to 0}\left(\chi^\perp_{ab}(\omega)-\chi^\perp_{ba}(\omega)\right)\\&=2\sum_{(m,p)\neq (0,0)}\frac{\text{Im}(\braket{0|\perp[A_a]|m,0,p}\braket{m,0,p|\perp[A_b]|0})}{(m\omega_X+p\omega_Z)^2}\nonumber\\&=2\sum_{(m,p)\neq (0,0)}\frac{\text{Im}(\braket{0|A_a|m,0,p}\braket{m,0,p|A_b|0})}{(m\omega_X+p\omega_Z)^2},\nonumber
\end{align}
where to get from the second to third line above we used Eqs.~(\ref{eq:projection function exp value 1}) and (\ref{eq:expectation projection property}). Therefore, we see
\begin{align}\label{eq: difference in projected Hall response functions}
&\perp[\chi^H]_{ab}-\chi^{H\perp}_{ab}\\&=\sum_{m,p}\sum_{n=1}^\infty \frac{\text{Im}\left(\braket{0|A_a|m,n,p}\braket{m,n,p|A_b|0}\right)}{(m\omega_X+n\omega_Y+p\omega_Z)^2}.\nonumber
\end{align}
Clearly, the denominator is on the order of $\omega_Y^2$. 
Furthermore, let us assume that
can be written in the form $A_a=F_a(X,Z)+G_a(X,Y,Z)$ such that 
$G_a\sim \omega_X\left(\frac{\omega_X}{\omega_Y}\right)^{\alpha_a}\left(\frac{\omega_Z}{\omega_Y}\right)^{\beta_a}$ with $\alpha_a,\beta_a>0$ (we have included an overall factor of $\omega_X$ because we make the unrestrictive assumption that $A_a$ has units of energy to make the response function dimensionless); we refer to such operators as \emph{planar}.
In that case, the difference between the two Hall linear response functions is of order $\left(\frac{\omega_X}{\omega_Y}\right)^{2+\alpha_a+\alpha_b}\left(\frac{\omega_Z}{\omega_Y}\right)^{\beta_a+\beta_b}$. To summarize, certain plausible conditions being satisfied, the antisymmetric linear response function that Alice calculates differs from the the one directly projected from the fully 3D result by perturbatively small corrections. Moreover, because both the original and projected Hamiltonians are gapped, the symmetric components of $\perp[\chi_{ab}]$ and $\chi^\perp_{ab}$ are both zero.

Finally, we can concisely formulate the motivation behind introducing the $\perp$ projection function: it lets us define the projected Hamiltonian living in the projected Hilbert space, which is equipped with only two ladder operators rather than three and thus describes a two-dimensional system rather than a three-dimensional one. From the projected Hamiltonian, it is possible to calculate important properties of the original three-dimensional system, like (i) the ground state expectation values of the variational operators exactly and (ii) the antisymmetric linear response functions with small corrections dependent on the energy scale of the third dimension of the original system. It is in this sense that the projection operator $\perp$ allows us to identify an effective system capturing the two-dimensional behavior of the three-dimensional system.
 
With these generalities in place, we now apply the projection formalism to the tilted-field system.

\subsubsection{Projecting the tilted field system}\label{sec:projected the tilted field system}

In the present section, we examine the 2D quantum mechanical system that results from projecting the tilted field system. We must first identify the projected Hamiltonian, including the manner in which it couples to the strain and electromagnetic fields. In the previous section, we assumed the Hamiltonian was diagonalized by ladder operators $X$, $Y$ and $Z$. In the tilted field system, we have the ladder operators $X$, $Y$ and $c$ playing the analogous roles. Thus, given the diagonalized tilted field Hamiltonian in Eq.~(\ref{eq:TF gauge indep Hamiltonian diagonalized}), the projected Hamiltonian is
\begin{align}\label{eq:projected TF Hamiltonian}
H^\perp\equiv \perp[H]=\frac{\omega_1}{2}(X^\dagger X+XX^\dagger)+\frac{\omega_2}{2},
\end{align}
which has the energy eigenstates $\ket{m,p}=\frac{1}{\sqrt{m!p!}}(X^\dagger)^m(c^\dagger)^p\ket{0}$ with energy eigenvalues $E_{mn}=\omega_1\left(m+\frac{1}{2}\right)+\omega_2/2$. Meanwhile, the projected Hamiltonian with the linear order coupling to the strain and electromagnetic fields is
\begin{align}\label{eq:projected TF perturbed Hamiltonian}
H^{\perp,\Lambda,A}&\equiv \perp[H^{\Lambda,A}]\\&=H^\perp+T^\perp_{ab}\lambda_{ab}+J^\perp_a\delta A_a+O(\lambda^2,A^2,\lambda A),\nonumber
\end{align}
where, as we argued in Eq.~(\ref{eq:projection commutes with derivation}), the operators $T^\perp_{ab}$ and $J^\perp_{ab}$ that couple to $\lambda_{ab}$ and $A_a$ are just the projections of the full stress tensor and current operator, $\perp[T_{\mu\nu}]$ and $\perp[J_{\mu}]$. To leading  order in terms of the $X$ and $c$ ladder operators, we have
\begin{align}
J_{x}^\perp&=-i\sqrt{\frac{\omega_z}{2m}}\left(1-\frac{k^2\ell^2}{4}\right)\left(X^\dagger-X\right),\label{eq:projected J_x}\\J_{y}^\perp&=-\sqrt{\frac{\omega_z}{2m}}\left(1-\frac{3k^2\ell^2}{4}\right)\left(X^\dagger+X\right),\label{eq:projected J_y}\\
T_{xx}^\perp&=-\frac{\omega_z}{2}\left(1-\frac{k^2\ell^2}{2}\right)\left(X^\dagger-X\right)^2,\label{eq:projected T_xx}\\T_{xy}^\perp&=\frac{i\omega_z}{2}\left(1-k^2\ell^2\right)\left((X^\dagger)^2-X^2\right),\label{eq:projected T_xy}\\T_{yy}^\perp&=\frac{\omega_z}{2}\left(1-\frac{3k^2\ell^2}{2}\right)\left(X^\dagger+X\right)^2+\frac{\omega_z}{2}k^2\ell.\label{eq:projected T_yy}
\end{align}

We next check whether the projected system yields a good approximation to the Hall viscosity and Hall conductivity of the tilted field system. Before we do so, however, let us examine the projected stress tensor, an operator whose strangeness we commented on in section \ref{subsec:TF stress tensor}, in greater detail.

There are two features of the projected stress tensor that warrant attention. The first is the constant term $\frac{\omega_z}{2}k^2\ell$ in $T_{yy}^\perp$. The second is the fact that $T_{xy}^\perp=T_{yx}^\perp$, which is not surprising when one considers that the three-dimensional stress tensor of the tilted field system is symmetric, but is surprising when one thinks of the projected system as being approximated by an effective anisotropic mass. Examining the projection of the momentum density continuity equation sheds light on the origin of these features of $T_{\mu\nu}^\perp$.

We begin with the continuity equation in (\ref{eq:TF continuity equation}) of the tilted field system. By working to leading order in the wavevector $q_\nu$, we arrive at the following expression for the stress tensor:
\begin{align}
T_{\mu\nu}&=\frac{1}{2}\partial_t\{\pi_\nu,x_\mu\}+\frac{1}{2m}\epsilon_{\nu\rho\sigma}B_\sigma\{\pi_\rho,x_\mu\}+m\omega_0^2zx_\mu \delta_{\nu 3}.
\end{align}
In words, the stress tensor accounts for the evolution of the momentum density that is unaccounted for by the Lorentz force and the confining force. We are interested in the projection of the continuity equation, given by
\begin{align}\label{eq:projected continuity equation}
\perp[T_{\mu\nu}]&=\frac{1}{2}\partial_t\perp[\{\pi_\nu,x_\mu\}]+\frac{1}{2m}\perp[\epsilon_{\nu\rho\sigma}B_\sigma \{\pi_\rho,x_\mu\}]\nonumber\\&+m\omega_0^2\perp[zx_\mu \delta_{\nu3}],
\end{align}
where we have used the fact that $P$ and $H$ commute to exchange the order of $\perp$ and $\partial_t$.
Setting $\mu\to a=1,2$ and $\nu\to b=1,2$, we see that the contribution from the confining potential drops out. Expressing $\pi_\nu$ and $x_\mu$ in terms of $X$, $Y$, and $c$ makes both the projection and time evolution of the two remaining terms simple. See appendix \ref{app:important results} for the explicit projections of $\{x_\mu,\pi_\nu\}$.

Let us first focus on the projection of $T_{yy}$:
\begin{align}\label{eq:Tyy continuity projection}
T_{yy}^\perp&=\frac{1}{2}\partial_t\perp[\{y,\pi_y\}]+\frac{1}{2m}\left(B_x\{\pi_z,y\}-B_z\{\pi_x,y\}\right).
\end{align}
Unsurprisingly, if we use the expressions for the projections of $\{x_\mu,\pi_\nu\}$, we recreate Eq.~(
\ref{eq:projected T_yy}). What is interesting, however, is that the projection of $\{\pi_z,y\}$ contains the constant term $k\ell$ which yields the $\frac{\omega_x^2}{2\omega_0}$ term in $T_{yy}^\perp$; thus, the exotic constant contribution to the stress tensor traces directly to the Lorentz-force in the momentum continuity equation and depends critically on $B_x$ and $\pi_z$ being non-zero. Examining Eq.~(\ref{eq:projected TF perturbed Hamiltonian}), we see that we can interpret this as a strain-dependence of the zero point energy $\omega_2/2$.

Next, let us consider the projection of $T_{xx}$, $T_{yx}$ and $T_{xy}$. We have
\begin{align}
T_{xx}^\perp&=\frac{1}{2}\partial_t[\{\pi_x,x\}]+\frac{1}{2m}B_z\perp[\{\pi_y,x\}],\label{eq:Txx continuity projection}
\\
T_{yx}^\perp&=\frac{1}{2}\partial_t \perp[\{\pi_x,y\}]+\frac{1}{2m}B_z\perp[\{\pi_y,y\}],\label{eq:Tyx continuity projection}\\T_{xy}^\perp&=\frac{1}{2}\partial_t\perp[\{\pi_y,x\}]\label{eq:Txy continuity projection}\\&+\frac{1}{2m}\left(B_x\perp[\{\pi_z,x\}]-B_z\perp[\{\pi_x,x\}]\right).\nonumber
\end{align}
It is noteworthy that the projection of $T_{xx}$ and $T_{yx}$ look like bona fide two-dimensional calculations; there is no $z$, $\pi_z$, or $B_x$ to be seen. The projection of $T_{xy}$, on the other hand, includes the term $B_x\perp[\{\pi_z,x\}]$, making the contribution to the stress from the perpendicular direction obvious. Indeed, we will demonstrate shortly that $T_{xx}^\perp$ and $T_{yx}^\perp$ are captured by an anisotropic mass tensor akin to the one in Eq.~(\ref{eq:effective mass tensor from Hall visc}), while $T_{xy}^\perp$ and $T_{yy}^\perp$ are not. From the perspective of the continuity equation, therefore, projection of the tilting field system seems to yield an effective anisotropic mass system with additional physical effects arising from the parallel component of the magnetic field. The parallel magnetic field's contribution to the Lorentz force is directly responsible not only for shifting $T_{yy}^\perp$ by a constant but also for keeping $T_{xy}^\perp$ and $T_{yx}^\perp$ precisely equal.

Finally, note that Eq.~(\ref{eq:projected continuity equation}) reads equally well as the projection of the tilted field continuity equation and as a continuity equation within the projected system. This follows from the fact that the time dependence of $X$ and $c$ is the same whether we consider the full Hamiltonian $H$ or the projected Hamiltonian $H^\perp$.

Moving on to the projection of the linear response functions, we first examine the leading order dependence of the unprojected planar current operators $J_x$ and $J_y$ on $Y$ and $Y^\dagger$. They are:
\begin{align}
J_x\sim J_x^\perp-\frac{\sqrt{B_z}}{m\sqrt{2}}k\ell^{3/2}(Y^\dagger+Y),\label{eq:J_x leading order Y dependence}\\J_y\sim J_y^\perp+i\frac{\sqrt{B_z}}{m\sqrt{2}}k\ell^{1/2}(Y^\dagger-Y).\label{eq:J_y leading order Y dependence}
\end{align}
Here, $J_x^\perp$ and $J_y^\perp$ play the role of the function $F(X,c)$ we introduced below Eq.~(\ref{eq: difference in projected Hall response functions}), while the remaining terms play the role of $G(X,Y,c)$. Therefore, according to our reasoning following Eq.~(\ref{eq:projected Hall response function of first type with css inserted}), $\perp[\sigma^H]_{ab}-\sigma_{ab}^{H\perp}$ has corrections of order $k^2\ell^5$ if $a=x$, $b=x$; of order $k^2\ell^3$ if $a=y$, $b=y$; and of order $k^2\ell^4$ if $a=x$, $b=y$. These corrections are all smaller than $k^2\ell^2$. Therefore, at the order in $k$ and $\ell$ that we have been working, we expect the Hall conductivity derived from the projected perturbed Hamiltonian to match the in-plane components of the three-dimensional result. And indeed, using Eq.~(\ref{eq:Hall conductivity with CSS inserted}), we find
\begin{align}\label{eq:Alice's Hall conductivity calculation}
\sigma_{xy}^{H \perp}&=\frac{2}{L^2}\sum_{n=1}^\infty \frac{\text{Im}\left(\braket{0|J_x^\perp|n}\braket{n|J_y^\perp|0}\right)}{n^2\omega_1^2}\nonumber\\&=-\frac{1}{L^2}\left(1-\frac{k^2\ell^2}{2}\right)\frac{1}{m\omega_1}=-\frac{1}{L^2B_z}
\end{align}
at leading order (note that the topological protection of the Hall conductivity ensures that the projected result matches the exact result to all orders). Inserting factors of $e$ and including the contributions of all $N$ electrons, we arrive at
\begin{align}
\sigma_{xy}^{H\perp}&=-\frac{e\rho}{B_z}.
\end{align}
Comparing with Eq.~(\ref{eq:TF Hall conductivity}), we see the Hall conductivity calculated from the projected Hamiltonian matches the in-plane result for the tilted field system, as long as the ground state electron densities at $\nu=1$ filling are the same.

Similarly, we examine the leading order dependence of the unprojected planar stress operators on $Y$ and $Y^\dagger$. They are:
\begin{align}
T_{xx}&\sim T_{xx}^\perp +i\omega_zk\ell^{3/2}(X^\dagger-X)(Y^\dagger+Y),\label{eq:T_{xx} leading order Y dependence}\\T_{xy}&\sim T_{xy}^\perp\nonumber\\&+\frac{\omega_z}{2}k\ell^{1/2}(X^\dagger Y^\dagger+XY-X^\dagger Y-XY^\dagger),\label{eq:T_{xy} leading order Y dependence}\\T_{yy}&\sim T_{yy}^\perp-i\omega_zk\ell^{1/2}(X^\dagger+X)(Y^\dagger-Y).\label{eq:T_{yy} leading order Y dependence}
\end{align}

We see that $\perp[\eta^H]_{abcd}-\eta^{H\perp}_{abcd}$ has corrections of order $k^2\ell^3$ or smaller. Therefore, at the order in $k$ and $\ell$ that we have been working, we expect the Hall viscosity derived from the projected perturbed Hamiltonian to match the in-plane components of the three-dimensional result. And indeed, using Eq.~(\ref{eq:Hall viscosity with CSS inserted}), we have
\begin{align}\label{eq:Alice's Hall viscosity calculation}
\eta^H_{abcd}&=-\frac{2}{L_xL_y}\sum_{n=1}^\infty \frac{\text{Im}\left(\braket{0|T_{ab}^\perp|n}\braket{n|T_{cd}^\perp|0}\right)}{(E_0-E_n)^2},
\end{align}
and for the particular components, with $\hbar$ and the contributions of $N$ electrons included, we find
\begin{align}
\eta_{1122}^{H\perp}&=0,\label{eq:Alice's calculation eta1122}\\\eta_{1112}^{H\perp}&=\frac{\hbar \rho}{4}\left(1-\frac{k^2\ell^2}{2}\right),\label{eq:Alice's calculation eta1112}\\\eta_{1222}^{H\perp}&=\frac{\hbar \rho}{4}\left(1-\frac{3k^2\ell^2}{2}\right).\label{eq:Alice's calculation eta1222}
\end{align}
at leading order. Comparing with Eqs.~(\ref{eq:TF Hall viscosity 1122})-(\ref{eq:TF Hall viscosity 1222}), we see that the Hall viscosity calculated from the projected Hamiltonian also matches the in-plane result for the tilted field system. Furthermore, from Eq.~(\ref{eq:Hall elastic modulus}) and the fact that $\braket{T_{ab}}_0=\braket{T^\perp_{ab}}_0$, it is clear that the projected system exactly recreates the Hall elastic modulus of the tilted field system. Therefore, although the projection formalism that we developed in the previous section does not guarantee that the linear response functions derived via the projected Hamiltonian match the in-plane components of the fully three-dimensional linear response functions, we find for the case of the tilted field system that the Hall viscosity and Hall conductivity are correctly reproduced by the projected Hamiltonian system at leading order. The Hall elastic modulus is recreated to all orders. For the Hall viscosity, we do not expect the agreement to be maintained at higher orders. Nonetheless, the agreement at the lowest non-trivial order is sufficient for our purposes.

To give the effective projected system more physical meaning, let us next equip it with coordinate and momentum operators, $\bar{x}$, $\bar{y}$, $\bar{\pi}_x$ and $\bar{\pi}_y$. We insist that these ``bar'' coordinates and momenta obey the commutation relations $[\bar{x}_a,\bar{\pi}_b]=\delta_{ab}$, $[\bar{\pi}_a,\bar{\pi}_b]=-i\epsilon_{ab}B^*$ and $[\bar{x}_a,\bar{x}_b]=0$. At present, $B^*$ is the unspecified strength of the perpendicular magnetic field, but because we want the electron density for the projected system to match the electron surface density of the tilted field system at $\nu=1$ filling, we require $B^*=B_z$. There is still ambiguity in how we define $\bar{x}_a$ and $\bar{\pi}_a$, however; given one definition, we may always give another using $\bar{x}_a'=\Xi_{ab}^T\bar{x}_b$ and $\bar{\pi}_a'=\Xi^{-1}_{ab}\bar{\pi}_b$ for any matrix $\Xi$ with unit determinant. Note that although the redefinition looks like tensor transformation, we do not simultaneously change $\lambda_{ab}$. Consequently, the form of the Hamiltonian, current and stress tensor looks different in different frames we choose to define. We may as well choose coordinates in which all three look particularly simple. A natural first choice is
\begin{align}
\bar{x}&=-\frac{1}{\sqrt{2B_z}}\left(X^\dagger+X\right)-\frac{i}{\sqrt{2B_z}}\left(c^\dagger-c\right),\label{eq:bar x}\\
\bar{y}&=\frac{i}{\sqrt{2B_z}}\left(X^\dagger-X\right)+\frac{1}{\sqrt{2B_z}}\left(c^\dagger+c\right),\label{eq:bar y}\\
\bar{\pi}_x&=-i\sqrt{\frac{B_z}{2}}\left(X^\dagger-X\right),\label{eq:bar pi_x}\\
\bar{\pi}_y&=-\sqrt{\frac{B_z}{2}}(X^\dagger+X).\label{eq:bar pi_y}
\end{align}
In terms of these operators, we have
\begin{align}
H^{\perp,\Lambda,A}&=\frac{1}{2m^*}\left(\bar{\pi}_x^2+\bar{\pi}_y^2\right)+T_{ab}^\perp\lambda_{ab}+J_a^\perp \delta A_a+\frac{\omega_2}{2},\label{eq:bar projected Hamiltonian}
\end{align}
where we have introduced the effective mass $m^*\equiv (1+k^2\ell^2/2)m$ and where the projected current and stress tensor are given by
\begin{align}
J_x^\perp&=\left(1+\frac{k^2\ell^2}{4}\right)\frac{\bar{\pi}_x}{m^*},\label{eq: bar J_x}\\J_y^\perp&=\left(1-\frac{k^2\ell^2}{4}\right)\frac{\bar{\pi}_y}{m^*},\label{eq: bar J_y}\\
T_{xx}^\perp&=\frac{\bar{\pi}_x^2}{m^*},\label{eq: bar T_xx}\\T_{xy}^\perp&=\left(1-\frac{k^2\ell^2}{2}\right)\frac{1}{2m^*}\{\bar{\pi}_x,\bar{\pi}_y\},\label{eq: bar T_xy}\\T_{yy}^\perp&=\left(1-k^2\ell^2\right)\frac{\bar{\pi}_y^2}{m^*}+\frac{\omega_x^2}{2\omega_0}.\label{eq: T_yy}
\end{align}
Thus, the new position and momenta in the effective projected system give the unperturbed Hamiltonian the usual form for a point particle of mass $m^*$ and charge $-|e|=-1$ in a uniform magnetic field and Euclidean background.

However, since $J_a^\perp\neq -\frac{\bar{\pi}_a}{m^*}$ and $T^\perp_{ab}\neq \frac{1}{2m^*}\{\bar{\pi}_a,\bar{\pi}_b\}$, the coupling of the Hamiltonian to perturbations in the electromagnetic field or the metric is not minimal. Alternatively stated, the particles of the projected fluid are not featureless point sources of charge and momentum in the bar frame; the conserved number and momentum densities associated with the Hamiltonian in Eq.~(\ref{eq:bar projected Hamiltonian}) are not simple delta functions peaked at the positions of the particles.

A  natural next step is to use the freedom given to us by the canonical transformations to try to define a frame in which the Hamiltonian couples minimally to the electromagnetic field, i.e.~where the number density is a delta function peaked at each particle postiion. Therefore, let us introduce new ``tilde'' coordinates and momenta, given by
\begin{align}
\tilde{x}_a=\Xi_{ab}^T\bar{x}_b,\label{eq: tilde x_a}\\\tilde{\pi}_a=\Xi^{-1}_{ab}\bar{\pi}_b,\label{eq: tilde pi_a}
\end{align}
where 
\begin{align}
\Xi^{-1}&=\left(\begin{array}{cc}1-\frac{k^2\ell^2}{4}&0\\0&1+\frac{k^2\ell^2}{4}\end{array}\right).\label{eq: bar to tilde transformation matrix}
\end{align}
In terms of these new operators, the perturbed Hamiltonian is
\begin{align}
H^{\perp,\Lambda,A}&=\frac{1}{2}\tilde{m}_{ab}\tilde{\pi}_a\tilde{\pi}_b+T^\perp_{ab}\lambda_{ab}+J_a^\perp \delta A_a+\frac{\omega_2}{2},\label{eq: tilde projected Hamiltonian}
\end{align}
where the inverse mass tensor is given by
\begin{align}
\tilde{m}_{ab}&=\frac{1}{m^*}\left(\begin{array}{cc}1+\frac{k^2\ell^2}{2}&0\\0&1-\frac{k^2\ell^2}{2}\end{array}\right),
\end{align}\label{eq: tilde inverse mass tensor}
and where the stress tensor and current operator take the form
\begin{align}
J_a^\perp&=\tilde{m}_{ab}\tilde{\pi}_b,\label{eq: tilde J_a}\\
T_{xx}^\perp&=\left(1+\frac{k^2\ell^2}{2}\right)\frac{\tilde{\pi}_x^2}{m^*},\label{eq: tilde T_xx}\\T_{xy}^\perp&=\left(1-\frac{k^2\ell^2}{2}\right)\frac{1}{2m^*}\{\tilde{\pi}_x,\tilde{\pi}_y\},\label{eq: tilde T_xy}\\T_{yy}^\perp&=\left(1-\frac{3k^2\ell^2}{2}\right)\frac{\tilde{\pi}_y^2}{m^*}+\frac{\omega_x^2}{2\omega_0}.\label{eq: tilde T_yy}
\end{align}

This is the main result of this section. 

This new coordinate system introduces an anisotropic mass tensor into the unperturbed Hamiltonian, but the payoff is that the electromagnetic field now couples minimally. Namely, we can define $\tilde{\pi}_a\equiv \tilde{p}_a+A_a$, where $\vec{A}=(A_x,A_y)$ satisfies $\epsilon_{ab}\frac{\partial A_b}{\partial \tilde{x}_a}=B_z$, and vary $A_a\to A_a+\delta A_a$ in the unperturbed Hamiltonian to derive $J_a^\perp$:
\begin{align}\label{eq: tilde projected Hamiltonian variation in A}
H^\perp&\to H^\perp+\tilde{m}_{ab}\tilde{\pi}_b\delta A_a+O(A^2),\\&\to H^\perp+J_a^\perp \delta A_a+O(A^2).
\end{align}
The coupling of the strain fields to the Hamiltonian is still not minimal, however. If the coupling were minimal, meaning that under a strain $\Lambda=1+\lambda+O(\lambda^2)$ the position and momenta change according to $\tilde{x}_a\to \tilde{x}_a+\lambda_{ba}\tilde{x}_b$ and $\tilde{\pi}_a\to \tilde{\pi}_a-\lambda_{ab}\tilde{\pi}_b$ at leading order, then we would find $H^\Lambda=H^0-\lambda_{ab}\frac{1}{2}\tilde{m}_{ab}\{\tilde{\pi}_a,\tilde{\pi}_b\}$. However, from Eqs.~(\ref{eq: tilde T_xx})-(\ref{eq: tilde T_yy}), it is clear that $T_{ab}^\perp =\frac{1}{2}\tilde{m}_{ab}\{\tilde{\pi}_a,\tilde{\pi}_b\}$ holds for $ab=xx,yx$ but not for $ab=xy,yy$, which is precisely what the projection of the continuity equation in Eqs.~(\ref{eq:Tyy continuity projection})-(\ref{eq:Txy continuity projection}) suggested. Therefore, although we have succeeded in centering the charges on the positions $\tilde{x}_i$ of the particles, the momentum distribution is still nonstandard.

Incidentally, one might ask why we choose to use the canonical transformations to make the fluid particles simple point charges rather than simple point sources of momentum. The first answer is primarily a pragmatic one: there does not appear to be a coordinate frame in which the coupling to the metric is minimal, in large part because of the constant term in $T_{yy}^\perp$. The second answer is that $U(1)$ gauge symmetry implies conservation of charge and therefore of number of electrons. The tilted field system conserves electron number and we'd like the projected system to do the same. Making the coupling of the projected Hamiltonian to the electromagnetic field minimal, and thus being able to write $\pi_a=p_a+A_a$, gives us $U(1)$ gauge symmetry and ensures that particle number in the projected and unprojected system coincide. Alternative approaches to minimal coupling have been considered recently in Ref.~\onlinecite{limtragool2016anomalous} in an unrelated context.

Finally, we should address a simple but important question regarding the projected system defined in Eq.~(\ref{eq: tilde projected Hamiltonian}): does it have strain generators whose commutators with the unperturbed Hamiltonian, $H^\perp=\frac{1}{2}\tilde{m}_{ab}\tilde{\pi}_a\tilde{\pi}_b$, yield the stress tensors? And if so, what do they look like? This question is especially important because our derivation of the Kubo formula for the Hall viscosity given in Eq.~(\ref{eq:Hall viscosity with CSS inserted}) relied on the Ward identity $T_{\mu\nu}=-\partial_t J_{\mu\nu}$. In other words, can Alice derive the correct form of the contact terms in the projected Kubo formula for viscosity? 

Some careful thought reveals that Alice can indeed identify strain generators for the projected system. We know it is possible to define strain generators satisfying Eqs.~(\ref{eq:J and x commutator}) and (\ref{eq:J and pi commutator}) for a two-dimensional system with magnetic field\cite{bradlyn2012kubo,bradlyn2015linear}; let's call these generators $J^{\text{2D}}_{ab}$. It is easy to check that the commutator of the 2D strain generators with the unperturbed projected Hamiltonian is given by
\begin{align}
-i[H^\perp,J_{ab}^{\text{2D}}]=\frac{1}{2}\tilde{m}_{ac}\{\tilde{\pi}_b,\tilde{\pi}_c\},
\end{align}
which is, unsurprisingly, the expression for the stress tensor we arrived at in section \ref{subsec:BM stress tensor and Hall viscosity} via the momentum density continuity equation. Therefore, we find
\begin{align}
T^\perp_{xx}&=-i[H^\perp,J_{xx}^{\text{2D}}],\\T^\perp_{xy}&=-i[H^\perp,(1-k^2\ell^2)J_{xy}^{\text{2D}}],\\T_{yx}^\perp&=-i[H^\perp,J_{yx}^{\text{2D}}],\\T^\perp_{yy}-\frac{\omega_x^2}{2\omega_0}&=-i[H^\perp,(1-k^2\ell^2)J_{yy}^{\text{2D}}],
\end{align}
and we can acceptably define three of the four strain generators of the projected system to be
\begin{align}
J_{xx}^\perp&=J_{xx}^{\text{2D}},\label{eq:projected strain 1}\\J_{xy}^\perp&=(1-k^2\ell^2)J_{xy}^{\text{2D}},\\J_{yx}^\perp&=J_{yx}^{\text{2D}}.
\end{align}
Defining $J_{yy}^\perp$ requires a little more care because of the $\frac{\omega_x^2}{2\omega_0}$ term. In particular, we must add a term, $\Delta$, to $(1-k^2\ell^2)J_{yy}^{\text{2D}}$ that satisfies $-i[H^\perp,\Delta ]=\frac{\omega_x^2}{2\omega_0}$. If $\omega_x^2$ and $\omega_0$ are taken to be c-numbers, such a $\Delta$ is forbidden by Pauli's theorem\cite{pauli_1980}--- to be precise, if $\frac{\omega_x^2}{2\omega_0}$ is a c-number, then $p_H=\frac{2\omega_0}{\omega_x^2}\Delta$ is Hermitian and satisfies $[H,p_H]=i$, implying that $U_H(E)\equiv e^{-ip_H E}$ is an energy translation operator that can lower energy eigenstates ad infinitum, an unacceptable result given that the Hamiltonian is bounded below. The way to get around Pauli's theorem is to promote $\omega_x$ to an operator and introduce its conjugate, $\Omega$, satisfying $[\omega_x,\Omega]=i$. We let $\omega_x$ and $\Omega$ commute with the $X$ and $c$ ladder operators. Since the projected Hamiltonian is now properly written $H^\perp=\left(1-\frac{\omega_x^2}{2\omega_0^2}\right)(X^\dagger X+XX^\dagger)$, we see that $\Delta\equiv -\frac{1}{4}\frac{\omega_0}{X^\dagger X+XX^\dagger}\{\omega_x,\Omega\}$ has the desired commutator with $H^\perp$; $\Delta$ is admittedly not pretty, but it gets the job done. Note that the eigenvalues of $X^\dagger X+XX^\dagger$ are the positive odd integers, justifying our use of its inverse. The fourth strain generator in the projected system is thus
\begin{align}
J^\perp_{yy}&=(1-k^2\ell^2)J^{\text{2D}}_{yy}+\Delta.\label{eq:projected strain 2}
\end{align}

Since we can define the strain generators, we are justified in applying the Kubo formula for the Hall viscosity in the projected system. Note, however, that the strain generators above do not satisfy $\mathfrak{gl}(2,\mathbb{R})$ commutation relations. Admittedly, there is a lot of freedom in the definition of the strain generators: we may add any term that commutes with the Hamiltonian to any of the strain generators and they will still satisfy $-i[H^\perp,J_{ab}^\perp]=T_{ab}^\perp$. This freedom means, firstly, that there is not one but rather a large family of projected systems that behave to leading order like the strongly-confined tilted field system. Secondly, we can use the freedom to try to make the strain-generators nicer -- for instance,  we may try to identify strain generators obeying the Lie algebra of the general linear group-- and to make the effect of the strain on the system, manifested in the commutators $[J^\perp_{ab},\tilde{x}_c]$ and $[J^\perp_{ab},\tilde{\pi}_c]$, simpler.

It is not solely for aesthetic reasons we would like to find well-behaved strain generators. Since the derivation of Eq.~(\ref{eq:Hall elastic modulus}) depends on the commutators of the strain generators obeying the Lie algebra of the general linear group, our claim that the projected system recreates the Hall elastic modulus of the tilted field system is not entirely in good faith. To apply Eq.~(\ref{eq:Hall elastic modulus}), we should identify strain generators with the general linear commutation relations; it is unclear at present whether it is possible to do so. These are interesting questions worth pursuing in future work.

\section{Conclusions}

In this work, we have pushed mapping between magnetic field tilt and mass anisotropy quite far. We have shown that by appropriately redefining the effective area and perpendicular magnetic field, the tilted field system can be mapped perturbatively (to order $k^2\ell^2$) onto a system with an anisotropic effective mass tensor, as regards the Hall conductivity and contracted Hall viscosity tensors. Nevertheless, several differences between the two systems emerge when we look deeper. By developing a rigorous projection procedure, we have shown that the stress tensor and elastic response of the titled field system (to order $k^2\ell^2$) display features that cannot be captured by a simple two-dimensional fluid of non-interacting point particles. We found that the projected fluid couples non-minimally to deformations of the systems. The stress tensor operator is symmetric (even though the mass is anisotropic), and that there is an anomalous ``vacuum stress'' $\langle T_{xx}-T_{yy}\rangle_0$ in the ground state. This leads to a nonvanishing and nondissipative \emph{Hall} elastic modulus, which is uncharacteristic for a fluid of point particles.

This, along with the difficulties in mapping the projected strain generators to the algebra $\mathfrak{gl}(2,\mathbb{R})$, strongly suggest that our projected fluid is made up of composite objects. We saw that both odd features of the stress tensor in the previous paragraph arose due to the perturbative action of the Lorentz force due to the in-plane magnetic field. This suggests that our projected fluid is composed of particles with \emph{quenched internal degrees of freedom}.

Indeed, we can look back at the Hamiltonian Eq.~(\ref{eq:TF gauge indep Hamiltonian diagonalized}) and choose an interpretation where the $X$ and $c$ oscillators correspond to a 2D Landau level problem with cyclotron frequency $\omega_1$. In this point of view, the $Y$ oscillator corresponds to an internal, vibrational degree of freedom of each particle, with frequency $\omega_2$, as depicted in Fig.~\ref{fig:projection schematic}. For nonzero tilt angles, this internal vibrational motion has some component of displacement in the $x-y$ plane. In our projection procedure, we quench the internal degree of freedom to its ground state, leaving behind only the zero-point motion. The zero-point displacement, however, still couples to strains of the system, leading to anomalous stresses in the ground state. In particular, the zero-point energy of the oscillator is sensitive to strains, leading to a c-number contribution in the stress tensor. From this perspective, it is natural then to expect the projected system to be composed of non-point particles extended over a length proportional to the oscillator length of the $Y$ oscillator. Furthermore, our definition of the projected response functions allows for virtual transitions to excited internal states, and therefore the projected response functions capture the transfer of stress between internal and ambient motion. We have borne out this expectation quantitatively.

In order to make these points, it was sufficient for us to examine the effects of tilted-field and mass anisotropy in isolation. We note, however, that there is no fundamental obstacle to evaluating the linear response coefficients for a system with both forms of anisotropy, provided the anisotropy in the mass tensor is confined to the $x-y$ plane. Indeed, in this case the canonical transformation in Eq.~(\ref{eq:BM momentum in terms of ladder ops}) can be used to put the Hamiltonian in isotropic tilted-field form of Eq.~(\ref{eq:tilted field Hamiltonian}), albeit with a rescaling of the in-plane component of the field.

Throughout this work, we have highlighted several points which warrant further investigation. First, we have highlighted a new non-dissipative susceptibility, the Hall elastic modulus, which could in principle be measured experimentally. Second, the role of the non-minimal coupling of field tilt to strain as it applies to tilt-induced nematicity and stripe phases is an interesting avenue for future research. Third, the form Eqs.~(\ref{eq:bimetric Hall viscosity}--\ref{eq:bimetric xi}) of the contracted Hall tensor suggests that it would be fruitful to investigate the Hall viscosity of systems with vector or pseudovector anisotropy. Finally, the failure of the naive projected strain generators of Eqs.~(\ref{eq:projected strain 1}--\ref{eq:projected strain 2}) to obey $\mathfrak{gl}(2,\mathbb{R})$ commutation relations points towards the need for a generalization of the Kubo formalism for viscosity to extended objects with internal degrees of freedom which can couple to the metric\cite{link2018elastic}. We are hopeful that this paper will lay the groundwork for some of these investigations.

\acknowledgments
The authors would like to thank E.~Fradkin, A.~Gromov, K.~Landsteiner, P.~W.~Phillips, N.~Read, S.~L.~Sondhi, V.~Vitelli, and M.~A.~H.~Vozmediano for fruitful discussions. This work was performed in part at the Aspen Center for Physics, which is supported by National Science Foundation grant PHY-1607611. B.O. was supported during the summer of 2018 by the Fred Fox Class of 1939 Fund and the Roundtable Fund.

\appendix

\section{Leading order expansions for important results}\label{app:important results}
In analyzing the tilted field system, many of our calculations reduce to expressing operators in terms of $X$, $Y$, and $c$ and then expanding in $k$ and $\ell$. The purpose of this appendix is to give explicit expressions for the most pertinent operators. These are the tilted field positions $x_\mu$ and momenta $\pi_\mu$; the tilted field stress tensor $T_{\mu\nu}$; the positions $\tilde{x}_a$ and momenta $\tilde{\pi}_a$ in the tilde frame of the projected system; and the projection of terms of the form $\{x_\mu,\pi_\nu\}$, which appear in the momentum continuity equation of the tilted field system.

In each of the expressions below, corrections satisfy one of the following: (i) they have at least two powers of $k$ and two powers of $\ell$ with the sum of the powers of $k$ and $\ell$ greater than or equal to five, (ii) they have at least 10 powers of $\ell$, or (iii) they have at least 10 powers of $k$. For instance, potential corrections include $k^2\ell^3, k^3\ell^2, k\ell^{10}, k^{10}$, and exclude $k\ell^3, k\ell^2, k^7$. The choice of using 10 as a cutoff was rather arbitrary. However, it does guarantee that $k^2\ell^2\sim k^{10}$ in a lab setting where $\ell\sim 10^{-2}$ and the tilt angle is $20^\circ$. Moreover, the expansions are not really sensitive to what cutoff we use. 

In the expressions below, we frequently use the shorthand 
\begin{align}\label{eq:definition of Delta}
\Delta &\equiv 1+\ell^2+\ell^4+\ell^6+\ell^8+\ell^{10}
\end{align}

\begin{enumerate}
\item Physical momenta and coordinates of the tilted field system
\begin{align}
\pi_x&=-i\sqrt{\frac{B_z}{2}}\left(1-\frac{k^2\ell^2}{4}\right)(X^\dagger-X)\nonumber\\&-\sqrt{\frac{m\omega_0}{2}}k\ell^2\Delta(Y^\dagger+Y)\label{eq:pi_x leading order expansion in X,Y,c}\\\pi_y&=-\sqrt{\frac{B_z}{2}}\left(1-\frac{3}{4}k^2\ell^2\right)\left(X^\dagger+X\right)\nonumber\\&+i\sqrt{\frac{m\omega_0}{2}}k\ell\Delta(Y^\dagger-Y)\label{eq:pi_y leading order expansion in X,Y,c}\\\pi_z&=-i\sqrt{\frac{B_z}{2}}k\ell^2\Delta\left(X^\dagger-X\right)\nonumber\\&+\sqrt{\frac{m\omega_0}{2}}\left(1+\frac{k^2\ell^2}{4}\right)(Y^\dagger+Y)\label{eq:pi_z leading order expansion in X,Y,c}\\z&=-\frac{1}{\sqrt{2m\omega_0}}k\ell^{3/2}\Delta(X^\dagger+X)\nonumber\\&-\frac{1}{\sqrt{2m\omega_0}}\left(1-\frac{1}{4}k^2\ell^2\right)(Y^\dagger-Y)\label{eq:z leading order expansion in X,Y,c}\\
    x&=\frac{i}{\sqrt{2B_z}}(c-c^\dagger)-\frac{1}{\sqrt{2B_z}}\left(1+\frac{k^2\ell^2}{4}\right)(X^\dagger+X)\nonumber\\&+\frac{ik\ell^{5/2}}{\sqrt{2B_z}}\Delta(Y^\dagger-Y)\\y&=\frac{1}{\sqrt{2B_z}}(c+c^\dagger)+\frac{i}{\sqrt{2B_z}}\left(1-\frac{k^2\ell^2}{4}\right)(X^\dagger-X)\nonumber\\&+\frac{k\ell^{3/2}}{\sqrt{2B_z}}\Delta (Y^\dagger+Y)
\end{align}

The leading order expansion of the current operators, $J_\mu$, follows from $J_\mu=-\frac{1}{m}\pi_\mu$ and Eqs.~(\ref{eq:pi_x leading order expansion in X,Y,c})-(\ref{eq:pi_z leading order expansion in X,Y,c}).

\item Stress tensor of the tilted field system
\begin{align}
T_{xx}&=-\frac{\omega_z}{2}\left(1-\frac{k^2\ell^2}{2}\right)(X^\dagger-X)^2+\frac{\omega_0}{2}k^2\ell^4(Y^\dagger+Y)^2\nonumber\\&+i\omega_zk\ell^{3/2}\Delta (X^\dagger-X)(Y^\dagger+Y)\label{eq:T_{xx} leading order expansion in X,Y,c}\\T_{xy}&=\frac{i\omega_z}{2}\left(1-k^2\ell^2\right)\left((X^\dagger)^2-X^2\right)-\frac{i\omega_z}{2}k^2\ell^2\left((Y^\dagger)^2-Y^2\right)\nonumber\\&+\frac{\omega_z}{2}k\ell^{1/2}\Delta\Big((1+\ell)(X^\dagger Y^\dagger+XY)\nonumber\\&\hspace{2.5cm}-(1-\ell)(XY^\dagger+X^\dagger Y)\Big)\label{eq:T_{xy} leading order expansion in X,Y,c}\\T_{yy}&=\frac{\omega_z}{2}\left(1-\frac{3k^2\ell^2}{2}\right)(X^\dagger+X)^2-\frac{\omega_0}{2}k^2\ell^2(Y^\dagger-Y)^2\nonumber\\&-i\omega_zk\ell^{1/2}\Delta (X^\dagger+X)(Y^\dagger-Y)\label{eq:T_{yy} leading order expansion in X,Y,c}\\T_{xz}&=-\frac{\omega_z}{2}k\ell^2\Delta(X^\dagger-X)^2-\frac{\omega_z}{2}k\ell\Delta(Y^\dagger+Y)^2\nonumber\\&-\frac{i\omega_z}{2}\ell^{-1/2}(X^\dagger-X)(Y^\dagger+Y)\label{eq:T_{xz} leading order expansion in X,Y,c}\\T_{yz}&=\frac{i\omega_z}{2}k\ell^2\Delta((X^\dagger)^2-X^2)+\frac{i\omega_z}{2}k\Delta((Y^\dagger)^2-Y^2)\nonumber\\&-\frac{\omega_z}{2}\ell^{-1/2}\left(1-\frac{k^2\ell^2}{2}\right)(X^\dagger+X)(Y^\dagger+Y)\label{eq:T_{yz} leading order expansion in X,Y,c}\\T_{zz}&=-\frac{\omega_z}{2}k^2\ell^4(X^\dagger-X)^2+\frac{\omega_0}{2}\left(1+\frac{k^2\ell^2}{2}\right)(Y^\dagger+Y)^2\nonumber\\&-i\omega_zk\ell^{3/2}\Delta (X^\dagger-X)(Y^\dagger+Y)\label{eq:T_{zz} leading order expansion in X,Y,c}
\end{align}
It follows that the projected forms for $T_{xz}$, $T_{yz}$ and $T_{zz}$ are given by
\begin{align}
T_{xz}^\perp&=-\frac{\omega_z}{2}k\ell^2 (X^\dagger-X)^2-\frac{\omega_z}{2}k\ell\\T_{yz}^\perp&=\frac{i\omega_z}{2}k\ell^2((X^\dagger)^2-X^2)\\T_{zz}^\perp&=-\frac{\omega_z}{2}k^2\ell^4(X^\dagger-X)^2+\frac{\omega_0}{2}\left(1+\frac{k^2\ell^2}{2}\right)
\end{align}
The ground state expectation values for $T_{xz}$, $T_{yz}$, and $T_{zz}$ are therefore
\begin{align}
\braket{T_{xz}}&=-\frac{\omega_z}{2}k\ell(1-\ell)\\\braket{T_{xz}}&=0\\\braket{T_{zz}}&=\frac{\omega_0}{2}\left(1+\frac{k^2\ell^2}{2}\right)
\end{align}
\item Coordinates and momenta of the projected system in the tilde frame
\begin{align}
\tilde{x}&=-\frac{1}{\sqrt{2B_z}}\left(1+\frac{k^2\ell^2}{4}\right)\left[(X^\dagger+X)+i(c^\dagger-c)\right]\\\tilde{y}&=\frac{1}{\sqrt{2B_z}}\left(1-\frac{k^2\ell^2}{4}\right)\left[i(X^\dagger-X)+(c^\dagger+c)\right]\\\tilde{\pi}_x&=-i\sqrt{\frac{B_z}{2}}\left(1-\frac{k^2\ell^2}{4}\right)(X^\dagger-X)\\\tilde{\pi}_y&=-\sqrt{\frac{B_z}{2}}\left(1+\frac{k^2\ell^2}{4}\right)(X^\dagger+X)
\end{align}

\item Projection of $\{x_\mu,\pi_\nu\}$ terms:
\begin{align}
    \perp[\{\pi_x,x\}]&=\left(1-\frac{k^2\ell^2}{4}\right)(X^\dagger-X)(c-c^\dagger)\nonumber\\&+\frac{i}{2}\{X^\dagger-X,X^\dagger+X\}\\\perp[\{\pi_x,y\}]&=-i\left(1-\frac{k^2\ell^2}{4}\right)(X^\dagger-X)(c+c^\dagger)\nonumber\\&+\left(1-\frac{k^2\ell^2}{2}\right)(X^\dagger-X)^2\\\perp[
    \{\pi_x,z\}]&=\frac{ik\ell^2\Delta}{2}\{X^\dagger-X,X^\dagger+X\}\\\perp[\{\pi_y,x\}]&=-i\left(1-\frac{3}{4}k^2\ell^2\right)(X^\dagger+X)(c-c^\dagger)\nonumber\\&+\left(1-\frac{k^2\ell^2}{2}\right)(X^\dagger+X)^2\\\perp[\{\pi_y,y\}]&=-\left(1-\frac{3k^2\ell^2}{4}\right)(c+c^\dagger)(X^\dagger+X)\nonumber\\&-\frac{i}{2}(1-k^2\ell^2)\{X^\dagger+X,X^\dagger-X\}\\\perp[\{\pi_y,z\}]&=k\ell^2\Delta (X^\dagger+X)^2-k\ell\Delta\\\perp[\{\pi_z,x\}]&=k\ell^2\Delta (c-c^\dagger)(X^\dagger-X)\nonumber\\&+\frac{ik\ell^2\Delta}{2} \{X^\dagger-X,X^\dagger+X\}\\\perp[\{\pi_z,y\}]&=-ik\ell^2\Delta (c+c^\dagger)(X^\dagger-X)\nonumber\\&+k\ell^2\Delta (X^\dagger-X)^2+k\ell\Delta\\\perp[\{\pi_z,z\}]&=\frac{ik\ell^3\Delta^2}{2}\{X-X^\dagger,X+X^\dagger\}
\end{align}

\item Nonvanishing out-of-plane components of the Hall viscosity of the tilted field system:
\begin{align}\label{eq:out-of-plane Hall viscosity components}
\eta^H_{1123}&=-\frac{1}{4}k\ell^2,&\eta^H_{1213}&=-\frac{k\ell^2}{4},&\eta^H_{1323}&=-\frac{\ell}{2}(1-2\ell),\nonumber\\\eta^H_{2223}&=-\frac{3k\ell^2}{4},&\eta_{2333}^H&=-\frac{k\ell}{4},&\eta^H_{1233}&=\frac{k^2\ell^3}{4}
\end{align}
\end{enumerate}

\bibliography{mybib}

\begin{thebibliography}{69}%
\makeatletter
\providecommand \@ifxundefined [1]{%
 \@ifx{#1\undefined}
}%
\providecommand \@ifnum [1]{%
 \ifnum #1\expandafter \@firstoftwo
 \else \expandafter \@secondoftwo
 \fi
}%
\providecommand \@ifx [1]{%
 \ifx #1\expandafter \@firstoftwo
 \else \expandafter \@secondoftwo
 \fi
}%
\providecommand \natexlab [1]{#1}%
\providecommand \enquote  [1]{``#1''}%
\providecommand \bibnamefont  [1]{#1}%
\providecommand \bibfnamefont [1]{#1}%
\providecommand \citenamefont [1]{#1}%
\providecommand \href@noop [0]{\@secondoftwo}%
\providecommand \href [0]{\begingroup \@sanitize@url \@href}%
\providecommand \@href[1]{\@@startlink{#1}\@@href}%
\providecommand \@@href[1]{\endgroup#1\@@endlink}%
\providecommand \@sanitize@url [0]{\catcode `\\12\catcode `\$12\catcode
  `\&12\catcode `\#12\catcode `\^12\catcode `\_12\catcode `\%12\relax}%
\providecommand \@@startlink[1]{}%
\providecommand \@@endlink[0]{}%
\providecommand \url  [0]{\begingroup\@sanitize@url \@url }%
\providecommand \@url [1]{\endgroup\@href {#1}{\urlprefix }}%
\providecommand \urlprefix  [0]{URL }%
\providecommand \Eprint [0]{\href }%
\providecommand \doibase [0]{http://dx.doi.org/}%
\providecommand \selectlanguage [0]{\@gobble}%
\providecommand \bibinfo  [0]{\@secondoftwo}%
\providecommand \bibfield  [0]{\@secondoftwo}%
\providecommand \translation [1]{[#1]}%
\providecommand \BibitemOpen [0]{}%
\providecommand \bibitemStop [0]{}%
\providecommand \bibitemNoStop [0]{.\EOS\space}%
\providecommand \EOS [0]{\spacefactor3000\relax}%
\providecommand \BibitemShut  [1]{\csname bibitem#1\endcsname}%
\let\auto@bib@innerbib\@empty
\bibitem [{\citenamefont {Avron}\ \emph {et~al.}(1995)\citenamefont {Avron},
  \citenamefont {Seiler},\ and\ \citenamefont
  {Zograf}}]{1995-AvronSeilerZograf}%
  \BibitemOpen
  \bibfield  {author} {\bibinfo {author} {\bibfnamefont {J.~E.}\ \bibnamefont
  {Avron}}, \bibinfo {author} {\bibfnamefont {R.}~\bibnamefont {Seiler}}, \
  and\ \bibinfo {author} {\bibfnamefont {P.~G.}\ \bibnamefont {Zograf}},\
  }\href
  {http://www.google.com/search?client=safari&rls=en-us&q=VISCOSITY+OF+QUANTUM+HALL+FLUIDS&ie=UTF-8&oe=UTF-8}
  {\bibfield  {journal} {\bibinfo  {journal} {Phys Rev Lett}\ }\textbf
  {\bibinfo {volume} {75}},\ \bibinfo {pages} {697} (\bibinfo {year}
  {1995})}\BibitemShut {NoStop}%
\bibitem [{\citenamefont {L\'{e}vay}(1995)}]{Levay1995}%
  \BibitemOpen
  \bibfield  {author} {\bibinfo {author} {\bibfnamefont {P.}~\bibnamefont
  {L\'{e}vay}},\ }\href {\doibase 10.1063/1.531066} {\bibfield  {journal}
  {\bibinfo  {journal} {J. Math. Phys.}\ }\textbf {\bibinfo {volume} {36}},\
  \bibinfo {pages} {2792} (\bibinfo {year} {1995})}\BibitemShut {NoStop}%
\bibitem [{\citenamefont {Read}(2009)}]{read2009non}%
  \BibitemOpen
  \bibfield  {author} {\bibinfo {author} {\bibfnamefont {N.}~\bibnamefont
  {Read}},\ }\href@noop {} {\bibfield  {journal} {\bibinfo  {journal} {Physical
  Review B}\ }\textbf {\bibinfo {volume} {79}},\ \bibinfo {pages} {045308}
  (\bibinfo {year} {2009})}\BibitemShut {NoStop}%
\bibitem [{\citenamefont {Read}\ and\ \citenamefont
  {Rezayi}(2011)}]{read2011hall}%
  \BibitemOpen
  \bibfield  {author} {\bibinfo {author} {\bibfnamefont {N.}~\bibnamefont
  {Read}}\ and\ \bibinfo {author} {\bibfnamefont {E.}~\bibnamefont {Rezayi}},\
  }\href@noop {} {\bibfield  {journal} {\bibinfo  {journal} {Physical Review
  B}\ }\textbf {\bibinfo {volume} {84}},\ \bibinfo {pages} {085316} (\bibinfo
  {year} {2011})}\BibitemShut {NoStop}%
\bibitem [{\citenamefont {Bradlyn}\ \emph {et~al.}(2012)\citenamefont
  {Bradlyn}, \citenamefont {Goldstein},\ and\ \citenamefont
  {Read}}]{bradlyn2012kubo}%
  \BibitemOpen
  \bibfield  {author} {\bibinfo {author} {\bibfnamefont {B.}~\bibnamefont
  {Bradlyn}}, \bibinfo {author} {\bibfnamefont {M.}~\bibnamefont {Goldstein}},
  \ and\ \bibinfo {author} {\bibfnamefont {N.}~\bibnamefont {Read}},\
  }\href@noop {} {\bibfield  {journal} {\bibinfo  {journal} {Physical Review
  B}\ }\textbf {\bibinfo {volume} {86}},\ \bibinfo {pages} {245309} (\bibinfo
  {year} {2012})}\BibitemShut {NoStop}%
\bibitem [{\citenamefont {Abanov}\ and\ \citenamefont
  {Gromov}(2014)}]{Gromov20141}%
  \BibitemOpen
  \bibfield  {author} {\bibinfo {author} {\bibfnamefont {A.}~\bibnamefont
  {Abanov}}\ and\ \bibinfo {author} {\bibfnamefont {A.}~\bibnamefont
  {Gromov}},\ }\href {\doibase 10.1103/PhysRevB.90.014435} {\bibfield
  {journal} {\bibinfo  {journal} {Phys. Rev. B}\ }\textbf {\bibinfo {volume}
  {90}},\ \bibinfo {pages} {014435} (\bibinfo {year} {2014})}\BibitemShut
  {NoStop}%
\bibitem [{\citenamefont {Tokatly}\ and\ \citenamefont
  {Vignale}(2007)}]{2007-TokatlyVignale}%
  \BibitemOpen
  \bibfield  {author} {\bibinfo {author} {\bibfnamefont {I.~V.}\ \bibnamefont
  {Tokatly}}\ and\ \bibinfo {author} {\bibfnamefont {G.}~\bibnamefont
  {Vignale}},\ }\href {\doibase 10.1103/PhysRevB.76.161305} {\bibfield
  {journal} {\bibinfo  {journal} {Phys Rev B}\ }\textbf {\bibinfo {volume}
  {76}},\ \bibinfo {pages} {161305} (\bibinfo {year} {2007})}\BibitemShut
  {NoStop}%
\bibitem [{\citenamefont {Tokatly}\ and\ \citenamefont
  {Vignale}(2009)}]{2009-TokatlyVignale-JPhys}%
  \BibitemOpen
  \bibfield  {author} {\bibinfo {author} {\bibfnamefont {I.~V.}\ \bibnamefont
  {Tokatly}}\ and\ \bibinfo {author} {\bibfnamefont {G.}~\bibnamefont
  {Vignale}},\ }\href {\doibase 10.1088/0953-8984/21/27/275603} {\bibfield
  {journal} {\bibinfo  {journal} {J Phys-Condens Mat}\ }\textbf {\bibinfo
  {volume} {21}},\ \bibinfo {pages} {275603} (\bibinfo {year}
  {2009})}\BibitemShut {NoStop}%
\bibitem [{\citenamefont {Wiegmann}\ and\ \citenamefont
  {Abanov}(2014)}]{wiegmann2014anomalous}%
  \BibitemOpen
  \bibfield  {author} {\bibinfo {author} {\bibfnamefont {P.}~\bibnamefont
  {Wiegmann}}\ and\ \bibinfo {author} {\bibfnamefont {A.~G.}\ \bibnamefont
  {Abanov}},\ }\href@noop {} {\bibfield  {journal} {\bibinfo  {journal}
  {Physical review letters}\ }\textbf {\bibinfo {volume} {113}},\ \bibinfo
  {pages} {034501} (\bibinfo {year} {2014})}\BibitemShut {NoStop}%
\bibitem [{\citenamefont {Haldane}(1983)}]{Haldane1983}%
  \BibitemOpen
  \bibfield  {author} {\bibinfo {author} {\bibfnamefont {F.~D.~M.}\
  \bibnamefont {Haldane}},\ }\href {\doibase 10.1103/PhysRevLett.51.605}
  {\bibfield  {journal} {\bibinfo  {journal} {Phys. Rev. Lett.}\ }\textbf
  {\bibinfo {volume} {51}},\ \bibinfo {pages} {605} (\bibinfo {year}
  {1983})}\BibitemShut {NoStop}%
\bibitem [{\citenamefont {Zaletel}\ \emph {et~al.}(2013)\citenamefont
  {Zaletel}, \citenamefont {Mong},\ and\ \citenamefont
  {Pollmann}}]{Zaletel-PhysRevLett.110.236801}%
  \BibitemOpen
  \bibfield  {author} {\bibinfo {author} {\bibfnamefont {M.~P.}\ \bibnamefont
  {Zaletel}}, \bibinfo {author} {\bibfnamefont {R.~S.~K.}\ \bibnamefont
  {Mong}}, \ and\ \bibinfo {author} {\bibfnamefont {F.}~\bibnamefont
  {Pollmann}},\ }\href {\doibase 10.1103/PhysRevLett.110.236801} {\bibfield
  {journal} {\bibinfo  {journal} {Phys. Rev. Lett.}\ }\textbf {\bibinfo
  {volume} {110}},\ \bibinfo {pages} {236801} (\bibinfo {year}
  {2013})}\BibitemShut {NoStop}%
\bibitem [{\citenamefont {Avron}(1998{\natexlab{a}})}]{1998-Avron}%
  \BibitemOpen
  \bibfield  {author} {\bibinfo {author} {\bibfnamefont {J.~E.}\ \bibnamefont
  {Avron}},\ }\href
  {http://www.google.com/search?client=safari&rls=en-us&q=Odd+viscosity&ie=UTF-8&oe=UTF-8}
  {\bibfield  {journal} {\bibinfo  {journal} {J Stat Phys}\ }\textbf {\bibinfo
  {volume} {92}},\ \bibinfo {pages} {543} (\bibinfo {year}
  {1998}{\natexlab{a}})}\BibitemShut {NoStop}%
\bibitem [{\citenamefont {Scaffidi}\ \emph {et~al.}(2017)\citenamefont
  {Scaffidi}, \citenamefont {Nandi}, \citenamefont {Schmidt}, \citenamefont
  {Mackenzie},\ and\ \citenamefont {Moore}}]{scaffidi2017hydrodynamic}%
  \BibitemOpen
  \bibfield  {author} {\bibinfo {author} {\bibfnamefont {T.}~\bibnamefont
  {Scaffidi}}, \bibinfo {author} {\bibfnamefont {N.}~\bibnamefont {Nandi}},
  \bibinfo {author} {\bibfnamefont {B.}~\bibnamefont {Schmidt}}, \bibinfo
  {author} {\bibfnamefont {A.~P.}\ \bibnamefont {Mackenzie}}, \ and\ \bibinfo
  {author} {\bibfnamefont {J.~E.}\ \bibnamefont {Moore}},\ }\href@noop {}
  {\bibfield  {journal} {\bibinfo  {journal} {Physical review letters}\
  }\textbf {\bibinfo {volume} {118}},\ \bibinfo {pages} {226601} (\bibinfo
  {year} {2017})}\BibitemShut {NoStop}%
\bibitem [{\citenamefont {Hoyos}\ and\ \citenamefont {Son}(2012)}]{Hoyos2012}%
  \BibitemOpen
  \bibfield  {author} {\bibinfo {author} {\bibfnamefont {C.}~\bibnamefont
  {Hoyos}}\ and\ \bibinfo {author} {\bibfnamefont {D.~T.}\ \bibnamefont
  {Son}},\ }\href {\doibase 10.1103/PhysRevLett.108.066805} {\bibfield
  {journal} {\bibinfo  {journal} {Phys. Rev. Lett.}\ }\textbf {\bibinfo
  {volume} {108}},\ \bibinfo {pages} {066805} (\bibinfo {year}
  {2012})}\BibitemShut {NoStop}%
\bibitem [{\citenamefont {Delacr{\'e}taz}\ and\ \citenamefont
  {Gromov}(2017)}]{delacretaz2017transport}%
  \BibitemOpen
  \bibfield  {author} {\bibinfo {author} {\bibfnamefont {L.~V.}\ \bibnamefont
  {Delacr{\'e}taz}}\ and\ \bibinfo {author} {\bibfnamefont {A.}~\bibnamefont
  {Gromov}},\ }\href@noop {} {\bibfield  {journal} {\bibinfo  {journal}
  {Physical review letters}\ }\textbf {\bibinfo {volume} {119}},\ \bibinfo
  {pages} {226602} (\bibinfo {year} {2017})}\BibitemShut {NoStop}%
\bibitem [{\citenamefont {Haldane}\ and\ \citenamefont
  {Shen}(2015)}]{haldane2015geometry}%
  \BibitemOpen
  \bibfield  {author} {\bibinfo {author} {\bibfnamefont {F.~D.~M.}\
  \bibnamefont {Haldane}}\ and\ \bibinfo {author} {\bibfnamefont
  {Y.}~\bibnamefont {Shen}},\ }\href@noop {} {\bibfield  {journal} {\bibinfo
  {journal} {arXiv preprint arXiv:1512.04502}\ } (\bibinfo {year}
  {2015})}\BibitemShut {NoStop}%
\bibitem [{\citenamefont {Gromov}\ \emph {et~al.}(2017)\citenamefont {Gromov},
  \citenamefont {Geraedts},\ and\ \citenamefont
  {Bradlyn}}]{gromov2017investigating}%
  \BibitemOpen
  \bibfield  {author} {\bibinfo {author} {\bibfnamefont {A.}~\bibnamefont
  {Gromov}}, \bibinfo {author} {\bibfnamefont {S.~D.}\ \bibnamefont
  {Geraedts}}, \ and\ \bibinfo {author} {\bibfnamefont {B.}~\bibnamefont
  {Bradlyn}},\ }\href@noop {} {\bibfield  {journal} {\bibinfo  {journal}
  {Physical review letters}\ }\textbf {\bibinfo {volume} {119}},\ \bibinfo
  {pages} {146602} (\bibinfo {year} {2017})}\BibitemShut {NoStop}%
\bibitem [{\citenamefont {Liu}\ \emph {et~al.}(2018)\citenamefont {Liu},
  \citenamefont {Gromov},\ and\ \citenamefont {Papi{\'c}}}]{liu2018geometric}%
  \BibitemOpen
  \bibfield  {author} {\bibinfo {author} {\bibfnamefont {Z.}~\bibnamefont
  {Liu}}, \bibinfo {author} {\bibfnamefont {A.}~\bibnamefont {Gromov}}, \ and\
  \bibinfo {author} {\bibfnamefont {Z.}~\bibnamefont {Papi{\'c}}},\ }\href@noop
  {} {\bibfield  {journal} {\bibinfo  {journal} {arXiv preprint
  arXiv:1803.00030}\ } (\bibinfo {year} {2018})}\BibitemShut {NoStop}%
\bibitem [{\citenamefont {Lapa}\ \emph {et~al.}(2018)\citenamefont {Lapa},
  \citenamefont {Gromov},\ and\ \citenamefont {Hughes}}]{lapa2018geometric}%
  \BibitemOpen
  \bibfield  {author} {\bibinfo {author} {\bibfnamefont {M.~F.}\ \bibnamefont
  {Lapa}}, \bibinfo {author} {\bibfnamefont {A.}~\bibnamefont {Gromov}}, \ and\
  \bibinfo {author} {\bibfnamefont {T.~L.}\ \bibnamefont {Hughes}},\
  }\href@noop {} {\bibfield  {journal} {\bibinfo  {journal} {arXiv preprint
  arXiv:1809.06386}\ } (\bibinfo {year} {2018})}\BibitemShut {NoStop}%
\bibitem [{\citenamefont {Haldane}(2009)}]{haldane2009hall}%
  \BibitemOpen
  \bibfield  {author} {\bibinfo {author} {\bibfnamefont {F.~D.~M.}\
  \bibnamefont {Haldane}},\ }\href@noop {} {\bibfield  {journal} {\bibinfo
  {journal} {arXiv preprint arXiv:0906.1854}\ } (\bibinfo {year}
  {2009})}\BibitemShut {NoStop}%
\bibitem [{\citenamefont {Fradkin}\ and\ \citenamefont
  {Kivelson}(1999)}]{fradkin1999liquid}%
  \BibitemOpen
  \bibfield  {author} {\bibinfo {author} {\bibfnamefont {E.}~\bibnamefont
  {Fradkin}}\ and\ \bibinfo {author} {\bibfnamefont {S.~A.}\ \bibnamefont
  {Kivelson}},\ }\href@noop {} {\bibfield  {journal} {\bibinfo  {journal}
  {Physical Review B}\ }\textbf {\bibinfo {volume} {59}},\ \bibinfo {pages}
  {8065} (\bibinfo {year} {1999})}\BibitemShut {NoStop}%
\bibitem [{\citenamefont {Fradkin}\ \emph {et~al.}(2010)\citenamefont
  {Fradkin}, \citenamefont {Kivelson}, \citenamefont {Lawler}, \citenamefont
  {Eisenstein},\ and\ \citenamefont {Mackenzie}}]{fradkin2010nematic}%
  \BibitemOpen
  \bibfield  {author} {\bibinfo {author} {\bibfnamefont {E.}~\bibnamefont
  {Fradkin}}, \bibinfo {author} {\bibfnamefont {S.~A.}\ \bibnamefont
  {Kivelson}}, \bibinfo {author} {\bibfnamefont {M.~J.}\ \bibnamefont
  {Lawler}}, \bibinfo {author} {\bibfnamefont {J.~P.}\ \bibnamefont
  {Eisenstein}}, \ and\ \bibinfo {author} {\bibfnamefont {A.~P.}\ \bibnamefont
  {Mackenzie}},\ }\href@noop {} {\bibfield  {journal} {\bibinfo  {journal}
  {Annu. Rev. Condens. Matter Phys.}\ }\textbf {\bibinfo {volume} {1}},\
  \bibinfo {pages} {153} (\bibinfo {year} {2010})}\BibitemShut {NoStop}%
\bibitem [{\citenamefont {You}\ \emph {et~al.}(2014)\citenamefont {You},
  \citenamefont {Cho},\ and\ \citenamefont {Fradkin}}]{you2014theory}%
  \BibitemOpen
  \bibfield  {author} {\bibinfo {author} {\bibfnamefont {Y.}~\bibnamefont
  {You}}, \bibinfo {author} {\bibfnamefont {G.~Y.}\ \bibnamefont {Cho}}, \ and\
  \bibinfo {author} {\bibfnamefont {E.}~\bibnamefont {Fradkin}},\ }\href@noop
  {} {\bibfield  {journal} {\bibinfo  {journal} {Physical Review X}\ }\textbf
  {\bibinfo {volume} {4}},\ \bibinfo {pages} {041050} (\bibinfo {year}
  {2014})}\BibitemShut {NoStop}%
\bibitem [{\citenamefont {Maciejko}\ \emph {et~al.}(2013)\citenamefont
  {Maciejko}, \citenamefont {Hsu}, \citenamefont {Kivelson}, \citenamefont
  {Park},\ and\ \citenamefont {Sondhi}}]{maciejko2013field}%
  \BibitemOpen
  \bibfield  {author} {\bibinfo {author} {\bibfnamefont {J.}~\bibnamefont
  {Maciejko}}, \bibinfo {author} {\bibfnamefont {B.}~\bibnamefont {Hsu}},
  \bibinfo {author} {\bibfnamefont {S.}~\bibnamefont {Kivelson}}, \bibinfo
  {author} {\bibfnamefont {Y.}~\bibnamefont {Park}}, \ and\ \bibinfo {author}
  {\bibfnamefont {S.}~\bibnamefont {Sondhi}},\ }\href@noop {} {\bibfield
  {journal} {\bibinfo  {journal} {Physical Review B}\ }\textbf {\bibinfo
  {volume} {88}},\ \bibinfo {pages} {125137} (\bibinfo {year}
  {2013})}\BibitemShut {NoStop}%
\bibitem [{\citenamefont {Eisenstein}\ \emph {et~al.}(1992)\citenamefont
  {Eisenstein}, \citenamefont {Boebinger}, \citenamefont {Pfeiffer},
  \citenamefont {West},\ and\ \citenamefont {He}}]{eisensteintiltedfqh}%
  \BibitemOpen
  \bibfield  {author} {\bibinfo {author} {\bibfnamefont {J.~P.}\ \bibnamefont
  {Eisenstein}}, \bibinfo {author} {\bibfnamefont {G.~S.}\ \bibnamefont
  {Boebinger}}, \bibinfo {author} {\bibfnamefont {L.~N.}\ \bibnamefont
  {Pfeiffer}}, \bibinfo {author} {\bibfnamefont {K.~W.}\ \bibnamefont {West}},
  \ and\ \bibinfo {author} {\bibfnamefont {S.}~\bibnamefont {He}},\ }\href
  {\doibase 10.1103/PhysRevLett.68.1383} {\bibfield  {journal} {\bibinfo
  {journal} {Phys. Rev. Lett.}\ }\textbf {\bibinfo {volume} {68}},\ \bibinfo
  {pages} {1383} (\bibinfo {year} {1992})}\BibitemShut {NoStop}%
\bibitem [{\citenamefont {Murphy}\ \emph {et~al.}(1994)\citenamefont {Murphy},
  \citenamefont {Eisenstein}, \citenamefont {Boebinger}, \citenamefont
  {Pfeiffer},\ and\ \citenamefont {West}}]{eisensteintiltedinteger}%
  \BibitemOpen
  \bibfield  {author} {\bibinfo {author} {\bibfnamefont {S.~Q.}\ \bibnamefont
  {Murphy}}, \bibinfo {author} {\bibfnamefont {J.~P.}\ \bibnamefont
  {Eisenstein}}, \bibinfo {author} {\bibfnamefont {G.~S.}\ \bibnamefont
  {Boebinger}}, \bibinfo {author} {\bibfnamefont {L.~N.}\ \bibnamefont
  {Pfeiffer}}, \ and\ \bibinfo {author} {\bibfnamefont {K.~W.}\ \bibnamefont
  {West}},\ }\href {\doibase 10.1103/PhysRevLett.72.728} {\bibfield  {journal}
  {\bibinfo  {journal} {Phys. Rev. Lett.}\ }\textbf {\bibinfo {volume} {72}},\
  \bibinfo {pages} {728} (\bibinfo {year} {1994})}\BibitemShut {NoStop}%
\bibitem [{\citenamefont {Engel}\ \emph {et~al.}(1992)\citenamefont {Engel},
  \citenamefont {Hwang}, \citenamefont {Sajoto}, \citenamefont {Tsui},\ and\
  \citenamefont {Shayegan}}]{shayegantilted}%
  \BibitemOpen
  \bibfield  {author} {\bibinfo {author} {\bibfnamefont {L.~W.}\ \bibnamefont
  {Engel}}, \bibinfo {author} {\bibfnamefont {S.~W.}\ \bibnamefont {Hwang}},
  \bibinfo {author} {\bibfnamefont {T.}~\bibnamefont {Sajoto}}, \bibinfo
  {author} {\bibfnamefont {D.~C.}\ \bibnamefont {Tsui}}, \ and\ \bibinfo
  {author} {\bibfnamefont {M.}~\bibnamefont {Shayegan}},\ }\href {\doibase
  10.1103/PhysRevB.45.3418} {\bibfield  {journal} {\bibinfo  {journal} {Phys.
  Rev. B}\ }\textbf {\bibinfo {volume} {45}},\ \bibinfo {pages} {3418}
  (\bibinfo {year} {1992})}\BibitemShut {NoStop}%
\bibitem [{\citenamefont {Pan}\ \emph {et~al.}(1999)\citenamefont {Pan},
  \citenamefont {Du}, \citenamefont {Stormer}, \citenamefont {Tsui},
  \citenamefont {Pfeiffer}, \citenamefont {Baldwin},\ and\ \citenamefont
  {West}}]{pantilted}%
  \BibitemOpen
  \bibfield  {author} {\bibinfo {author} {\bibfnamefont {W.}~\bibnamefont
  {Pan}}, \bibinfo {author} {\bibfnamefont {R.~R.}\ \bibnamefont {Du}},
  \bibinfo {author} {\bibfnamefont {H.~L.}\ \bibnamefont {Stormer}}, \bibinfo
  {author} {\bibfnamefont {D.~C.}\ \bibnamefont {Tsui}}, \bibinfo {author}
  {\bibfnamefont {L.~N.}\ \bibnamefont {Pfeiffer}}, \bibinfo {author}
  {\bibfnamefont {K.~W.}\ \bibnamefont {Baldwin}}, \ and\ \bibinfo {author}
  {\bibfnamefont {K.~W.}\ \bibnamefont {West}},\ }\href {\doibase
  10.1103/PhysRevLett.83.820} {\bibfield  {journal} {\bibinfo  {journal} {Phys.
  Rev. Lett.}\ }\textbf {\bibinfo {volume} {83}},\ \bibinfo {pages} {820}
  (\bibinfo {year} {1999})}\BibitemShut {NoStop}%
\bibitem [{\citenamefont {Cs\'athy}\ \emph {et~al.}(2005)\citenamefont
  {Cs\'athy}, \citenamefont {Xia}, \citenamefont {Vicente}, \citenamefont
  {Adams}, \citenamefont {Sullivan}, \citenamefont {Stormer}, \citenamefont
  {Tsui}, \citenamefont {Pfeiffer},\ and\ \citenamefont
  {West}}]{csathytiltsecondll}%
  \BibitemOpen
  \bibfield  {author} {\bibinfo {author} {\bibfnamefont {G.~A.}\ \bibnamefont
  {Cs\'athy}}, \bibinfo {author} {\bibfnamefont {J.~S.}\ \bibnamefont {Xia}},
  \bibinfo {author} {\bibfnamefont {C.~L.}\ \bibnamefont {Vicente}}, \bibinfo
  {author} {\bibfnamefont {E.~D.}\ \bibnamefont {Adams}}, \bibinfo {author}
  {\bibfnamefont {N.~S.}\ \bibnamefont {Sullivan}}, \bibinfo {author}
  {\bibfnamefont {H.~L.}\ \bibnamefont {Stormer}}, \bibinfo {author}
  {\bibfnamefont {D.~C.}\ \bibnamefont {Tsui}}, \bibinfo {author}
  {\bibfnamefont {L.~N.}\ \bibnamefont {Pfeiffer}}, \ and\ \bibinfo {author}
  {\bibfnamefont {K.~W.}\ \bibnamefont {West}},\ }\href {\doibase
  10.1103/PhysRevLett.94.146801} {\bibfield  {journal} {\bibinfo  {journal}
  {Phys. Rev. Lett.}\ }\textbf {\bibinfo {volume} {94}},\ \bibinfo {pages}
  {146801} (\bibinfo {year} {2005})}\BibitemShut {NoStop}%
\bibitem [{\citenamefont {Jungwirth}\ \emph {et~al.}(1999)\citenamefont
  {Jungwirth}, \citenamefont {MacDonald}, \citenamefont
  {Smr\ifmmode~\check{c}\else \v{c}\fi{}ka},\ and\ \citenamefont
  {Girvin}}]{girvintilt1999}%
  \BibitemOpen
  \bibfield  {author} {\bibinfo {author} {\bibfnamefont {T.}~\bibnamefont
  {Jungwirth}}, \bibinfo {author} {\bibfnamefont {A.~H.}\ \bibnamefont
  {MacDonald}}, \bibinfo {author} {\bibfnamefont {L.}~\bibnamefont
  {Smr\ifmmode~\check{c}\else \v{c}\fi{}ka}}, \ and\ \bibinfo {author}
  {\bibfnamefont {S.~M.}\ \bibnamefont {Girvin}},\ }\href {\doibase
  10.1103/PhysRevB.60.15574} {\bibfield  {journal} {\bibinfo  {journal} {Phys.
  Rev. B}\ }\textbf {\bibinfo {volume} {60}},\ \bibinfo {pages} {15574}
  (\bibinfo {year} {1999})}\BibitemShut {NoStop}%
\bibitem [{\citenamefont {Maan}(1984)}]{maan1984combined}%
  \BibitemOpen
  \bibfield  {author} {\bibinfo {author} {\bibfnamefont {J.}~\bibnamefont
  {Maan}},\ }in\ \href@noop {} {\emph {\bibinfo {booktitle} {Two-Dimensional
  Systems, Heterostructures, and Superlattices}}}\ (\bibinfo  {publisher}
  {Springer, Berlin},\ \bibinfo {year} {1984})\ pp.\ \bibinfo {pages}
  {183--191}\BibitemShut {NoStop}%
\bibitem [{\citenamefont {Halonen}\ \emph {et~al.}(1990)\citenamefont
  {Halonen}, \citenamefont {Pietil{\"a}inen},\ and\ \citenamefont
  {Chakraborty}}]{halonen1990subband}%
  \BibitemOpen
  \bibfield  {author} {\bibinfo {author} {\bibfnamefont {V.}~\bibnamefont
  {Halonen}}, \bibinfo {author} {\bibfnamefont {P.}~\bibnamefont
  {Pietil{\"a}inen}}, \ and\ \bibinfo {author} {\bibfnamefont {T.}~\bibnamefont
  {Chakraborty}},\ }\href@noop {} {\bibfield  {journal} {\bibinfo  {journal}
  {Physical Review B}\ }\textbf {\bibinfo {volume} {41}},\ \bibinfo {pages}
  {10202} (\bibinfo {year} {1990})}\BibitemShut {NoStop}%
\bibitem [{\citenamefont {Wang}\ \emph {et~al.}(2003)\citenamefont {Wang},
  \citenamefont {Demler},\ and\ \citenamefont {Sarma}}]{wang2003spontaneous}%
  \BibitemOpen
  \bibfield  {author} {\bibinfo {author} {\bibfnamefont {D.-W.}\ \bibnamefont
  {Wang}}, \bibinfo {author} {\bibfnamefont {E.}~\bibnamefont {Demler}}, \ and\
  \bibinfo {author} {\bibfnamefont {S.~D.}\ \bibnamefont {Sarma}},\ }\href@noop
  {} {\bibfield  {journal} {\bibinfo  {journal} {Physical Review B}\ }\textbf
  {\bibinfo {volume} {68}},\ \bibinfo {pages} {165303} (\bibinfo {year}
  {2003})}\BibitemShut {NoStop}%
\bibitem [{\citenamefont {Papi{\'c}}(2013)}]{papic2013fractional}%
  \BibitemOpen
  \bibfield  {author} {\bibinfo {author} {\bibfnamefont {Z.}~\bibnamefont
  {Papi{\'c}}},\ }\href@noop {} {\bibfield  {journal} {\bibinfo  {journal}
  {Physical Review B}\ }\textbf {\bibinfo {volume} {87}},\ \bibinfo {pages}
  {245315} (\bibinfo {year} {2013})}\BibitemShut {NoStop}%
\bibitem [{\citenamefont {Eisenstein}\ \emph {et~al.}(2000)\citenamefont
  {Eisenstein}, \citenamefont {Lilly}, \citenamefont {Cooper}, \citenamefont
  {Pfeiffer},\ and\ \citenamefont {West}}]{eisensteinhigher}%
  \BibitemOpen
  \bibfield  {author} {\bibinfo {author} {\bibfnamefont {J.}~\bibnamefont
  {Eisenstein}}, \bibinfo {author} {\bibfnamefont {M.}~\bibnamefont {Lilly}},
  \bibinfo {author} {\bibfnamefont {K.}~\bibnamefont {Cooper}}, \bibinfo
  {author} {\bibfnamefont {L.}~\bibnamefont {Pfeiffer}}, \ and\ \bibinfo
  {author} {\bibfnamefont {K.}~\bibnamefont {West}},\ }\href {\doibase
  https://doi.org/10.1016/S1386-9477(99)00043-0} {\bibfield  {journal}
  {\bibinfo  {journal} {Physica E: Low-dimensional Systems and Nanostructures}\
  }\textbf {\bibinfo {volume} {6}},\ \bibinfo {pages} {29 } (\bibinfo {year}
  {2000})}\BibitemShut {NoStop}%
\bibitem [{\citenamefont {Xia}\ \emph {et~al.}(2010)\citenamefont {Xia},
  \citenamefont {Cvicek}, \citenamefont {Eisenstein}, \citenamefont
  {Pfeiffer},\ and\ \citenamefont {West}}]{xia52tilt}%
  \BibitemOpen
  \bibfield  {author} {\bibinfo {author} {\bibfnamefont {J.}~\bibnamefont
  {Xia}}, \bibinfo {author} {\bibfnamefont {V.}~\bibnamefont {Cvicek}},
  \bibinfo {author} {\bibfnamefont {J.~P.}\ \bibnamefont {Eisenstein}},
  \bibinfo {author} {\bibfnamefont {L.~N.}\ \bibnamefont {Pfeiffer}}, \ and\
  \bibinfo {author} {\bibfnamefont {K.~W.}\ \bibnamefont {West}},\ }\href
  {\doibase 10.1103/PhysRevLett.105.176807} {\bibfield  {journal} {\bibinfo
  {journal} {Phys. Rev. Lett.}\ }\textbf {\bibinfo {volume} {105}},\ \bibinfo
  {pages} {176807} (\bibinfo {year} {2010})}\BibitemShut {NoStop}%
\bibitem [{\citenamefont {Yang}\ \emph {et~al.}(2017)\citenamefont {Yang},
  \citenamefont {Lee}, \citenamefont {Zhang},\ and\ \citenamefont
  {Hu}}]{yang2017anisotropic}%
  \BibitemOpen
  \bibfield  {author} {\bibinfo {author} {\bibfnamefont {B.}~\bibnamefont
  {Yang}}, \bibinfo {author} {\bibfnamefont {C.~H.}\ \bibnamefont {Lee}},
  \bibinfo {author} {\bibfnamefont {C.}~\bibnamefont {Zhang}}, \ and\ \bibinfo
  {author} {\bibfnamefont {Z.-X.}\ \bibnamefont {Hu}},\ }\href@noop {}
  {\bibfield  {journal} {\bibinfo  {journal} {Physical Review B}\ }\textbf
  {\bibinfo {volume} {96}},\ \bibinfo {pages} {195140} (\bibinfo {year}
  {2017})}\BibitemShut {NoStop}%
\bibitem [{\citenamefont {Gromov}\ and\ \citenamefont
  {Son}(2017)}]{gromov2017bimetric}%
  \BibitemOpen
  \bibfield  {author} {\bibinfo {author} {\bibfnamefont {A.}~\bibnamefont
  {Gromov}}\ and\ \bibinfo {author} {\bibfnamefont {D.~T.}\ \bibnamefont
  {Son}},\ }\href@noop {} {\bibfield  {journal} {\bibinfo  {journal} {Physical
  Review X}\ }\textbf {\bibinfo {volume} {7}},\ \bibinfo {pages} {041032}
  (\bibinfo {year} {2017})}\BibitemShut {NoStop}%
\bibitem [{\citenamefont {Forster}(1975)}]{forster1975hydrodynamic}%
  \BibitemOpen
  \bibfield  {author} {\bibinfo {author} {\bibfnamefont {D.}~\bibnamefont
  {Forster}},\ }in\ \href@noop {} {\emph {\bibinfo {booktitle} {Reading, Mass.,
  WA Benjamin, Inc.(Frontiers in Physics. Volume 47), 1975. 343 p.}}},\
  Vol.~\bibinfo {volume} {47}\ (\bibinfo {year} {1975})\BibitemShut {NoStop}%
\bibitem [{\citenamefont {Bradlyn}(2015)}]{bradlyn2015linear}%
  \BibitemOpen
  \bibfield  {author} {\bibinfo {author} {\bibfnamefont {B.~J.}\ \bibnamefont
  {Bradlyn}},\ }\href@noop {} {\emph {\bibinfo {title} {Linear response and
  Berry curvature in two-dimensional topological phases}}}\ (\bibinfo
  {publisher} {Yale University, New Haven},\ \bibinfo {year}
  {2015})\BibitemShut {NoStop}%
\bibitem [{\citenamefont {Gromov}\ and\ \citenamefont
  {Abanov}(2014)}]{abanov2014}%
  \BibitemOpen
  \bibfield  {author} {\bibinfo {author} {\bibfnamefont {A.}~\bibnamefont
  {Gromov}}\ and\ \bibinfo {author} {\bibfnamefont {A.}~\bibnamefont
  {Abanov}},\ }\href {\doibase 10.1103/PhysRevLett.113.046803} {\bibfield
  {journal} {\bibinfo  {journal} {Phys. Rev. Lett.}\ }\textbf {\bibinfo
  {volume} {113}},\ \bibinfo {pages} {266802} (\bibinfo {year}
  {2014})}\BibitemShut {NoStop}%
\bibitem [{\citenamefont {Fetter}\ and\ \citenamefont
  {Walecka}(2012)}]{fetter2012quantum}%
  \BibitemOpen
  \bibfield  {author} {\bibinfo {author} {\bibfnamefont {A.~L.}\ \bibnamefont
  {Fetter}}\ and\ \bibinfo {author} {\bibfnamefont {J.~D.}\ \bibnamefont
  {Walecka}},\ }\href@noop {} {\emph {\bibinfo {title} {Quantum theory of
  many-particle systems}}}\ (\bibinfo  {publisher} {Courier Corporation,
  Mineola NY},\ \bibinfo {year} {2012})\BibitemShut {NoStop}%
\bibitem [{\citenamefont {Gromov}(2015)}]{gromovtalk}%
  \BibitemOpen
  \bibfield  {author} {\bibinfo {author} {\bibfnamefont {A.}~\bibnamefont
  {Gromov}},\ }\href {url link to talk abstract if any} {\enquote {\bibinfo
  {title} {Geometric response at the edge},}\ } (\bibinfo {year} {2015}),\
  \bibinfo {note} {presented at \emph{Geometric Aspects of the Quantum Hall
  Effect}, Cologne}\BibitemShut {NoStop}%
\bibitem [{\citenamefont {Ganeshan}\ and\ \citenamefont
  {Abanov}(2017)}]{sriramhydro}%
  \BibitemOpen
  \bibfield  {author} {\bibinfo {author} {\bibfnamefont {S.}~\bibnamefont
  {Ganeshan}}\ and\ \bibinfo {author} {\bibfnamefont {A.~G.}\ \bibnamefont
  {Abanov}},\ }\href {\doibase 10.1103/PhysRevFluids.2.094101} {\bibfield
  {journal} {\bibinfo  {journal} {Phys. Rev. Fluids}\ }\textbf {\bibinfo
  {volume} {2}},\ \bibinfo {pages} {094101} (\bibinfo {year}
  {2017})}\BibitemShut {NoStop}%
\bibitem [{\citenamefont {Niu}\ \emph {et~al.}(1985)\citenamefont {Niu},
  \citenamefont {Thouless},\ and\ \citenamefont {Wu}}]{niu1985quantized}%
  \BibitemOpen
  \bibfield  {author} {\bibinfo {author} {\bibfnamefont {Q.}~\bibnamefont
  {Niu}}, \bibinfo {author} {\bibfnamefont {D.~J.}\ \bibnamefont {Thouless}}, \
  and\ \bibinfo {author} {\bibfnamefont {Y.-S.}\ \bibnamefont {Wu}},\
  }\href@noop {} {\bibfield  {journal} {\bibinfo  {journal} {Physical Review
  B}\ }\textbf {\bibinfo {volume} {31}},\ \bibinfo {pages} {3372} (\bibinfo
  {year} {1985})}\BibitemShut {NoStop}%
\bibitem [{\citenamefont {Bradlyn}\ and\ \citenamefont
  {Read}(2015)}]{bradlyn2014low}%
  \BibitemOpen
  \bibfield  {author} {\bibinfo {author} {\bibfnamefont {B.}~\bibnamefont
  {Bradlyn}}\ and\ \bibinfo {author} {\bibfnamefont {N.}~\bibnamefont {Read}},\
  }\href@noop {} {\bibfield  {journal} {\bibinfo  {journal} {Phys. Rev. B}\
  }\textbf {\bibinfo {volume} {91}},\ \bibinfo {pages} {125303} (\bibinfo
  {year} {2015})}\BibitemShut {NoStop}%
\bibitem [{\citenamefont {Son}(2013)}]{son2013newton}%
  \BibitemOpen
  \bibfield  {author} {\bibinfo {author} {\bibfnamefont {D.~T.}\ \bibnamefont
  {Son}},\ }\href@noop {} {\bibfield  {journal} {\bibinfo  {journal} {arXiv
  preprint arXiv:1306.0638}\ } (\bibinfo {year} {2013})}\BibitemShut {NoStop}%
\bibitem [{\citenamefont {Luttinger}(1964)}]{luttinger1964}%
  \BibitemOpen
  \bibfield  {author} {\bibinfo {author} {\bibfnamefont {J.}~\bibnamefont
  {Luttinger}},\ }\href {\doibase 10.1103/PhysRev.135.A1505} {\bibfield
  {journal} {\bibinfo  {journal} {Phys. Rev.}\ }\textbf {\bibinfo {volume}
  {135}},\ \bibinfo {pages} {1505} (\bibinfo {year} {1964})}\BibitemShut
  {NoStop}%
\bibitem [{Note1()}]{Note1}%
  \BibitemOpen
  \bibinfo {note} {We follow the standard terminology here, where indices are
  raised and lowered with the metric $g_{\mu \nu }$}\BibitemShut {NoStop}%
\bibitem [{\citenamefont {Bertlmann}(2000)}]{bertlmann2000anomalies}%
  \BibitemOpen
  \bibfield  {author} {\bibinfo {author} {\bibfnamefont {R.~A.}\ \bibnamefont
  {Bertlmann}},\ }\href@noop {} {\emph {\bibinfo {title} {Anomalies in quantum
  field theory}}},\ Vol.~\bibinfo {volume} {91}\ (\bibinfo  {publisher} {Oxford
  University Press, Oxford},\ \bibinfo {year} {2000})\BibitemShut {NoStop}%
\bibitem [{\citenamefont {Parodi}(1970)}]{parodi}%
  \BibitemOpen
  \bibfield  {author} {\bibinfo {author} {\bibfnamefont {O.}~\bibnamefont
  {Parodi}},\ }\href@noop {} {\bibfield  {journal} {\bibinfo  {journal}
  {Journal de Physique}\ }\textbf {\bibinfo {volume} {31}},\ \bibinfo {pages}
  {581} (\bibinfo {year} {1970})}\BibitemShut {NoStop}%
\bibitem [{\citenamefont {Banerjee}\ \emph {et~al.}(2017)\citenamefont
  {Banerjee}, \citenamefont {Souslov}, \citenamefont {Abanov},\ and\
  \citenamefont {Vitelli}}]{vitelli2017odd}%
  \BibitemOpen
  \bibfield  {author} {\bibinfo {author} {\bibfnamefont {D.}~\bibnamefont
  {Banerjee}}, \bibinfo {author} {\bibfnamefont {A.}~\bibnamefont {Souslov}},
  \bibinfo {author} {\bibfnamefont {A.~G.}\ \bibnamefont {Abanov}}, \ and\
  \bibinfo {author} {\bibfnamefont {V.}~\bibnamefont {Vitelli}},\ }\href@noop
  {} {\bibfield  {journal} {\bibinfo  {journal} {Nature Communications}\
  }\textbf {\bibinfo {volume} {8}},\ \bibinfo {pages} {1573} (\bibinfo {year}
  {2017})}\BibitemShut {NoStop}%
\bibitem [{\citenamefont {Yang}\ \emph {et~al.}(2012)\citenamefont {Yang},
  \citenamefont {Papi{\'c}}, \citenamefont {Rezayi}, \citenamefont {Bhatt},\
  and\ \citenamefont {Haldane}}]{yang2012band}%
  \BibitemOpen
  \bibfield  {author} {\bibinfo {author} {\bibfnamefont {B.}~\bibnamefont
  {Yang}}, \bibinfo {author} {\bibfnamefont {Z.}~\bibnamefont {Papi{\'c}}},
  \bibinfo {author} {\bibfnamefont {E.}~\bibnamefont {Rezayi}}, \bibinfo
  {author} {\bibfnamefont {R.}~\bibnamefont {Bhatt}}, \ and\ \bibinfo {author}
  {\bibfnamefont {F.~D.~M.}\ \bibnamefont {Haldane}},\ }\href@noop {}
  {\bibfield  {journal} {\bibinfo  {journal} {Physical Review B}\ }\textbf
  {\bibinfo {volume} {85}},\ \bibinfo {pages} {165318} (\bibinfo {year}
  {2012})}\BibitemShut {NoStop}%
\bibitem [{\citenamefont {Qiu}\ \emph {et~al.}(2012)\citenamefont {Qiu},
  \citenamefont {Haldane}, \citenamefont {Wan}, \citenamefont {Yang},\ and\
  \citenamefont {Yi}}]{qiu2012model}%
  \BibitemOpen
  \bibfield  {author} {\bibinfo {author} {\bibfnamefont {R.-Z.}\ \bibnamefont
  {Qiu}}, \bibinfo {author} {\bibfnamefont {F.~D.~M.}\ \bibnamefont {Haldane}},
  \bibinfo {author} {\bibfnamefont {X.}~\bibnamefont {Wan}}, \bibinfo {author}
  {\bibfnamefont {K.}~\bibnamefont {Yang}}, \ and\ \bibinfo {author}
  {\bibfnamefont {S.}~\bibnamefont {Yi}},\ }\href@noop {} {\bibfield  {journal}
  {\bibinfo  {journal} {Physical Review B}\ }\textbf {\bibinfo {volume} {85}},\
  \bibinfo {pages} {115308} (\bibinfo {year} {2012})}\BibitemShut {NoStop}%
\bibitem [{\citenamefont {Adachi}(1985)}]{adachi1985gaas}%
  \BibitemOpen
  \bibfield  {author} {\bibinfo {author} {\bibfnamefont {S.}~\bibnamefont
  {Adachi}},\ }\href@noop {} {\bibfield  {journal} {\bibinfo  {journal}
  {Journal of Applied Physics}\ }\textbf {\bibinfo {volume} {58}},\ \bibinfo
  {pages} {R1} (\bibinfo {year} {1985})}\BibitemShut {NoStop}%
\bibitem [{\citenamefont {Dingle}\ \emph {et~al.}(1974)\citenamefont {Dingle},
  \citenamefont {Wiegmann},\ and\ \citenamefont {Henry}}]{gaaswell}%
  \BibitemOpen
  \bibfield  {author} {\bibinfo {author} {\bibfnamefont {R.}~\bibnamefont
  {Dingle}}, \bibinfo {author} {\bibfnamefont {W.}~\bibnamefont {Wiegmann}}, \
  and\ \bibinfo {author} {\bibfnamefont {C.~H.}\ \bibnamefont {Henry}},\ }\href
  {\doibase 10.1103/PhysRevLett.33.827} {\bibfield  {journal} {\bibinfo
  {journal} {Phys. Rev. Lett.}\ }\textbf {\bibinfo {volume} {33}},\ \bibinfo
  {pages} {827} (\bibinfo {year} {1974})}\BibitemShut {NoStop}%
\bibitem [{\citenamefont {Klitzing}\ \emph {et~al.}(1980)\citenamefont
  {Klitzing}, \citenamefont {Dorda},\ and\ \citenamefont
  {Pepper}}]{klitzing1980new}%
  \BibitemOpen
  \bibfield  {author} {\bibinfo {author} {\bibfnamefont {K.~v.}\ \bibnamefont
  {Klitzing}}, \bibinfo {author} {\bibfnamefont {G.}~\bibnamefont {Dorda}}, \
  and\ \bibinfo {author} {\bibfnamefont {M.}~\bibnamefont {Pepper}},\
  }\href@noop {} {\bibfield  {journal} {\bibinfo  {journal} {Physical Review
  Letters}\ }\textbf {\bibinfo {volume} {45}},\ \bibinfo {pages} {494}
  (\bibinfo {year} {1980})}\BibitemShut {NoStop}%
\bibitem [{\citenamefont {Tsui}\ \emph {et~al.}(1982)\citenamefont {Tsui},
  \citenamefont {Stormer},\ and\ \citenamefont {Gossard}}]{tsui1982two}%
  \BibitemOpen
  \bibfield  {author} {\bibinfo {author} {\bibfnamefont {D.~C.}\ \bibnamefont
  {Tsui}}, \bibinfo {author} {\bibfnamefont {H.~L.}\ \bibnamefont {Stormer}}, \
  and\ \bibinfo {author} {\bibfnamefont {A.~C.}\ \bibnamefont {Gossard}},\
  }\href@noop {} {\bibfield  {journal} {\bibinfo  {journal} {Physical Review
  Letters}\ }\textbf {\bibinfo {volume} {48}},\ \bibinfo {pages} {1559}
  (\bibinfo {year} {1982})}\BibitemShut {NoStop}%
\bibitem [{\citenamefont {Zak}(1964)}]{zak1964magnetic}%
  \BibitemOpen
  \bibfield  {author} {\bibinfo {author} {\bibfnamefont {J.}~\bibnamefont
  {Zak}},\ }\href@noop {} {\bibfield  {journal} {\bibinfo  {journal} {Physical
  Review}\ }\textbf {\bibinfo {volume} {134}},\ \bibinfo {pages} {A1602}
  (\bibinfo {year} {1964})}\BibitemShut {NoStop}%
\bibitem [{Note2()}]{Note2}%
  \BibitemOpen
  \bibinfo {note} {Strictly speaking, all we can deduce is that the degeneracy
  of each level is an integer multiple of $M$, since $W_x$ may have degenerate
  eigenstates. However, in the current case of interest, the eigenstates of
  $W_x$ will be nondegenerate.}\BibitemShut {Stop}%
\bibitem [{\citenamefont {Fradkin}(2013)}]{fradkin2013field}%
  \BibitemOpen
  \bibfield  {author} {\bibinfo {author} {\bibfnamefont {E.}~\bibnamefont
  {Fradkin}},\ }\href@noop {} {\emph {\bibinfo {title} {Field theories of
  condensed matter physics}}}\ (\bibinfo  {publisher} {Cambridge University
  Press, Cambridge},\ \bibinfo {year} {2013})\BibitemShut {NoStop}%
\bibitem [{\citenamefont {Landsteiner}\ \emph {et~al.}(2016)\citenamefont
  {Landsteiner}, \citenamefont {Liu},\ and\ \citenamefont
  {Sun}}]{landsteiner-weyl-visc}%
  \BibitemOpen
  \bibfield  {author} {\bibinfo {author} {\bibfnamefont {K.}~\bibnamefont
  {Landsteiner}}, \bibinfo {author} {\bibfnamefont {Y.}~\bibnamefont {Liu}}, \
  and\ \bibinfo {author} {\bibfnamefont {Y.-W.}\ \bibnamefont {Sun}},\ }\href
  {\doibase 10.1103/PhysRevLett.117.081604} {\bibfield  {journal} {\bibinfo
  {journal} {Phys. Rev. Lett.}\ }\textbf {\bibinfo {volume} {117}},\ \bibinfo
  {pages} {081604} (\bibinfo {year} {2016})}\BibitemShut {NoStop}%
\bibitem [{\citenamefont {Arjona}\ and\ \citenamefont
  {Vozmediano}(2018)}]{vozmediano-rotational-strain}%
  \BibitemOpen
  \bibfield  {author} {\bibinfo {author} {\bibfnamefont {V.}~\bibnamefont
  {Arjona}}\ and\ \bibinfo {author} {\bibfnamefont {M.~A.~H.}\ \bibnamefont
  {Vozmediano}},\ }\href {\doibase 10.1103/PhysRevB.97.201404} {\bibfield
  {journal} {\bibinfo  {journal} {Phys. Rev. B}\ }\textbf {\bibinfo {volume}
  {97}},\ \bibinfo {pages} {201404} (\bibinfo {year} {2018})}\BibitemShut
  {NoStop}%
\bibitem [{\citenamefont {Gromov}\ \emph {et~al.}(2016)\citenamefont {Gromov},
  \citenamefont {Jensen},\ and\ \citenamefont {Abanov}}]{gromov2016boundary}%
  \BibitemOpen
  \bibfield  {author} {\bibinfo {author} {\bibfnamefont {A.}~\bibnamefont
  {Gromov}}, \bibinfo {author} {\bibfnamefont {K.}~\bibnamefont {Jensen}}, \
  and\ \bibinfo {author} {\bibfnamefont {A.~G.}\ \bibnamefont {Abanov}},\
  }\href@noop {} {\bibfield  {journal} {\bibinfo  {journal} {Physical review
  letters}\ }\textbf {\bibinfo {volume} {116}},\ \bibinfo {pages} {126802}
  (\bibinfo {year} {2016})}\BibitemShut {NoStop}%
\bibitem [{\citenamefont {Simon}(1983)}]{simon1983homotopy}%
  \BibitemOpen
  \bibfield  {author} {\bibinfo {author} {\bibfnamefont {B.}~\bibnamefont
  {Simon}},\ }\href {\doibase 10.1103/PhysRevLett.51.2167} {\bibfield
  {journal} {\bibinfo  {journal} {Phys. Rev. Lett.}\ }\textbf {\bibinfo
  {volume} {51}},\ \bibinfo {pages} {2167} (\bibinfo {year}
  {1983})}\BibitemShut {NoStop}%
\bibitem [{\citenamefont {Avron}(1998{\natexlab{b}})}]{avron1995adiabatic}%
  \BibitemOpen
  \bibfield  {author} {\bibinfo {author} {\bibfnamefont {J.~E.}\ \bibnamefont
  {Avron}},\ }in\ \href
  {http://phsites.technion.ac.il/avron/wp-content/uploads/sites/3/2013/08/leshouches.pdf}
  {\emph {\bibinfo {booktitle} {Quantum Symmetries/Symetries Quantiques:
  Proceedings of the Les Houches Summer School, Session Lxiv, Les Houches,
  France, 1 August - 8 September, 1995 (Les Houches Summer School
  proceedings)}}}\ (\bibinfo  {publisher} {Elsevier, Amsterdam},\ \bibinfo
  {year} {1998})\ \bibinfo {note}
  {http://phsites.technion.ac.il/avron/wp-content/uploads/sites/3/2013/08/leshouches.pdf}\BibitemShut
  {NoStop}%
\bibitem [{\citenamefont {Limtragool}\ and\ \citenamefont
  {Phillips}(2016)}]{limtragool2016anomalous}%
  \BibitemOpen
  \bibfield  {author} {\bibinfo {author} {\bibfnamefont {K.}~\bibnamefont
  {Limtragool}}\ and\ \bibinfo {author} {\bibfnamefont {P.~W.}\ \bibnamefont
  {Phillips}},\ }\href@noop {} {\bibfield  {journal} {\bibinfo  {journal}
  {arXiv preprint arXiv:1601.02340}\ } (\bibinfo {year} {2016})}\BibitemShut
  {NoStop}%
\bibitem [{\citenamefont {Pauli}(1980)}]{pauli_1980}%
  \BibitemOpen
  \bibfield  {author} {\bibinfo {author} {\bibfnamefont {W.}~\bibnamefont
  {Pauli}},\ }\href@noop {} {\emph {\bibinfo {title} {General principles of
  quantum mechanics}}}\ (\bibinfo  {publisher} {Springer, Berlin},\ \bibinfo
  {year} {1980})\BibitemShut {NoStop}%
\bibitem [{\citenamefont {Link}\ \emph {et~al.}(2018)\citenamefont {Link},
  \citenamefont {Sheehy}, \citenamefont {Narozhny},\ and\ \citenamefont
  {Schmalian}}]{link2018elastic}%
  \BibitemOpen
  \bibfield  {author} {\bibinfo {author} {\bibfnamefont {J.~M.}\ \bibnamefont
  {Link}}, \bibinfo {author} {\bibfnamefont {D.~E.}\ \bibnamefont {Sheehy}},
  \bibinfo {author} {\bibfnamefont {B.~N.}\ \bibnamefont {Narozhny}}, \ and\
  \bibinfo {author} {\bibfnamefont {J.}~\bibnamefont {Schmalian}},\ }\href@noop
  {} {\bibfield  {journal} {\bibinfo  {journal} {Physical Review B}\ }\textbf
  {\bibinfo {volume} {98}},\ \bibinfo {pages} {195103} (\bibinfo {year}
  {2018})}\BibitemShut {NoStop}%
\end{thebibliography}%
\end{document}